\documentclass{cimento}

\usepackage{amssymb}
\usepackage{amsmath,amsfonts,amsthm,bm,amssymb,comment} 

\usepackage{graphicx}  
\usepackage{appendix}
\usepackage{xcolor}
\newtheorem{theorem}{Theorem}

\newcommand{\dif}{\mathrm{d}}
\newcommand{\mathd}{\mathrm{d}}

\title{Applications of large deviation theory\\ in geophysical fluid dynamics and climate science}
\shorttitle{Large deviation theory in GFD and climate science}
\author{V.M.~Galfi\from{ins:1},\from{ins:1b}\ETC,
V.~Lucarini\from{ins:2}\from{ins:4},
F.~Ragone\from{ins:3}\from{ins:5}
        \atque
J.~Wouters\from{ins:2}\from{ins:4}}
\instlist{\inst{ins:1} Meteorological Institute, CEN, University of Hamburg, Hamburg, Germany
\inst{ins:1b} Department of Earth Sciences, Uppsala University, Uppsala, Sweden
  \inst{ins:2} Department of Mathematics and Statistics, University of Reading,  Reading, UK
    \inst{ins:4} Centre for the Mathematics of Planet Earth, University of Reading, Reading, UK
  \inst{ins:3} Georges Lemaitre Centre for Earth and Climate Research, Earth and Life Institute, Universit\'e catholique de Louvain, Louvain-la-Neuve, Belgium
    \inst{ins:5} Royal Meteorological Institute of Belgium, Bruxelles, Belgium}

\begin{document}

\maketitle

\begin{abstract}
The climate system is a complex, chaotic system with many degrees of freedom and variability on a vast range of temporal and spatial scales. Attaining a deeper level of understanding of its dynamical processes is a scientific challenge of great urgency, especially given the ongoing climate change and the evolving climate crisis. 
In statistical physics, complex, many-particle systems are studied successfully using the mathematical framework of Large Deviation Theory (LDT). A great potential exists for applying LDT to problems relevant for geophysical fluid dynamics and climate science, both of more fundamental and of more applied nature. In particular, LDT allows for understanding the fundamental properties of persistent deviations of climatic fields from the long-term averages and for associating them to low-frequency, large scale patterns of climatic variability. Additionally, LDT can be used in conjunction with so-called rare events algorithms to explore rarely visited regions of the phase space and thus to study special dynamical configurations of the climate. These applications are of key importance to improve our understanding of high-impact weather and climate events. Furthermore, LDT provides powerful tools for evaluating the probability of noise-induced transitions between competing metastable states of the climate system or of its components. This in turn essential for improving our understanding of the global stability properties of the climate system and of its predictability of the second kind in the sense of Lorenz.   
The goal of this review is manifold. First, we want to provide an introduction to the derivation of large deviation laws  in the context of stochastic processes. We then relate such results to the existing literature showing the current status of applications of LDT in climate science and geophysical fluid dynamics. Finally, we propose some possible lines of future investigations. We hope that this paper will set the ground for a series of future studies applying LDT to solve problems encountered in climate science and geophysical fluid dynamics.
\end{abstract}

\tableofcontents

\section{Introduction and motivation}
\subsection{{\color{black}The Climate Crisis: Extreme Events in a Changing Climate}}\label{sec:intro}
The climate is a forced and dissipative nonlinear heterogeneous system composed by several subdomains, namely the atmosphere, the hydrosphere, the cryosphere, the soil, and the biosphere. The climate evolves under the action of a primary forcing given by the incoming solar radiation and modulating factors such as the atmospheric composition, the optical properties of the surface of the planet,  gravity, and the rotation of the Earth around its vertical axis. Each of these subsystems features complex nonlinear physical and chemical processes, and the various subsystems interact among themselves through exchanges of energy, momentum, and chemical species. As a result of the interplay between forcing, dissipation, and internal nonlinear dynamics, the climate system  features variability of a vast range of spatial and temporal scales. The climate  can be seen as a prominent example of nonequilibrium system where an approximate steady state is reached as the inhomogeneous absorption of solar radiation occurring throughout its domain is compensated by a variety of physical mechanisms, including thermal emission in the infrared and complex patterns of transport of sensible and latent heat. 
{\color{black} 
Lorenz \cite{Lorenz1967} provided a first comprehensive theory of the dynamics and thermodynamics of climate able to bring together the main mechanisms of forcing, energy conversion, and dissipation. The large-scale flows of the ocean and of the atmosphere ultimately result from the conversion of available potential into kinetic energy performed by the climatic engine. The conversion takes place through various mechanisms of instability fuelled by the presence of spatial temperature gradients. Such instabilities allow for energy conversion between the background state and the fluctuations of the climatic field and lead to chaotic conditions that are associated with heterogenous turbulence in the geophysical fluids. Additionally, these instabilities establish negative feedbacks, because they tend to reduce, via transport and mixing, the temperature gradients that support them. See \cite{Peixoto1992,Lucarini2014} for an extensive discussion of these mechanism. } We remark that an exact steady state is never achieved because of the fluctuations in the incoming solar radiation and in the processes, both natural and anthropogenic, that alter the atmospheric composition and the surface of the planet \cite{IPCC13,Ghil2020}.

Improving our understanding of the dynamical and statistical properties of the climate system, of the links between its response to anthropogenic and natural forcings and of its natural variability is key to provide scientific tools for anticipating, predicting, and possibly addressing the ongoing \textit{climate crisis}. The current popularity of the expression \textit{climate crisis} as opposed to - the more usual one - \textit{climate change} is motivated by the desire to focus on the understanding of how changing climate conditions will unfold as variations in the higher moments of the distribution of the climatic variables, able to better capture the properties of extreme events and tipping points \cite{Ghil2020}. 

Indeed, the study of extreme events is essential for addressing the natural hazards associated with  climate variability and climate change and affecting in a potentially catastrophic way human and environmental welfare. As the resilience of any system and the incurred damages due to an unusual, high-impact event change drastically when certain thresholds in the intensity and/or in the duration of the hazard are reached, it is clear that understanding the fate of extreme events in the context of the changing climate is essential for accurately factoring in future losses and damages and prepare for them \cite{Ipcc2012}. High-impact events affecting human and environmental welfare are sometimes associated with the presence of long temporal persistence of a large anomaly in the field of interest, as resilience  - the ability of any system to resist - against anomalous environmental conditions does not last indefinitely \cite{Easterling2000,Who2004,Poumadere2005}.  
Meteo-climatic extremes characterised by persistence are usually referred to as slow onset events, as opposed to the fast onset ones, which, instead, are associated with fast processes.
Prominent examples of slow onset events are droughts, heatwaves, and cold spells, while flash floods and intense snowfalls lead to hazards in the category of fast onset events \cite{Ipcc2012}. {\color{black}We remark that it might be worth considering a revision of such a classical terminology, because of the multiscale nature of some of the so-called slow onset events: the onset of a blocking responsible for a heatwave takes usually up to a couple of days, while its duration can be much much longer, and ranging up to several weeks. Also the exit from long lasting events can be quite rapid and, anyhow, considerably  shorter than the duration \cite{Tibaldi1990,Tibaldi2018}.}

{\color{black}An important remark is also needed regarding the problem of defining persistent extreme events; we refer here, for the sake of clarity, to the specific case of heatwaves. While the general understanding is that a heatwave is a period of extreme and unusual warmth, there is no rigorous nor commonly accepted definition for it in terms of intensity and persistence of the anomalous weather conditions, despite several attempts in this direction; see, e.g. \cite{Robinson2001}. As noted in \cite{Perkins2015}, \textit{... it seems that almost, if not every climatological study that looks at heatwaves uses a different metric.}; see also \cite{Smith2013,mccarthy2019} for a discussion on the lack of consensus for a shared definition of heatwaves. The confusion around the definition of heatwave has serious implications as it hinders attempts at mitigating their impacts \cite{WMOWHO2015,Radovic2019,Casanueva2019}.}

Changing climate conditions can lead to dramatic changes in the statistics of 
extreme events. As mentioned above, this is one of the main manifestations of the climate crisis. In the future climate, changes in the statistics of heatwaves are  worrying, as more persistent  and larger temperature fluctuations are possible as a result of changes in the properties of the low frequency variability of the atmosphere and of the properties of the soil. This effect compounds with the trend in the average temperature, leading to a greatly increased risk of such catastrophic events \cite{Seneviratne2006,Coumou2013,Pfleiderer2019}. Climate change additionally leads to a reduced winter  weather variability, as a result of the reduction in the temperature difference between low and high latitudes \cite{IPCC13}. Therefore, the probability of occurrence of cold spells  is likely to greatly decrease \cite{Smith2020}, even if structural changes in the dynamics of the atmosphere can rarely create special conditions that facilitate their occurrence \cite{Kretschmer2018,Cohen2018}. Looking instead at flash floods, consensus exists that they will become more likely and more intense in the future because the  higher retention of water vapour in the atmosphere - made possible by warmer  conditions as a result of the Clausius-Clapeyron relation - compounds with strengthened convective motions, possibly leading to disproportionately enhanced extreme precipitation events in specific locales \cite{Yin2018,Fowler2020}. {\color{black}However, we remark that, due to the complexity of the (microscopic and macroscopic) dynamical and thermodynamical processes involved, the precipitation is not distributed in space and time in direct proportion to the available precipitable water \cite{Yano2021}.}

A somewhat separate research agenda tries, instead, to relate in a direct way climate change and \textit{individual} extreme events, thus blurring the distinction between climate and weather, statistics and analysis of specific case studies. Following the landmark paper \cite{Allen2003}, considerable efforts have been directed at developing tools for assessing to what extent climate change has impacted and is impacting either the frequency, or the likelihood, or the intensity of \textit{individual} extreme events, with the goal of providing the basis for science-based liability for the impacts of such extremes. The scientific debate around extreme events attribution has relevant implications in terms of climate adaptation, risk assessment, public policy, infrastructural design, insurance instruments design, international relations, and even migration policies.  \cite{Jezequel2018,Otto2019,Swain2020,Sippel2020}. 
\subsection{Quest for Universality of Extreme Events}\label{sec:introue}
Empirical frequentist approaches aimed at the study of extremes applied to actual climate records or to the output of climate models are essential for keeping track of the observed events, but face the unavoidable problem of being unable to say anything about the probability of occurrence of out-of-sample events. 
In order to achieve predictive power in a statistical sense - i.e. being able to estimate the probability of occurrence of events that are more extreme than the observed ones - one needs to interpret data through mathematical approaches that provide some form of universality. 

Extreme Value Theory (EVT) provides a powerful framework for studying extreme events in a multitude of applications. It is based on limit theorems mimicking the central limit theorem that allow one, under rather general hypotheses and taking suitable limits, to define universal laws describing the probability of occurrence of events generated according to a given stochastic process above a sufficiently high threshold. Alternatively, one can develop the theory in order to describe the distribution of the maximum of a set of independent and identically distributed stochastic variables in the limit of large sets \cite{Coles2001}. The theory can be adapted for dealing with correlations between the variables \cite{LR98} and for treating the outputs of chaotic dynamical systems \cite{Lucarini2016extremes}, in such a way that deep connections emerge between the statistical properties of suitably defined extremes and the geometry of the attractor of the system \cite{Lucarini2014,Galfi2017,Bodai2017,PonsJSP}. EVT has received a great deal of attention in geosciences  \cite{Felici2007,Yiou2008,Vitolo2009,Ghil2011,Franzke2017,Wang2016,Wehner2020} and is extremely influential especially in hydrology \cite{Katz2002,katz2005statistics,Deidda2013}. In the context of climate dynamics, the analysis of extremes has proved very fruitful for providing a new viewpoint for understanding atmospheric predictability by looking at the recurrence of weather patterns \cite{Faranda2017,Messori2018}.
Persistence, as mentioned above, is a key factor in determining the impact of large climatic fluctuations. 
EVT can deal with time correlations in time series, through the introduction of the extremal index, which allows one to quantify the average size of clusters of consecutive extreme events \cite{Ferro2003,Moloney2019}. The extremal index encodes important information on the dynamics of the system \cite{VaientiJSP}. 

A more direct line to attack the problem of studying persistent extreme events can be taken through the use of Large Deviation Theory (LDT) \cite{varadhanLargeDeviationsApplications1984}. In a nutshell, in one of its most basic formulations, LDT aims at providing limit laws for the average of $n$ (typically identically distributed) stochastic variables, where $n$ is large. Similarly to the case of EVT, a unified approach for LDT can be used on stochastic processes and chaotic dynamical systems \cite{Kifer1990,Young2002}. It is hard to overestimate the importance of LDT in contemporary physics and mathematics \cite{Touchette2009,Vulpiani2014,Dembo1998,Hollander2000}. Establishing a large deviation principle for an observable - see below - leads to gaining predictive power of the process. While in EVT such a power is aimed at being able to predict the probability of occurrence of events larger  than those already observed, in the case of LDT the predictive power is twofold, as it is directed towards predicting the property of occurrence of events that are larger and/or more persistent than the observed ones.  Drawing an example from climate science, EVT is better suited for studying the probability of occurrence of extremely hot days, whereas LDT is better suited for studying the probability of occurrence of heatwaves. 

\begin{figure}
    \centering
    \includegraphics[width=1\textwidth]{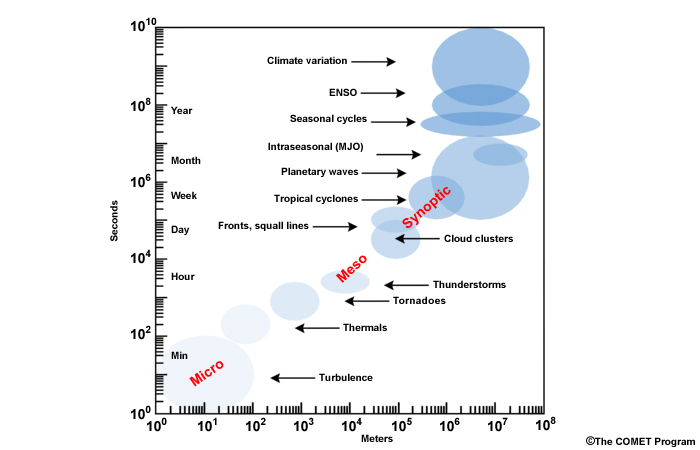}
    \caption{{\color{black}Idealized power spectra
for the atmosphere indicating the relationship between the spatial and temporal scales of atmospheric flows. The source of this material is the COMET$^\copyright$ Website at http://meted.ucar.edu/ of the University Corporation for Atmospheric Research (UCAR), sponsored in part through cooperative agreement(s) with the National Oceanic and Atmospheric Administration (NOAA), U.S. Department of Commerce (DOC). $\copyright$1997-2017 University Corporation for Atmospheric Research. All Rights Reserved. }}
    \label{fig:scales}
\end{figure}

A fascinating aspect of looking at the properties of long time-averages of climatic fields is the following. The theory of low-frequency variability of the atmosphere indicates that long temporal persistence  and large spatial extent of the anomalous patterns go hand in hand \cite{Ghil2001d,Ghil2020}; {\color{black} see Fig. \ref{fig:scales}. In the mid-latitudes, it is customary and indeed scientific meaningful to distinguish between synoptic variability, due to mid-latitude eastward-moving weather systems and associated with temporal scales of  3-7 days and spatial scales of the order of 1000 Km, and low-frequency variability, whose temporal and spatial scales are typically larger, amounting to 1-3 months and several thousands of Km, respectively. \cite{Ghil2001d,Speranza1983}. The main manifestations of low-frequency variability in the mid-latitudes are the so-called blocking events, which are  persistent, large-scale  departures from the approximately zonally symmetric flow associated
with the presence of large-amplitude, almost-stationary
pressure anomalies \cite{Egger1978,Tibaldi1990,Speranza1983,Lindzen1986,Ghil2001d,Ghil2001d,Tibaldi2018,LucariniG2020}. The difference between synoptic and low-frequency variability is clarified when performing a spectral analysis of the atmospheric fields: the former is associated with eastward propagating waves, while the latter is characterised by stationary or weakly propagating planetary waves \cite{DellAquila2005}. Persistence is key to creating conditions conducive to long-lasting extreme events, and, indeed, it is well-known that the  anomalies of the flow due to occurrence of blockings can lead to long-duration warm  \cite{Dole2011} as well as cold extreme events \cite{Buehler2011}. Given their long time duration and large spatial extent, blockings can lead, in a cascade process, to the onset of extreme events also at considerable geographic distance from the core of the blocking, as in the case of the summer 2010 floods in Pakistan resulting from the large scale flow associated with the blocking - and ensuing heatwave - in Russia \cite{Lau2012}. Advancing our understanding of the low-frequency variability of the atmosphere would be very beneficial because, despite continuous improvements, our ability to perform accurate extended-range (beyond 7–10 days) weather forecast in the mid-latitudes is still limited \cite{Ghil2001d}, and because attaining a convincing representation of the statistical and dynamical properties of blocking events is still challenging for both  numerical weather forecast models \cite{Ferranti2015} and climate models \cite{DAndrea&al2016}.

The hope is that, by focusing on suitably defined large deviations of the atmospheric fields, one could distill information on the  low-frequency variability of the atmosphere}. Roughly speaking, as discussed below, it can be proven rigorously that \textit{any large deviation is realized in the least unlikely of all the unlikely ways} \cite{Hollander2000}. Let’s clarify this important concept using again an example drawn from climate science. Let’s assume that we have established a large deviation law describing the probability of occurrence of heatwaves in a given location. In principle, the corresponding rare events can take place as a result of a variety of large scale atmospheric configurations; see a recent analysis of heatwaves in France \cite{Alvarez2018}. Nonetheless, LDT imposes that, in fact, if we look at \textit{true} extremes, with overwhelming probability the heatwaves we observe will take place, apart from small-scale spatio-temporal fluctuations, as a result of a well-defined large-scale  atmospheric configuration, which is very rare in the standard statistics, but is \textit{typical} \textcolor{black}{if we consider the multitude of possible heatwaves with same intensity}. {\color{black}By typical here we mean that the  probability of the occurrence of a large scale atmospheric pattern that is very close to such a configuration, conditional on the occurrence of heatwave at the reference location, is very high, and gets closer to one as we consider more stringer criteria - in terms of intensity and duration - for the occurrence of (rarer) heatwaves.}
In dynamical terms, one has that selecting events associated with large deviations amounts to considering a very small portion of the phase space. The property above implies that the (rarely occurring) approach to this very special region overwhelmingly occurs through a well-defined set of paths, that are singled out by LDT, even if much unlikelier paths are still possible.

Indeed, looking at the specific case of the catastrophic 2010 Russian heatwave, one does find that the observed extreme event is in some sense typical \cite{Dole2011,Galfi_Lucarini2020}. This does not exclude the possibility of more exotic atmospheric configurations on the scale of Eurasia, but their occurrence is much more unlikely than those, already extremely rare, described by LDT. These exotic events might be interpreted as \textit{dragon kings} \cite{Sornette2012}.   

Of course the possibility of practically using  LDT in a complex and multiscale system like the climate is far from being an obvious task for all possible climatic observables. {\color{black}The mathematical foundations for using LDT in the context of the climate lay on taking into account, on the one hand, its chaoticity, and, on the other hand, the fact that stochastic effects emerge as a result of considering its coarse-grained evolution. Indeed, most of the results we present below are a natural extension of the scientific programme aimed at developing and analysing stochastic climate models pioneered by Hasselmann \cite{Hasselmann1976}; see later developments in \cite{saltzman_dynamical,Majda2001,Imkeller2001,Imkeller2002a,Penland2003}.} Additionally, one needs to take into account that while most LDT results require stationarity of the time series, the climate system is only approximately stationary, because of the periodicity in the solar forcing and the natural and anthropogenic forcings to the atmospheric composition  (e.g. change in greenhouse gases and in aerosols) and to the properties of the land surface (e.g. forest fires; agriculture; deforestation). Therefore, one might need to pre-process the data (e.g. removing the seasonal cycle; removing trends) before being able to apply LDT. Clearly, since the climate is a nonlinear system, the previous pre-processing aimed at removing part of the time-dependence is in principle partly arbitrary and definitely non uniquely defined. Nonetheless, one needs to resort to reasonable pragmatism in treating observational or model-generated data that do not conform exactly to the demands of the mathematical theory, and possibly derive nonetheless useful information, as often in fact done in physical sciences.

Another aspect to be kept in consideration is the presence of serial correlations in the time series of the observables. If one considers, for example, the serial correlation of {\color{black}the anomalies of } the surface temperature (obtained after removing seasonal cycle and long-term trends) somewhere in the middle of a continent, like Central Europe, and the serial correlation of the same observable over an oceanic region, like the North Atlantic (not far away from the first location), one would notice that the strength of the serial correlation is much weaker and the auto-correlation function decays substantially faster over the continent as compared to the oceanic region.
In the latest case, the decay of correlations will be slower than exponential (at least on a vast range of scales), as a result of the presence of long-term memory in the system. {\color{black}Large differences in the heat capacity of land surface vs water, and the dynamical link between  surface waters and deeper levels of the ocean explain such a discrepancy between the two cases. The fact that the same climatic field - anomalies of the  surface temperature - features such fundamentally different properties, in terms of stochasticity, depending on the geographical location of interest} provides a good example of the complexity of the climate system. Note that, as we will discuss below, while in the former case one is able to establish large deviation laws to describe accurately long and persistent temperature fluctuations behind heatwaves and cold spells, LDT will not apply in the latter case.

\subsection{Paths and Transitions}
LDT can be used for different scopes than looking at persistent deviation of fields. Indeed, it provides tools for studying how such special configurations of the climate are dynamically realised. One can use a more general definition of events that encompasses trajectories in the phase space, and adapt LDT to study rare trajectories leading to target extreme events. In this settings the dynamical equations contain a small parameter, describing either a weak noise strength or an inverse time scale separation. Under such conditions the path probabilities collapse onto one single path as the small parameter goes to zero, either the deterministic zero-noise path for weak noise systems or the averaged equation for system with a time scale separation. Also here the principle holds that the unlikely event is reached in the least unlikely way. Such paths, called instanton paths, can be seen as minimizers of an action describing the cost of going against the natural tendency of the system to revert to the limiting path \cite{Freidlin2012}. Take, for example, a particle in a double well potential with weak noise. The particle can transition from one well to another, but in the weak noise limit such transitions will be rare. LDT then gives us not only an approximation of the transition probability, but also of the mean exit time and the transition trajectory.

Such a knowledge can furthermore be used to tackle challenges in numerically sampling unlikely paths to rare events. In rare event simulation methods, a model is dynamically driven in such a way that otherwise very rarely visited paths are overpopulated \cite{rubinoRareEventSimulation2009,bouchet_rare_2019}. This can be done either by manipulating the dynamical equations of the system, or by implementing  genetic algorithms on top of the system, which selectively kill and clone parallel realizations of the model. Hence, such trajectories become statistically tractable without resorting to ultralong numerical integrations. Enriching the statistics, while retaining the correct dynamics, makes it possible to explore the dynamical processes behind the extreme event of interest.

The previously mentioned fact that LDT allows one to select \textit{typical} extreme events is key for interpreting some recent results on so-called rogue waves in the ocean \cite{Didenkulova2011,Slunyaev2011,Adcock2014}. Rogue waves are extremely dangerous hazards impacting the marine and coastal environment, and manifest themselves as hard-to-predict surface waves that can have surprisingly high destructive power and that, apparently, materialize out of nothing  \cite{Nikolkina2011,Didenkulova2020}. A novel viewpoint has been recently proposed for finding a comprehensive theoretical framework on rogue waves, able to generalise earlier theories. The idea has been to use LDT to study the properties of the solutions of the one-dimensional nonlinear Schr\"odinger equation starting from suitably defined random initial conditions constructed in accordance with observations taken from an oceanographic campaign. Both numerical and experimental evidence strongly suggest that rogues waves can be seen as hydrodynamic instantons, whose precursors can be clearly identified, and that can be computed by minimizing a suitably defined action  \cite{Dematteis2018,Dematteis2019}. 

A related area of investigation is the study of - rarely occurring - noise-induced transitions between  metastable states associated with alternative configurations of geophysical flows or actual competing climatic states. In this case, along the lines of the classical Freidlin-Wentzell theory \cite{freidlin1984}, the target region in the phase space for the endpoints of the desired paths is a special portion in the basin boundary separating the competing basins of attraction, which corresponds to a saddle in the classical case of motion in an energy landscape. 

The multistability of the climate system manifests itself both locally and globally. By local we mean that the  difference between the competing metastable states is, in fact, geographically confined and associated with one of the so-called tipping elements \cite{Lenton2008}, representing features of the climate system that can go through critical transitions if forced beyond the point of no return. These include the dieback of the Amazon forest \cite{Boers2017}, the shut-down of the thermohaline circulation of the Atlantic ocean \cite{Rahmstorf2005}, the methane release resulting from the melting of the permafrost \cite{Walter2006}, and the collapse of the atmospheric circulation regime associated to the Indian monsoon \cite{Levermann2009}. 

A hierarchically higher level of multistability is present in the Earth as our planet is well known to have at least two possible steady climatic states in the current and past astronomical configuration, the  warm climate, and a frozen one, termed snowball, which features global glaciation, extremely low temperatures and limited climatic variability. This is confirmed by geological and paleomagnetic evidence \cite{Pierrehumbert2011,Hoffman1998} and well understood in terms of relevant dynamical processes \cite{Budyko1969,sellers1969,Ghil1976,Lucarini2010}. Despite the presence of chaotic dynamics in the competing attractors and of a complex geometrical structure in the basin boundary \cite{Lucarini2017N}, suitable generalizations of the Freidlin-Wentzell theory  
proposed in \cite{Graham1986,Graham1987,Graham1991} allow one to establish large deviation laws able to describe in the weak noise limit the transitions between the competing metastable states. Indeed, one can define a generalised quasipotential, whose local minima correspond to the competing attractors, while the transition paths cross preferentially the basin boundaries in special locations, which are saddles also termed Melancholia states \cite{Lucarini2017N,Lucarini2019,Lucarini2020,Bodai2020basin}. There are good reasons to believe that, in fact, the climate system allows for the presence of additional competing meastable states on top of the warm and snowball climate \cite{Lewis2007,Abbot2011,Brunetti2019}. This leads to a more complex pattern of possible transition paths between them and requires a careful statistical examination when noise is added into the system \cite{Margazoglou2021}. Finally, one can interpret the localised tipping elements described above as being associated with smaller local minima and saddles, which define the multiscale nature of the quasi-potential. Therefore, an adequate use of LDT might be key for making a more careful assessment of the risk coming from irreversible transitions for present-day tipping elements, and then for more precisely evaluating the risk of going beyond the so-called global planetary boundaries \cite{Steffen2015,Steffen2018}.  
	
\subsection{This Review}
 {\color{black}The goals of this paper are to provide an informal mathematical introduction to LDT and then to lead the reader to explore some relevant applications of the theory for analysing properties of geophysical flows and of the climate system. The range of topics covered by this paper is somewhat broader and more targeted to real-life applications as compared to the excellent and more theoretically inclined earlier contribution by Bouchet and Vernaille on the statistical mechanics of two-dimensional and geophysical flows \cite{Bouchet2012}.}
 
 Depending on the observable and on the scales of interest, and specifically on the strength of correlations, one can rely on different stochastic models to approximate the behaviour of climatic observables: independent, identically distributed random variables, Markov chains, dependent sequences. The theoretical overview of LDT presented in Sec. \ref{sec:theo} is organized according to this line of thoughts. Subsequently, Sec. \ref{sec:ldds} {\color{black}introduces the concept of coarse-graining for the dynamics of geophysical flows, presents the general framework of stochastic climate models, and discusses} the establishment of large deviation laws in stochastic and deterministic systems. The analysis of large deviation laws for stochastic dynamical systems will provide key tools for understanding the dynamical and statistical properties of transition paths between competing metastable states and for studying rare paths, rather than just rare events. Instead, the results presented for deterministic dynamical systems will be useful for understanding the reason why Markov chain models are of general interest for modelling the statistical properties of chaotic dynamical systems. Section \ref{sec:appl} will then present a range of applications of LDT in various areas of geophysical fluid dynamics and climate science. We will showcase its use for understanding persistent climatic fluctuations, for characterising the fluctuations of the predictability of geophysical flows on different time scales, for providing a unified viewpoint for the understanding of rogue waves in the ocean, as well as for explaining special dynamical features associated with transitions between competing metastable states, thus mirroring the theoretical framework presented in the previous sections. This sections contains also novel, previously unpublished results. Finally, Sec. \ref{conclusions} presents our conclusions together with a discussion regarding opportunities and challenges for future applications of LDT in climate science.



\section{A Summary of Large Deviation Theory}\label{sec:theo}

In this section, we recapitulate the main elements of LDT for two stochastic models applied often successfully to geophysical data: independent, identically distributed (i.i.d.) random variables and Markov chains, or more generally dependent sequences. This summary is far from being complete and does not make use of much mathematical sophistication either. Hence readers experienced in mathematics are referred to \cite{Hollander2000}, whereas readers versed in physics are referred to \cite{Touchette2009}. These are at the same time the main sources we follow.

\subsection{Independent, identically distributes random variables}\label{sec:theo_iid}
The first basic results of LDT is known as Cram\'{e}r's theorem \cite{Cramer1938} and describes the large deviation behaviour of empirical sample averages $\frac{1}{n}\sum_{i=1}^{n}X_i=\frac{1}{n} S_n$.
\begin{theorem}
Let $(X_i)$ be i.i.d. $\mathbb{R}$-valued random variables with a finite moment generating function in a region around the origin, i.e. $0 \in \mathrm{int} \left( \mathcal{D}_\varphi \right)$ with $ \mathcal{D}_\varphi = \lbrace t \in \mathbb{R} : \varphi(t) := \mathbb{E}[e^{tX_1}] < \infty \rbrace $  
, where $\mathbb{E}[f(X_1)]$ is the expectation value of $f(X_1)$.
Let $S_n=\sum_{i=1}^{n}X_i$. Then, for all $a > \mathbb{E}[X_1]$,
\begin{equation}\label{eq:ldp0}
    \lim_{n \to \infty}\frac{1}{n}\log \mathbb{P}\left(\frac{1}{n}S_n \geq a\right) = -I(a)
\end{equation}
where
\begin{equation}\label{eq:rf0}
    I(z)=\sup_{t \in \mathbb{R}}[zt-\log \varphi(t)]
\end{equation}
\label{th:cramer}
\end{theorem}
According to (\ref{eq:ldp0}), which can be written in the form  $\mathbb{P}\left(S_n/n \geq a\right) \asymp  \exp \left( - n I (a)\right)$\footnote{We have that $a^{\varepsilon} \asymp b^{\varepsilon}$ if $\lim_{\varepsilon
\rightarrow 0} \frac{\ln (a^{\varepsilon})}{\ln (b^{\varepsilon})} = 1$; here  $1/\varepsilon=n$.}, the probability of empirical averages deviating from the mean decays exponentially with the averaging length $n$, as $n$ increases.  If this is the case, we say that we have found a large deviation principle. The speed of decay is described by the rate function $I$. The rate function  in Theorem~\ref{th:cramer} has some important and useful properties, such as
compact level sets, 
lower semi-continuity and convexity on $\mathbb{R}$ 
as well as continuity, strict convexity and smoothness on the interior of $\mathcal{D}_I=\{z \in \mathbb{R}: I(z) < \infty\}$. $I(z) \geq 0$ with equality if and only if $z=\mu$, with $\mu=\mathbb{E}[X_1]$. Thus, the minimum of the rate function is located at the expectation value of the random variable suggesting that the sample averages converge to the expected value, as stated by the the law of large numbers. Furthermore, $I''(\mu)=1/\sigma^2$, the second derivative of the rate function at its minimum is the inverse of the variance of the random variable $X_1$, which goes back to the central limit theorem.

As shown by (\ref{eq:rf0}), the rate function is the Legendre transform of the cumulant generating function $\log \varphi$. We will discuss this relationship in more detail below. Equations (\ref{eq:ldp0}) and (\ref{eq:rf0}) describe two different methods to estimate the rate function in case of applications: a direct method based on the probability density function (pdf) of averages and an indirect one based on the cumulant generating function, as discussed in detail in Section \ref{sec:ldds_ta}.

Considering that the rate function is lower semi-continuous and convex, and attains its unique minimum at the expectation value $\mu$, if $a>\mu$, then $I(z)\geq I(a)$ for all $z \geq a$. Thus, equation (\ref{eq:ldp0}) can be rewritten for $a>\mu$ as
\begin{equation}\label{eq:ldp0.1}
    \lim_{n \to \infty}\frac{1}{n}\log \mathbb{P}\left(\frac{1}{n}S_n \in A\right) = -\inf_{z \in A} I(z) \qquad \mathrm{with\ }  A=[a,\infty).
\end{equation}
Similarly, if, instead, $a<\mu$, one obtains:
\begin{equation}\label{eq:ldp0.1b}
    \lim_{n \to \infty}\frac{1}{n}\log \mathbb{P}\left(\frac{1}{n}S_n \in A\right) = -\inf_{z \in A} I(z) \qquad \mathrm{with\ }  A=(-\infty,a].
\end{equation}
This indicates one of the basic principles of LDT that we have hinted in the introduction. The occurrence of a large deviation \{$\frac{S_n}{n} \in A$\} is closely associated with the specific event corresponding to the lowest value of the rate function $I$ taken in $A$, 
as the probability of this event is exponentially larger than the probability of all the other events compatible with the conditions \{$\frac{S_n}{n} \in A$\}.  
The rate function can then be interpreted as a cost function, and we have that 
\textit{any large deviation is done in the least unlikely of all the unlikely ways} \cite{Hollander2000}.

In the following, we discuss some generalizations of Theorem \ref{th:cramer} by going from large deviations of empirical averages to large deviations of empirical measures. From the more general setting of Cram\'er's theorem we go now to a finite state space, where the i.i.d. random variables $X_1, X_2, ...$ take values in a finite set $X_i \in \Gamma =\{1,...,r\} \subset \mathbb{N}$ and obey the marginal law $\rho=(\rho_s)_{s \in \Gamma}$, $\rho_s > 0$. The empirical measure $L_n=\frac{1}{n}\sum_{i=1}^n \delta_{X_i}$ is a random probability measure on $\Gamma$. We denote the set of probability measures on $\Gamma$ by $M(\Gamma)=\{\nu =(\nu_1,...,\nu_r) \in [0,1]^r: \sum_{s=1}^r\nu_s=1\}$, where the total variation distance between two measures $\mu$ and $\nu$ is defined as $d(\mu,\nu)=\frac{1}{2}\sum_{s=1}^r|\mu_s-\nu_s|$. The following theorem, which goes back to Sanov \cite{Sanov1957}, contains a large deviation law of $L_n$ with respect to $\rho$.

\begin{theorem}\label{th:sanov}
Let ($X_i$) be i.i.d. random variables satisfying the conditions above, and $L_n=\frac{1}{n} \sum_{i_1}^n\delta_{X_i}$. Then, for all $a>0$,
\begin{equation}\label{eq:ldp1}
    \lim_{n \to \infty}\frac{1}{n}\log \mathbb{P}(L_n \in B_a^c(\rho)) = -\inf_{\nu \in B_a^c(\rho)} I_\rho(\nu),
\end{equation}
where $B_a(\rho)=\{\nu \in M(\Gamma): d(\nu,\rho)\leq a\}$, $B_a^c(\rho)=M(\Gamma) \setminus B_a(\rho)$, and
\begin{equation}\label{eq:rf1}
    I_\rho(\nu)=\sum_{s=1}^r\nu_s\log\left(\frac{\nu_s}{\rho_s}\right):= H(\nu|\rho)
\end{equation}
\end{theorem}

When comparing (\ref{eq:ldp0.1}) with (\ref{eq:ldp1}), it becomes clear that Theorem~\ref{th:sanov} is nothing more than a higher dimensional version of Theorem~\ref{th:cramer}. Instead of looking at deviations of the empirical averages away from the mean, we consider now deviations of the empirical measure $L_n$ away from the true measure $\rho$. The rate functions depends in this case on the different measures $\nu$ on $\Gamma$ and on how similar they are to $\rho$. The quantity $H(\nu|\rho)$ is the relative entropy of the measure $\nu$ with respect to the measure $\rho$ \cite{Kullback1951}. By applying Jensen's inequality to $I_\rho(\nu)=-\sum_s \nu_s \left(\log (\rho_s/\nu_s)\right)$, we have that  $I_\rho(\nu)\geq - \log \sum_s \nu_s (\rho_s/\nu_s) = 0$, with the  equality being realised if and only if $\nu=\rho$. 

In other terms, Sanov’s theorem states that the exponential rate of decay of the probability of a large deviation of size $\geq a$ between the empirical measure and the marginal distribution $\rho$ is controlled by the element of all measures on $\Gamma$ whose distance from $\rho$ is $\geq a $ that is closest to $\rho$ in the sense of relative entropy.

The contents of Theorem \ref{th:sanov} allow us to reinterpret and extend the results discussed in (\ref{eq:ldp0.1})-(\ref{eq:ldp0.1b}). Let's consider a function $f$
with $\sum_{s=1}^r f_s \rho_s = \mu_f \in \mathbb{R}$.
We define $\Phi_{f,a}=\{\phi\in M(\Gamma)| \sum_s f_s \phi_s \geq a\}$. We also define
$\Psi_{f,a}=\{\phi\in M(\Gamma)| \sum_s f_s \phi_s = a \}$. Clearly, one has $\Phi_{f,a}=\cup_{b\geq a} \Psi_{f,b}$.

We have that $\mathbb{P}(L_n\in\Phi_{f,a})= \mathbb{P}\left(\frac{1}{n}\sum_{j=1}^n f(X_j)\geq a\right)$, where $a\geq \mu_f$ and we consider the empirical measure $L_n=\frac{1}{n}\sum_{i=1}^n\delta_{X_i}$ introduced before. One then derives that:
\begin{align}\label{eq:ldp1s}
    \lim_{n \to \infty}\frac{1}{n}\log \mathbb{P}(L_n \in\Phi_{f,a}) &= \lim_{n \to \infty}\frac{1}{n}\log \mathbb{P} \left( \frac{1}{n}\sum_{j=1}^n f(X_j)\geq a \right) \\ &= -\inf_{\nu \in \Phi_{f,a}} I_\rho(\nu)=-\inf_{z\geq a}\inf_{\nu \in \Psi_{f,z}} I_\rho(\nu) \nonumber
\end{align}
Let's now consider the case $f(x)=x$. The empirical average $S_n/n$ is connected to the empirical measure $L_n$ through the formula $S_n/n=\sum_{s=1}^r s L_n(s)$. The rate function in (\ref{eq:ldp0.1}) can be obtained from (\ref{eq:ldp1s}) by 
\begin{equation}
    I(z)=\inf_{\nu \in \Psi_{f,z}} I_\rho(\nu).
\end{equation}
Thus, the rate function of the empirical average $z$ is equal to the infimum of the rate function for the empirical measure $\nu$ if the infimum is taken over all the measures $\nu$ with mean $\mu=\sum_{s=1}^r s \nu_s = z$. In other words, there is an equivalence between the large deviations of the empirical average $z$ and the large deviations of the least unlikely empirical measure $\nu$ with mean equal to $z$. This is an example of the contraction principle that we state now.

\begin{theorem}
Contraction Principle. Let $A_n$ be a family of random variables such that 
\begin{equation}\label{eq:c1}
    \lim_{n \to \infty}\frac{1}{n}\log \mathbb{P}\left(A_n \in \mathcal{A}\right) = -\inf_{z \in \mathcal{A}} I_A(z)
\end{equation}
and let's consider another family of random variables $B_n=T(A_n)$ where $T$ is a continuous function. It is possible to establish a LDP for $B_n$ as follows:
\begin{equation}\label{eq:c2}
    \lim_{n \to \infty}\frac{1}{n}\log \mathbb{P}\left(B_n \in \mathcal{B}\right) = -\inf_{z \in \mathcal{B}} I_B(z), \quad I_B(z)=\inf_{y=T^{-1}z}(I_A(y)).
\end{equation}

 \end{theorem}

Theorem~\ref{th:sanov} can be generalized further to large deviations of pair empirical measures as well as of measures with higher dimensions. Higher level large deviation laws imply the ones for lower levels, the downward link being provided by the contraction principle. The interested reader can find a short summary of the generalizations to higher dimension in Appendix~\ref{app1}, for a detailed discussion of this topic we refer to \cite{Hollander2000}. 

\subsection{Dependent sequences}\label{sec:theo_ge}

We continue with a generalization of Cram\'{e}r's Theorem for random sequences that have a form of moderate dependence, which goes back to \cite{Gaertner1977} and \cite{Ellis1984}. A rigorous derivation of the G\"{a}rtner-Ellis (GE) theorem would go beyond the scope if this paper, thus we concentrate on the main results. As above, we follow here the work of \cite{Touchette2009}.

We consider the sequence $(Z_n)$ of random variables on the probability space ($\mathbb{R}^d,\mathcal{B}(\mathbb{R}^d),\mathbb{P}$), where $\mathcal{B}(\mathbb{R}^d)$ is the Borel sigma-field on $\mathbb{R}^d$ with moment generating functions
\begin{equation}
    \varphi_n(t)=\mathbb{E}[e^{\langle t,Z_n\rangle}], \qquad t \in \mathbb{R}^d, n \in \mathbb{N}
\end{equation}
with $\langle \cdot,\cdot \rangle$ denoting the standard inner product. It can be useful to think of $(Z_n)$ as an empirical average, but this doesn't have to be the case. We assume that the limit 
\begin{equation}\label{eq:coge1}
    \lim_{n \to \infty} \frac{1}{n} \log \varphi_n(nt) = \Lambda(t) \in [-\infty,\infty]   
\end{equation}
exists and 
\begin{equation}\label{eq:coge2}
  0 \in \mathrm{int}(D_\Lambda), \mathrm{with\ } D_\Lambda=\{t \in \mathbb{R}^d: \Lambda(t)<\infty\}.
\end{equation}
 We also assume that $\Lambda$ is convex and differentiable on $\mathrm{int}(D_\Lambda)$. Furthermore, we assume that $\Lambda$ is lower semi-continuous on $\mathbb{R}$, and either $D_\Lambda=\mathbb{R}^d$ or $\Lambda$ is steep at $\partial D_\Lambda$.
 Let $P_n(\cdot)=\mathbb{P}(Z_n \in \cdot)$.
 Under the above conditions, the GE theorem states that $(P_n)$ satisfies a large deviations principle on $\mathbb{R}^d$ with rate $n$ and with rate function
\begin{equation}\label{eq:legtr}
    I(x)=\sup_{t \in \mathbb{R}^d}[\langle x,t\rangle - \Lambda(t)], \qquad x \in \mathbb{R}^d.
\end{equation}
Thus, the rate function $I$ is the Legendre-transform of $\Lambda$, also called the scaled cumulant generating function. The rate function $I$ is convex. Note that (\ref{eq:legtr}) is a generalized form of (\ref{eq:rf0}).

If $Z_n=\frac{1}{n} \sum_{i=1}^n X_i$ with $X_i$ a stationary random sequence, then conditions (\ref{eq:coge1}) and (\ref{eq:coge2}) can be interpreted as a kind of moderate dependence assumption on $(X_i)$. However, in case of strong dependence, the theorem would fail because the strict convexity of $\Lambda$ would be violated.

We have seen that by using the GE theorem one obtains a large deviations principle under fairly mild regularity assumptions. As mentioned above, it is not necessary that $Z_n$ represents sample averages. In fact, the large deviation principles presented in Sec.~\ref{sec:theo_iid} and \ref{sec:theo_ge}  for sample averages, empirical measures, pair empirical measures (see Appendix~\ref{app1}), and so on, can all be obtained by following the route given by the GE theorem as well. Below, we derive based on \cite{Hollander2000,Touchette2009} the rate functions of sample averages for i.i.d. random variables and for Markov chains, by using the GE theorem. 

1) Let ($X_i$) be i.i.d. $\mathbb{R}$-valued random variables satisfying $\varphi(t)=\mathbb{E}[e^{tX_1}] < \infty$, for all $t \in \mathbb{R}$. Let us consider the empirical average $Z_n=\frac{1}{n}\sum_{i=1}^n X_i$. Then,
\begin{equation}
    \varphi_n(nt)=\mathbb{E}[e^{ntZ_n}]=\mathbb{E}\left[e^{t\sum_{i=1}^n X_i}\right]=[\varphi(t)]^n,
\end{equation}
with $\varphi$ the moment generating function of $X_1$. Hence $\Lambda(t) = \log\varphi(t)$ and the GE theorem reduces to Cr\'amer's theorem (Theorem \ref{th:cramer}).

2) Let ($X_i$) be a stationary $\Gamma$-valued Markov chain. Let $Z_n=\frac{1}{n}\sum_{i=1}^n f(x_i)$, where $f: \Gamma \to \mathbb{R}^d,\ d \ge 1$. Then,
  \begin{multline*}
    \varphi_n(nt)=\mathbb{E}[e^{ntZ_n}]=\mathbb{E}\left[ e^{t\sum_{i=1}^n f(x_i)}\right]=\\
    \sum_{ x_1, \ldots , x_n \in \Gamma} \pi(x_1)e^{tf(x_1)} P(x_2|x_1)e^{tf(x_2)} \cdot\cdot\cdot P(x_n|x_{n-1})e^{tf(x_n)},
  \end{multline*}
where $\pi(x_1)$ denotes the probability of the initial state $x_1$, and $P(x_i|x_{i-1})$ denotes the conditional probability of state $x_{i}$ given $x_{i-1}$, $i=1,...,n$. By defining $\pi_{t}(x_1)=\pi(x_1)e^{tf(x_1)}$ and $P_t(x_i|x_{i-1})=P(x_i|x_{i-1})e^{tf(x_{i})}$, we have that
\begin{equation*}
    \varphi_n(nt)=\sum_{j \in \Gamma} (\Pi_t^{n-1}\pi_t)_j,
\end{equation*}
where $\pi_t$ is the vector of probabilities for which $(\pi_t)_i=\pi_t(x_1=i)$, and $\Pi_t$ denotes the matrix with elements $(\Pi_t)_{ji}=P_t(j|i)$. Based on Perron-Frobenius theory for positive matrices we get that $\lim_{n\rightarrow \infty}\log\varphi_n(nt)=\log\lambda(t)$, with $\lambda(t)$ denoting the unique largest eigenvalue of $\Pi_t$. Hence $\Lambda(t)=\log\lambda(t)$, and the rate function is given by the Legendre transform
\begin{equation}\label{eq:rfmcav}
    I(z)=\sup_{t\in \mathbb{R}}[zt-\log \lambda(t)].
\end{equation}
Please note that (\ref{eq:rfmcav}) can be used to obtain the rate function only if $\Pi$ has a unique stationary distribution $\pi$. If $\Pi$ has several stationary distributions, $\Lambda(t)$ exists, but depends on the initial  distribution $\pi(x_1)$. If $\Pi$ has no stationary distribution, generally no large deviation principle can be found and the law of large numbers does not even hold \cite{Touchette2009}.

\section{Large deviations in dynamical systems}\label{sec:ldds}

At this point, we leave the idealized world of i.i.d. random variables and discrete time processes, and turn our attention to systems evolving continuously in time, as we want to look into mathematical models that are more relevant for capturing the dynamical properties of the climate system. Instead of empirical measures and sample averages, we consider in the following probabilities of trajectories or paths of deterministic dynamical systems and finite time averages along these trajectories. However, the main ingredients leading to a large deviation result stay the same. One needs basically the attracting effect of an asymptotic limit leading to an exponential decay of probabilities of finite time estimates. By taking into consideration the dynamics in time and including the temporal dimension into the large deviation analysis, the methods presented below are directly relevant for geophysical applications. We will present some basic results pertaining to stochastic and to deterministic chaotic dynamical systems, for the sake of completeness, and because the modelling of geophysical flows follows both dynamical paradigms. 

{\color{black}First, we motivate the use of stochastic dynamics for investigating the properties of geophysical flows by introducing the concept of filtering and the development of evolution equations based on dynamical balances and specialised for specific scales of motion \cite{Speranza2005,Klein2010,Ghil2020}. The introduction of stochastic parametrizations \cite{wouters_multi-level_2013,MSM2015,Franzke.ea.2015,Berner2017,Ghil2020} is motivated through the use of the Mori-Zwanzig formalism \cite{mori_transport_1965,zwanzig_memory_1961}. When suitable limits are considered, the stochastic component, which provides a surrogate representation of the effects of the scales we are unable to describe explicitly, can be written as multiplicative white noise \cite{Pavliotis2008}. This provides the basis for a large class of stochastic climate models of very widespread use and great physical relevance \cite{Hasselmann1976,Imkeller2001,Majda2001,saltzman_dynamical,Imkeller2002a,Penland2003}.
Such stochastic models are amenable to being studied using the Freidlin-Wentzell theory \cite{freidlin1984}, which allows to derive powerful 
large deviation results.} 
Additionally, one should keep in mind that the climate undergoes actual stochastic forcing due to random fluctuations in the incoming solar radiation and other astrophysical factors.  More in general, the use of stochastic dynamics for describing nonequilibrium statistical mechanical systems has reached a high level of popularity and has shown a great potential for deriving results of great theoretical and practical relevance \cite{Kurchan1998,Baiesi2013,Livi2017,Ottinger2021}.



In case of the Freidlin-Wentzell theory the zero-noise limit of stochastic evolution law is given by its purely deterministic component. 
Hence, one obtains the probability of random paths deviating from the deterministic path in terms of large deviation laws. The probabilities of deviation of finite time averages from their asymptotic values 
can be obtained from the large deviation results for random paths using the contraction principle. A more general and pragmatic approach, however, which can be followed even in case of unknown model equations, 
is related to the fact that finite time averages of weakly correlated observables are (nearly) independent. Thus, one can model finite time averages of correlated observables as resulting from i.i.d. random variables or Markov chains. Consequently, the theorems presented in Sec.~\ref{sec:theo} can be applied in a similar way with the difference that the large deviations parameter $n$ is now related to time. In Sec.~\ref{sec:ldds_ta} we discuss a modified version of the GE theorem (\ref{eq:legtr}) acting on time averaged observables. 

Later on, we consider special chaotic dynamical systems, so-called Axiom A systems \cite{bowenruelle1975}, and discuss the emerging large deviation laws for finite time averages of given observables. 
The framework of Axiom A systems - which are essentially the closest deterministic relatives of the \textit{truly} stochastic systems - blurs the distinction between statistical mechanics and dynamical systems theory, mainly as a result of the fact that Axiom A systems possess a rather special ergodic invariant measure that has a clear physical interpretation  \cite{Eckmann1985}. Another remarkable property of Axiom A systems is that they admit a Markov partition, i.e. a partition of the attractor such that one can put in a one-to-one correspondence the actual orbit of the system with an infinite sequence of symbols describing the history of occupancy of the various elements of the partition by the orbit. Accordingly, the original map can be associated with a shift map, i.e. a finite-state Markov chain describing the probability of transition between the various elements of the partition \cite{Ruelle1989,Gallavotti2014}. The possibility of establishing the so-called symbolic dynamics guarantees that the results presented in Sec. \ref{sec:theo_ge} for finite-state Markov chain apply also for Axiom A systems. Nonetheless, there is no \textit{free lunch}: it is in general far from trivial to actually construct the Markov partition. As discussed below, while Axiom A systems are very special dynamical objects, the chaotic hypothesis \cite{Gallavotti1995,Gallavotti2014} makes them very relevant for providing a framework for studying large deviations laws in high-dimensional geophysical systems.


{\color{black}
\subsection{Stochastic Climate Models}\label{sclimatem}
The state of the climate system can be described using the continuum approximation, introducing field variables that depend on three spatial dimensions and time. The partial differential equations that describe the evolution of the field variables are based on the budget of mass (including different chemical species), momentum and energy. Since the climate system features variability on a vast range of spatial and temporal scales, as mentioned above, a key procedure one needs to apply, both on theoretical grounds and for reasons of defining efficient numerical models, is to specialise the evolution equations to a desired range of spatial and temporal scales of interest by the use of suitable approximations based on the validity of approximate dynamical balances \cite{Speranza2005,Klein2010,Ghil2020}. Additionally, when constructing an actual numerical model, the three-dimensional fields are discretized on a lattice, either in the physical space, or in the reciprocal space via spectral projection, or in a suitable combination of the two. Hence, the impact of the physical processes occurring in the unresolved spatio-temporal scales on those taking place in the resolved ones can be represented only through approximate parametrizations \cite{Franzke.ea.2015,Berner2017}. The Mori-Zwanzig coarse-graining  based on the projection operator  \cite{mori_transport_1965,zwanzig_memory_1961} clarifies that such parametrizations have in general a deterministic, a stochastic, and a non-markovian component \cite{wouters_disentangling_2012,wouters_multi-level_2013,MSM2015,CLW15a}.

Let us assume, for simplicity, that the \textit{true} evolution equation for the climate system can be written as a system of autonomous ordinary differential equations\footnote{We are here neglecting the - very important - presence of explicit time-dependence and stochastic forcing in the dynamics; see \cite{Chekroun2011,Ghil2015,Ghil2020} for a detailed discussion of this aspect.} of the form
\begin{equation}
\frac{\mathrm{d}z}{\mathrm{d}t}=G(z)\label{evolution}
\end{equation}
where $z\in\mathbb{R}^N$. The procedure of coarse-graining, associated with specialising the equations for a specific range of time and spatial scales, implies that we rewrite the state vector $z$ as $z=(x,y)$, where  $x\in\mathbb{R}^n$ and  $y\in\mathbb{R}^{N-n}$ and we aim at deriving approximate equations of the variables of interest $x$. It is reasonable to assume that $n\ll N$. Note that, alternatively, $x$ can correspond to the variables describing the state of a portion of the climate system (e.g. the atmosphere), and $y$ can instead describe the rest of the system. One does not need to assume \textit{a priori} the presence of a very large time-scale separation between the dynamics of the $x$ and $y$ components. One can then rewrite (\ref{evolution}) as:
\begin{align}
\begin{split}
\frac{\mathrm{d}x}{\mathrm{d}t}&=f(x)+\delta f_x(x,y)\\
\frac{\mathrm{d}y}{\mathrm{d}t}&=\frac{1}{\epsilon}g(y)+\frac{\delta}{\epsilon} g_y(x,y)
\end{split}\label{wl2013}
\end{align}
where $f$ and $g$ define the autonomous dynamics of the $x$ and $y$ components, respectively, $\delta$ is a constant controlling the intensity of the coupling, and $\epsilon$ defines the time scale separation between the two sets of variables. The Mori-Zwanzig theory indicates that one can in general write the dynamics of the $x$ variables in an implicit form as follows:
\begin{equation}
\frac{\mathrm{d}x}{\mathrm{d}t}=f_{\epsilon,\delta}(x)+\dot\sigma_{\epsilon,\delta}(x)+\int\mathrm{d}s K_{\epsilon,\delta}(x,t-s)x(s),\label{mz}
\end{equation}
where the three terms of the right hand side correspond to the deterministic drift, to a noise contribution, and to the memory term. In the weak-coupling limit ($\delta\rightarrow 0$), it is possible to derive via perturbative approach an explicit expression for these three terms that is valid up to order $\delta^2$ \cite{wouters_disentangling_2012,wouters_multi-level_2013}; see a practical implementation of this theory for the development of parametrizations in geophysical fluid dynamical models in \cite{Wouters2016,Demaeyer2017,Vissio2018a}. Note that one can derive an expression for the Mori-Zwanzig projected dynamics using data-driven approaches \cite{MSM2015,CLW15a}. Very recently, it has been shown \cite{Gutierrez2020} that the data-driven and the top-down approach presented in \cite{wouters_disentangling_2012,wouters_multi-level_2013} are fundamentally equivalent. 

Instead, if the two sets of variables $x$ and $y$ have an infinite time scale separation ($\epsilon\rightarrow0$), the dynamics of the variable $x$ converges to a deterministic averaged equation (for more details, see Sec. \ref{dynoise} below). Via homogenization theory \cite{Pavliotis2008} deviations from this averaged equation can be modeled by a stochastic differential equation without memory, and with multiplicative white noise, so that the evolution of the $x\in\mathbb{R}^n$ variables is controlled by:
\begin{equation}
  \dif x_t  =  F (x_t) \dif t +  \Sigma (x_t)  \dif W_t,  \label{eq:diffusiona}
\end{equation}
where it possible to derive explicit formulas for the renormalised drift term $F:\mathbb{R}^n\rightarrow\mathbb{R}^n$, and the diffusion matrix 
$\Sigma: \mathbb{R}^n \rightarrow \mathbb{R}^{n \times m}$, while $W_t$ is an $m$-dimensional Brownian motion. In a nutshell, the impact of the neglected scales of motions corresponding to the $y$ variables is twofold: it leads to a) a change in the deterministic contribution to the evolution of the $x$ variables; and to b) the inclusion of a random forcing. Stochasticity is essentially due to the lack of information on the state of the $y$ variables in the projected $x$ space; see a detailed discussion of this in \cite{CLW15a}.

Equation (\ref{eq:diffusiona}) is at the basis of stochastic climate models, whose investigation was initiated by Hasselmann \cite{Hasselmann1976}; see a comprehensive analysis of this viewpoint and further developments in \cite{Majda2001,Imkeller2001,saltzman_dynamical,Imkeller2002a,Penland2003}. Traditionally, the deterministic component of (\ref{eq:diffusiona}) features one or more fixed points, and the noise allows for the the system to explore regions of the phase space far from the deterministic solutions, and to perform transitions between competing metastable states. We will provide a broader view point on this in Sec. \ref{sec:metastability}, where we will consider more general competing asymptotic states. Stochastic climate models have been key for discovering fundamental physical processes like stochastic resonance \cite{Benzi1981,Benzi1982,Nicolis1981,Nicolis1982,Gammaitoni1998}, and have provided key insights for studying the transitions between different weather regimes in the atmosphere \cite{Charney1979,Benzi1986,Mo.Ghil.1987,Itoh1996,Kondrashov2004,Ruti2006}; see discussion in Sec. \ref{block}. Equation (\ref{eq:diffusiona}) is probably the most convenient starting point for discussing the use of LDT in geophysical flows even if, as shown in Sec. \ref{sec:chaotic}, LDT can be introduced also in the context of deterministic chaos. 
}

\subsection{Dynamical systems perturbed by weak noise}\label{dynoise}
We now focus on the stochastic climate models introduce in the previous subsection and aim at deriving large deviation laws. The Freidlin-Wentzell theory 
{\cite{wentzellRandomPerturbationsDynamical1998,varadhanLargeDeviationsApplications1984}}, allows one to study the convergence of
probability measures on the path-space of a
stochastic differential equation $X$ in $\mathbb{R}^n$
\begin{eqnarray}
  \dif X_t^{\varepsilon} & = & b (X_t^{\varepsilon}) \dif t + \sqrt{\varepsilon}
  \sigma (X^{\varepsilon}_t)  \dif W_t \quad X_0^{\varepsilon} = x, t
  \geqslant 0 .  \label{eq:diffusion}
\end{eqnarray}
where, as in (\ref{eq:diffusiona}) $b : \mathbb{R}^n \rightarrow \mathbb{R}^n$ is a deterministic drift,
$\sigma : \mathbb{R}^n \rightarrow \mathbb{R}^{n \times m}$ is the diffusion
function, $W_t$ is an $m$-dimensional Brownian motion, and we introduce here the parameter $\varepsilon>0$ that controls the intensity of the stochastic forcing. 

For bounded and Lipschitz $b$ and $\sigma$, it can be shown that as the noise
intensity goes to zero ($\varepsilon \rightarrow 0$), the distribution of
paths of $X_t^{\varepsilon}$ converges to the deterministic path determined by
$\dif x_t = b (x_t) \dif t$ {\cite{wentzellRandomPerturbationsDynamical1998}}. For all $T > 0$ and
$\delta > 0$
\begin{eqnarray*}
  \lim_{\varepsilon \rightarrow 0} \mathbb{P} \{ \max_{0 \leqslant t \leqslant
  T} | X_t^{\varepsilon} - x_t | > \delta \} & = & 0 .
\end{eqnarray*}

We may wonder, of course, about the probability of observing a given path $f
(t) \neq x_t$ when $\varepsilon \neq 0$. It can be shown that a large
deviation principle holds for $X^{\varepsilon}_t$, with a rate function or
action functional
\begin{eqnarray}
  I_T (f) & = & \frac{1}{2} \int_0^T \left\langle \dot{f} (t) - b (f (t)), a^{- 1}
  (f (t)) (\dot{f} (t) - b (f (t))) \right\rangle \dif t,  \label{eq:action}
\end{eqnarray}
where $a (x) = \sigma (x) \sigma (x)^{T}$ is the noise covariance. We have
that
\begin{eqnarray}
  \mathbb{P} \left\lbrace \sup_{t \in [0, T]} | X_t^{\varepsilon} - f (t) | < \delta \right\rbrace
  & \asymp & \exp \left( - \frac{1}{\varepsilon} I_x (f) \right) 
\end{eqnarray}
Similarly as integrals of the form $\int_a^b e^{- \frac{1}{\varepsilon} h (x)}
k (x)  \dif x$ are dominated by the minimum $x_0$ of $h (x)$ in Lagrange's
methods, as $\varepsilon \rightarrow 0$, the probability in a set $F \subset C
[0, T]$ of trajectories concentrates on the trajectory $f^{\star}$ with the
smallest rate function $I_x$:
\begin{eqnarray}
  I_T (f^{\star}) & = & \inf_{f \in F} I_T (f) . 
  \label{eq.instanton_minimization}
\end{eqnarray}
Such path is called the minimum action path or instanton.

\paragraph{The exit problem}

In the limit $\varepsilon \rightarrow 0$, the dynamics of
{\eqref{eq:diffusion}} is determined by the drift field $b$. When $b$ has
an attractor, the trajectory will never escape from it in the absence of noise.
The situation is markedly different when noise is added. The system can make
excursions away from the attractor, exit from its surrounding and possibly
transition to another attractor.

LDT provides a way to describe the exit from regions
containing an attractor, e.g. if $\Omega$ is a bounded set containing a stable
fixed point $\bar{x}$ of $\dif x_t = b (x_t) \dif t$, then the exit from the
domain will happen close to the point minimizing the action
{\eqref{eq:action}}. For more details see \cite{Freidlin2012}.

\paragraph{Instanton calculation}

The minimization of the action functional for problems of interest in
geophysics can usually not be done analytically. In such cases the instanton needs to be calculated numerically. 

Arguably the most direct way of finding the instanton is by minimizing the action \eqref{eq:action}.
In the minimum action method
{\cite{eMinimumActionMethod2004}}, the instanton $f^\star$ on a finite time interval $[0, T]$ is approximated by $f^\star(t_i)$ on a discrete temporal grid, a discrete approximation to the action is derived and a quasi-Newton method is then applied to minimize the discretized action.

Another fairly simple method of numerically finding the instanton is solving the Hamilton equation connected to this minimization problem. A difficulty arises here in that we are often looking for a minimization with fixed start and end points for $f^\star$ at $t=0$ and $t=T$. To solve the Hamilton equation we need to specify initial values for the coordinates and their conjugate momenta, however. A shooting method can be applied to find the initial values of the momenta, but this is in general difficult to apply in high dimensions.

Both these methods can only be applied to finite time
intervals, while in many cases we will want to allow for infinite time lengths
of transition. In the special case where the drift term is a gradient, i.e. $b=-\nabla U$, and $\sigma$ is the identity, these problems can be circumvented by using the string method \cite{eStringMethodStudy2002} which uses that the instanton is always parallel to the drift. The method alternates relaxation along the drift with a redistribution of the discretization points along the instanton curve.

This principle has been further generalized to non-gradient systems in the geometric minimum action method \cite{VdE_geo}. Here the action is reformulated in a geometric way that doesn't involve the time parameterization of the instanton. In this way the problem of infinite transition times can be circumvented.

An overview of numerical methods to
calculate the instanton is given in {\cite{grafkeNumericalComputationRare2019}}.

\paragraph{Systems with a time scale separation}

As mentioned above, in some geophysical settings, we may be interested in the evolution of a number of slowly evolving variables $x$ in interaction with other variables $y$ that evolve on a much faster time scale. {\color{black} We then consider a slightly modified version of (\ref{wl2013}):
\begin{align}
    \dif x &= f(x,y) \dif t \label{eq:slowfastx} \\
    \dif y &= \frac{1}{\epsilon} g(x,y) \dif t + \frac{1}{\sqrt{\epsilon}} \sigma(x,y) \dif W . \label{eq:slowfasty}
\end{align}
where we introduce a white noise forcing term for the fast (as we consider $\epsilon \rightarrow 0$) variables $y$. Note that  $\mathbb{E}\left((W(t)-W(s))^2\right) =(t-s)$, hence the scaling with $\sqrt{\epsilon}$.} Intuitively speaking, in the limit $\epsilon \rightarrow 0$, in any time interval of order $1$, no matter how short, the $y$ variable will explore the invariant measure of the equation for $y$ for fixed $x$ determined by
\begin{align}
    \dif \tilde{y} = g_x(\tilde{y}) \dif t + \sigma_x (\tilde{y}) \dif W, \label{eq:frozenx}
\end{align}
where $g_x(y)=g(x,y)$ and $\sigma_x(y)=\sigma(x,y)$. As a result, we get a law-of-large-numbers-like result for the slow variable $x$. As $\epsilon \rightarrow 0$, the path $x(t)$ converges to $X(t)$, the solution of
\begin{align*}
    \dif X = F(X) \dif t,
\end{align*}
where $F(X) = \int f(X,y) \tilde{\mu}_x(\dif y)$ with $\tilde{\mu}_x(\dif y)$ the invariant measure of \eqref{eq:frozenx}.

As with the law of large numbers for averages of i.i.d. random variables, we
may expect a large deviation result to hold here as well. To derive, at a heuristic level, the rate function for the path probabilities of the slow variable
$x$ we consider a discrete time approximation of
{\eqref{eq:slowfastx}}-{\eqref{eq:slowfasty}}. We approximate $x (t)$ for $t
\in [0, T]$ by discrete $x_i$ at times $i \Delta t_x$ with $i \in \{ 0,
\ldots, N_x \}$ with $N_x = \lfloor T / \Delta t_x \rfloor$.

Since $x (t + t_x) = x (t) + \int_t^{t + \Delta t_x} f (x (\tau), y (\tau))
\mathd \tau$ we approximate the increment of $x$ between two subsequent
discrete times by
\begin{eqnarray*}
  x_{i + 1} - x_i & = & \int_t^{t + \Delta t_x} f (x_i,
  \tilde{y}^{(\epsilon)}_{x_i} (\tau)) \mathd \tau,\\
  & = & \epsilon \int_0^{\Delta t_x / \epsilon} f (x_i, \tilde{y}_{x_i}
  (\tau)) \mathd \tau,
\end{eqnarray*}
which we can express in terms of a time average as
\begin{eqnarray*}
  \frac{x_{i + 1} - x_i}{\Delta t_x} & = & \frac{\epsilon}{\Delta t_x}
  \int_0^{\Delta t_x / \epsilon} f (x_i, \tilde{y}_{x_i} (\tau)) \mathd
  \tau .
\end{eqnarray*}
The probability of the slow process going from some given value $\varphi_i$ at
time $i \Delta t_x$ to $\varphi_{i + 1}$ at time $(i + 1) \Delta t_x$ can
therefore be estimated via the large devations of the time average of $f (x_i,
\tilde{y}_{x_i} (t))$ as
\begin{eqnarray*}
  \mathbb{P} (x_{i + 1} = \varphi_{i + 1} |x_i = \varphi_i) & = & \mathbb{P}
  \left( \left. \frac{\Delta x_i}{\Delta t_x} = \frac{\Delta \varphi_i}{\Delta
  t_x} \right| x_i = \varphi_i \right)\\
  & = & \mathbb{P} \left( \left. \frac{\epsilon}{\Delta t_x}
  \int_0^{\Delta t_x / \epsilon} f (x_i, \tilde{y}_{x_i} (\tau)) \mathd
  \tau = \frac{\Delta \varphi_i}{\Delta t_x} \right| x_i = \varphi_i \right)\\
  & \approx & e^{- \frac{\Delta t_x}{\epsilon} \Lambda_{\varphi_i}^{\ast}
  \left( \frac{\Delta \varphi_i}{\Delta t_x} \right)}
\end{eqnarray*}
where $\Lambda^{\ast}_{\varphi}$ is the rate function for the time averages of
$f (\varphi, \tilde{y}_{\varphi} (t))$, the Legendre-Fenchel transform of the
scaled cumulant generating function
\begin{eqnarray*}
  \lambda_{\varphi} (\theta) & = & \lim_{T \rightarrow \infty} \frac{1}{T} \ln
  \mathbb{E} \left( \exp \left( \theta \int_0^T f (\varphi,
  \tilde{y}_{\varphi} (\tau)) \mathd \tau \right) \right) .
\end{eqnarray*}
Hence, assuming Markovianity for $x_i$ in the limit $\epsilon \rightarrow
0$ due to rapid decorrelation of the $y$ process, the path probability for $x$
can be approximated as
\begin{eqnarray*}
  \mathbb{P} \left( x (t) = \varphi (t) \text{ for } t \in [0, T] \right) &
  \approx & \mathbb{P} (x_0 = \varphi_0, \ldots, x_{N_x} = \varphi_{N_x})\\
  & = & \mathbb{P} (x_0 = \varphi_0) \mathbb{P} (x_1 = \varphi_1 |x_0 =
  \varphi_0) \ldots \mathbb{P} (x_{N_x} = \varphi_{N_x} |x_{N_x - 1} =
  \varphi_{N_x - 1})\\
  & \approx & \mathbb{P} (x_0 = \varphi_0) \prod_{i = 0}^{N_x - 1} \exp
  \left( - \frac{\Delta t_x}{\epsilon} \Lambda_{\varphi_i}^{\ast} \left(
  \frac{\Delta \varphi_i}{\Delta t_x} \right) \right)\\
  & \rightarrow & \mathbb{P} (x_0 = \varphi_0) \exp \left( -
  \frac{1}{\epsilon} \int_0^T \Lambda_{\varphi (\tau)}^{\ast}
  (\dot{\varphi} (\tau)) \mathd \tau \right) .
\end{eqnarray*}
From this very non-rigorous derivation we can expect that the rate function for
the slow process $x$ as $\epsilon \rightarrow 0$ is $\int_0^T
\Lambda_{\varphi (\tau)}^{\ast} (\dot{\varphi} (\tau)) \mathd \tau$. The same result has been derived in a more rigorous manner in \cite{bouchetLargeDeviationsFast2016}.

\subsection{Time averaged observables}\label{sec:ldds_ta}

In this section we consider large deviation results for time averages of observables of dynamical systems. The large deviation parameter is in this case the inverse of the time length $T$ over which the average is taken. To illustrate the main results, let us consider a Markov process $X(t) \in \mathbb R^n$, and an observable $A: \mathbb R^n \rightarrow \mathbb R$. We have a large deviation principle for the time average $a=\frac{1}{T}\int_0^T A(X(t))\, \mathrm{d}t$ if its probability distribution scales for large $T$ as
\begin{equation}
\rho(a) \underset{T\rightarrow\infty}{\asymp}\mbox{e}^{-TI\left[a\right]}\label{eq:LD_DV}
\end{equation}
with rate function $I(a)$. Similarly to what discussed in the previous sections, one can define the scaled cumulant generating function
\begin{equation}\label{scaledcumulant}
\lambda(k)=\underset{T\rightarrow +\infty}{\lim}\frac{1}{T}\log{\mathbb{E}\left[e^{k\int_0^T A(t)\, \mathrm{d}t} \right]},
\end{equation}
and the G\"artner-Ellis theorem relates rate function $I(a)$ and scaled cumulant generating function $\lambda(k)$ through  Legendre transformation. In particular, when the rate function $I(a)$ is convex and differentiable, or equivalently when $\lambda(k)$ is differentiable, the Legendre transform can be inverted, and the rate function can be computed as solution of the variational problem as $I(a)=k(a)a-\lambda(k(a))$, where $k(a)$ is given by $a=\lambda'(k(a))$.

Large deviation results of this kind hold in general for mixing dynamics and for observables for which the tails of the distribution decay sufficiently fast. Mathematically, sufficient conditions are given by \cite{Donsker1975,Donsker1975a,Donsker1976,Donsker1983}; see also the discussion in \cite{Touchette2009}. In most applications to geophysical fluid dynamics or climate sciences we either use stochastic models which guarantee the conditions, or we consider deterministic chaotic systems of sufficient complexity that we expect the conditions to hold (see discussion in the introduction of Sec. \ref{sec:ldds} and more in details in Sec. \ref{sec:chaotic}). However, it is important to keep in mind that the existence of a large deviation result is in general not guaranteed, and must be proved or validated empirically.

Several physical systems have been reported featuring anomalous large deviation scalings \cite{Nickelsen2018}, that is scalings where the large deviation parameter appears to a power different from one ($T^\alpha$ with $\alpha \neq 1$ in the present case). Typically these are systems featuring non-Markovian dynamics or long-range correlations \cite{Nickelsen2018,Gradenigo_et_al_2013,Zeitouni_2006,Harris_and_Touchette_2009,Krapivsky_et_al_2014,Sadhu_et_al_2015,Imamura_et_al_2017,Doussal_et_al_2016,Sasorov_et_al_2017,Corwin_et_al_2018,Louidor_et_al_2015,Derrida_et_al_2017}{\color{black}; see a detailed treatment of the problem in the case of deterministic dynamical systems in \cite{rey-bellet_young_2008,Chazottes2015}}.  However, it has been shown that even in the case of a system as simple and well behaved as the Ornstein-Uhlenbeck process, simply considering as observable the third moment or higher of the state of the system leads to anomalous large deviation scalings \cite{Nickelsen2018,rey-bellet_young_2008,Chazottes2015}. It is therefore important to proceed carefully when testing large deviation scalings in complex systems like the ones typically analysed in climate science. 

If valid, a large deviation result for the time average of an observable gives an extension of the central limit theorem that allows to take into considerations fluctuations of order $T$ rather than $\sqrt{T}$. For ergodic systems the time average of an observable $A$ converges in the limit of large $T$ to the ergodic average $\mu=\mathbb{E}[A]$. Under mixing hypotheses, the central limit theorem gives that for large $T$, typical fluctuations of the time average are of order $\sqrt{T}$ and Gaussian distributed, that is $(a-\mu)/\sqrt{T}\sim N(0,\sigma^2 \tau_c)$, where $\sigma^2$ is the variance of $A$ and $\tau_c$ its integral autocorrelation time $\tau_c=\sigma^{-2}\int_{-\infty}^{+\infty} C(\tau,0)d\tau$, with $C(t,s)=\mathbb{E}[(A(t)-\mu)(A(s)-\mu)]$ the covariance of $A$. In certain applications however it is of interest to consider fluctuations more rare than those handled by the central limit theorem, and that instead scale with $T$. A large deviation result allows to have a limit distribution for these large fluctuations.

Being a more general result, the large deviation scaling allows to obtain informations about the Gaussian fluctuations directly from the knowledge of $I(a)$. Let us first note that in the large deviation limit of large $T$ the distribution function concentrates around the most probable value $a_m$ for which $I'(a_m)=0$, and that in the limit of large $T$ this value corresponds also to the average, that is $a_m=\mu=\mathbb{E}[A]$. Expanding the rate function in $a$ around this value, one finds that in the large deviation limit neglecting terms $O((a-\mu)^3)$ the distribution is $\rho(a) {\asymp}\mbox{e}^{-T(a-\mu)I^{''}(\mu)/2}$, that is a Gaussian distribution with variance given by the curvature of the rate function around the most probable value $1/I^{''}(\mu)$.

The specific form of the distribution can be obtained expanding the scaled cumulant generating function for small values of $k$ and using the Gartner-Ellis theorem (see Appendix~\ref{app2}), which results in the following quadratic form for the rate function
\begin{equation}\label{eq:approximation_rate_function_main}
I(a)\approx\frac{(a-\mu)^2}{2\sigma^2\tau_c}.
\end{equation}
We see that this corresponds to the Gaussian scaling predicted by the central limit theorem for the time average of a correlated process. A necessary condition for the approximation to hold is ${|a-\mu|}/{\sqrt{2\sigma^2 \tau_c}<1/\sqrt{T}}$, which is consistent with the expected scaling of Gaussian fluctuations. 

For an observable that is Gaussian distributed the quadratic form above is exact. For more general processes however the rate function contains more information than just the Gaussian fluctuations. The higher order  derivatives of the rate function correspond to higher order cumulants, and describe fluctuations beyond the Gaussian approximation. The most interesting aspect of studying the rate function is the reconstruction of the tails beyond the Gaussian bulk. Equation \ref{eq:approximation_rate_function_main} however can still be of interest, as discussed in section \ref{sec:appl} and for different applications in  \cite{bouchetLargeDeviationsFast2016,Ragone&al2018,Galfi2019,Ragone_Bouchet2020,Galfi_Lucarini2020}.

If the process under study is an ergodic Markov process, the large deviation functions can be computed using the Donsker–Varadhan theory of additive Markov processes, essentially extending to continuous time the results presented in  Sec.~\ref{sec:theo}. For most applications in geophysical fluid dynamics or the climate sciences however, the picture is much more complex, as the system typically has an extremely complex (deterministic) dynamics, whose equations are in some cases not even known (e.g., real world climatic observations, or even climate models data, given the complexity and relative opacity of the code of these numerical models). Part of the problem can be bypassed by taking the assumptions discussed in the introduction of Sec. \ref{sec:ldds} and more in details in Sec. \ref{sec:chaotic}, and treating the output of the system as an effective ergodic Markov process. However, it is still necessary to understand how one can compute the large deviation functions empirically, in the (frequent) cases when this is the only alternative.

Here we give a summary of a possible procedure, presented in more details in \cite{Rohwer2015,Ragone_Bouchet2020}. Let us assume that we have a time series of an observable $A(t)$ from time 0 to time $T$. The idea is to proceed with a block-averaging approach, and divide the time series in $N_b=T/\tau_b$ blocks of length $\tau_b>>\tau_c$. Since the length of the block is much larger than the autocorrelation time, the time averages 
\begin{equation}
a^j_{b} =\frac{1}{\tau_b}\int_{j\tau_b}^{(j+1)\tau_b} A(t)\, \mathrm{d}t,\,\,\,\,\,\,\,\,\,\,\,\,j=1,..,N
\end{equation}
can be considered as sum of $\tau_b/\tau_c$ independent values, possibly leading to convergence to a large deviation result. Additionally, the $N_b$ values of $a^j_{b}$ can be considered as independent realizations of the process $a_{b}=\frac{1}{\tau_b}\int_0^{\tau_b}A(t)\, \mathrm{d}t$, and they can be used to compute the expectation values in the definitions of the large deviation functions as ensemble averages, and to study the convergence to the limit for $\tau_b\rightarrow +\infty$.

The large deviation functions for large but finite values of $\tau_b$ can be computed in two ways. One way is to attempt to estimate directly the rate function by computing  
\begin{equation}\label{eq:rftau}
 I_{b}(a)=-\frac{\ln \rho(a_{b}=a)}{\tau_b},
\end{equation}
that gives an estimate of the rate function for finite $\tau_b$, up to an additive constant due to the prefactor in the large deviation scaling. The convergence of $I_b(a)$ to a limit function $I(a)$ can then be evaluated for each value of $a$ by increasing $\tau_b$ until the value reaches a plateau up to a given tolerance error. This approach has the advantage of being extremely straightforward. However, it requires a large amount of data to obtain a relative error that is constant with $\tau_b$ \cite{Rohwer2015}, and it suffers from the drawback that it is not easy to study precisely the convergence of the estimators, as it is based on the estimate of a probability density.

A second way consists of computing the scaled cumulant generating function first, and then use the Gartner-Ellis theorem to obtain the rate function \cite{Rohwer2015,Ragone_Bouchet2020}. This method has the advantage of not involving the estimate of a probability distribution and of requiring less data to achieve a similar precision. The method however suffers from the problem of only being able to define upper bounds to the statistical errors on the values of $I(a)$.

In typical climate applications both methods require a substantial amount of data to go beyond the Gaussian fluctuations \cite{Galfi2019,Ragone_Bouchet2020}. In the direct method the estimate of the non-Gaussian tails of the rate function is corrupted by the inability of properly computing the probability density function in ranges dominated by sparse data and outliers. In the indirect method the estimate of the tails of the scaled cumulant generating function becomes artificially linear for large values of $k$. For these values the estimate of the generating function is dominated by the contribution of the outliers of $a$. Both methods therefore fail to provide reliable estimates more or less for the same range of values of the fluctuations. A solution to the problem of estimating the tails of the large deviation functions in numerical models is given by the use of rare event algorithms, as discussed in \cite{Ragone_Bouchet2020} and Sec. \ref{sec:applres}.

Independently from the method chosen, there is always a delicate interplay between the autocorrelation time of the process, the mixing time, the time scale of the block averaging, and the time necessary to converge to the large deviation limit, that has to be considered. First of all, for a simple process with exponential autocorrelation function, the integral autocorrelation time and the mixing time are of the same order of magnitude. However, in more complex dynamics the picture can be more complicated, and the integral autocorrelation time may not be a good ``time unit" to estimate the time scales of convergence to the large deviation limit \cite{Ragone_Bouchet2020}. Secondly, for time series of finite length there is a practical trade off between $\tau_b$ and $N_b$. Larger values of $\tau_b$ mean better convergence to the large deviation limit, but a smaller number of samples $N_b$ and larger statistical errors. On the other hand, larger values of $N_b$ mean good statistics and small statistical errors, but poor convergence (if at all) to the limit values. 

It is therefore important when performing these analysis to provide a systematic study of the convergence and of the statistical errors, to identify the best compromise and assess the robustness of the statistical estimators used. As a general rule it would be probably better to use both approaches side by side, as suggested by \cite{Kwasniok2019}. See Sec. \ref{sec:appl} for a discussion on different analysis performed on climate data by \cite{Ragone&al2018,Galfi2019,Ragone_Bouchet2020,Galfi_Lucarini2020}.

\subsection{Large Deviation Laws in Chaotic Systems}\label{sec:chaotic}

In the previous subsections we have shown how large deviation laws emerge when looking at the statistical properties of stochastic dynamical systems. 
A different point of view on the problem suggests that it is possible to establish foundations for the study of nonequilibrium systems by taking advantage of the framework of chaotic dynamics \cite{Ruelle1999,Gallavotti2014}. More specifically, the idea is that nonequilibrium ensembles can be described by the Sinai-Ruelle-Bowen (SRB) measure supported by the attractors of Axiom A systems \cite{Eckmann1985,Ruelle1989}. These concepts are briefly and informally recapitulated below.


Let's consider a flow on a smooth compact manifold $\mathcal{M}$ of dimension $n$ such that $S^t x_0$ is the evolution at  time $t$ of the  $t_0=0$ initial condition $x_0\in \mathcal{M}$. Such evolution can be represented in differential form as $\dot{x}=b(x)$, where we have removed the stochastic component from  (\ref{eq:diffusion}). We assume that the flow has a {\color{black}compact} invariant set $\Lambda$ such that $S^t \Lambda=\Lambda$ for all $t\geq 0$. 
We also assume that $\Lambda$ is not decomposable in two sets that are also invariant and that there is a neighborhood $U$  of $\Lambda$ such that $U\supset \Lambda$, $S^t U\subset U$  $\forall t\geq 0$ and $\Lambda=\cap_{t\geq 0} S^t U$. {\color{black}$U$ is also called the forward isolating neighborhood of $\Lambda$. We assume that $0<m_n(U)<\infty$, where $m_n$ is the $n-$Lebesgue measure. At practical level, one can think $U$ as a finite-precision approximation of the true attractor $\Lambda$ and is the asymptotic set that is \textit{de facto} experimentally accessible in numerical simulations and experiments. Indeed, one can assume that, if an orbit is initialised in the basin of attraction of $\Lambda$ (the union of all orbits which converge towards $\Lambda$), its forward evolution enters after a possibly long transient the set $U$. $U$ is contained in the basin of attraction.}  

We now assume that on $\Lambda$ the flow is hyperbolic, which means that in $\Lambda$ we can continuously split the tangent space as the sum of three nontrivial subspaces $T_\Lambda \mathcal{M}=E^s+E^u+E^n$, where there are constant $c$ and $\lambda$ such that $D S^t(v)\leq c\exp(-\lambda t)|v|$ if $v\in E^s$ and $D S^{-t}(v)\geq c\exp(-\lambda t)|v|$ if $v\in E^u$; additionally, the dimensionality of $E^s$ and $E^u$ is constant in $\Lambda$. In simpler terms, infinitesimal perturbations grow if initialised along $E^u$ (unstable component) and shrink if initialised along $E^s$ (stable component). Finally, we assume that $E^n$ - the neutral space - is one-dimensional and associated with the direction of the flow. No contraction nor expansion takes place along $E^n$. We finally assume that $\Lambda$ is densely populated by (unstable) periodic orbits \footnote{This last hypothesis, which seems unnatural, has important consequences both at dynamical and statistical level, see \cite{svita88}.}. We then have that $\Lambda$ is an Axiom A attractor and the evolution law $S^t$ defines an Axiom A system. {\color{black}Note that if the flow is on the average contractive ($\nabla b<0$), the Hausdorff dimension of the $\Lambda$ is strictly smaller than $n$ \cite{Eckmann1985}. Therefore, choosing an initial condition randomly (with respect to the natural Lebesgue measure) in the set $U$, there is zero probability to choose a point belonging to $\Lambda$. This further clarifies the experimental relevance of $U$.} 

In general, any invariant measure $\nu$ is such that
\begin{equation}
\int \psi(y) \nu(\mathrm{d}y)=\int \psi(S^t y) \nu(\mathrm{d}y)= \int \psi(y) \Pi^t\nu(\mathrm{d}y)
\end{equation}
where $\Pi^t$ is the so-called transfer operator, which pushes measures forward in time \cite{Baladi2000}. Note that the previous equation establishes the transfer operator as the adjoint of the evolution operator, so that $\Pi^t=\left(S^t\right)^T$. An invariant measure is a fixed point of the transfer operator for all $t\geq 0$: $\Pi^t\nu=\nu$. For Axiom A systems one can define a special SRB ergodic measure  $\mu_{SRB}$ with support on $\Lambda$ such that for almost all (with respect to the measure $m_n$) $x\in U$ and for each continuous observable $\psi$, we have that 
\begin{equation}\label{srb}
\lim_{T\rightarrow\infty}\frac{1}{T} \int_{0}^T \psi(S^\tau x) \, \mathrm{d}\tau = \int \psi(y) \, \mu_{SRB}(\mathrm{d}y) =\mu_{SRB}(\psi).
\end{equation}
In other terms, long time averages computed from initial conditions in $U$ give the expectation value computed according to the invariant measure $\mu_{SRB}$ on $\Lambda$. It is very important to note that we are not requesting that the initial condition is on the attractor $\Lambda$, but in its neighborhood $U$, which has finite measure, and that is, physically speaking, experimentally accessible. The previous equation implies that after a certain transient almost any trajectory initialised in $U$ explores the attractor $\Lambda$ according to invariant measure $\mu_{SRB}$. This measure is, indeed, the one that is selected by any finite-precision operation on the system. The physical relevance of $\mu_{SRB}$ is further supported by the fact that it coincides with the zero-noise limit of the invariant measure realised when one consider stochastic perturbations of the system above \cite{Ruelle1989}. 

The mathematical setting given above of strange Axiom A attractors gives a possible (yet restrictive) setting for studying chaotic systems. Chaos is usually associated with the \textit{negative} property that divergence of nearby trajectories leads to having a limited time horizon of deterministic prediction. This is the celebrated butterfly effect first discussed by Lorenz \cite{Lorenz1963}. The limits posed by the butterfly effect provide the fundamental reason why improving the skill of a numerical weather forecast system is excruciatingly difficult; see \cite{kalnay2003} for an example of application of the splitting between stable, unstable, and neutral portions of the tangent space in the context of atmospheric predictability. On the other hand, (\ref{srb}) shows that chaos makes it possible to reconstruct ensemble averages even if we start outside the attractor (but, clearly, within its basin of attraction). Hence, we are able to collect the statistical properties of the system without knowing where precisely its attractor is. Therefore, chaos makes it possible to define at all the climate as the set of statistical properties of the climate system, and makes it operationally feasible to run climate models and interpret their results \cite{Ghil2020}.

Unfortunately, Axiom A systems are far from being generic or even typical, as more general, weaker notions of hyperbolicity have to be used to deal with real-life chaotic systems. Recently, it has been shown that much larger classes of dynamical system  possess SRB measure \cite{Young2002,Climenhaga2017}, thus providing further support to the so-called \textit{chaotic hypothesis} \cite{Gallavotti1995,Gallavotti2014}, which states, roughly speaking, that a chaotic system with many degrees of freedom \textit{de facto} behaves as an Axiom A system and in particular possesses a physically relevant SRB-like invariant measure. 


\subsubsection{Large Deviation Laws for Axiom A systems}\label{sec:ld_A}

We are now able to formulate more precisely the problem of estimating the probability of large deviations of a smooth observable $\psi$ for the system defined above. We want to study the rate of convergence of the average $\psi_{T,x}=\frac{1}{T} \int_{0}^{T} \psi(S^\tau x) \, \mathrm{d}\tau$ for $x\in U$. {\color{black}In other terms, we want to understand the probability of deviations of finite time averages with the respect to the asymptotic result given in (\ref{srb}) which is valid for almost all $x\in U$.}  We adapt below - in a very simplified way - for the case of flows the treatment of the problem presented in \cite{Young1990,Chazottes2015} for maps.

We choose a value $a\in\mathbb{R}$ 
and define the set {\color{black}$B_{T,\psi}^{a,+}=\{x\in U |\psi_{T,x}\geq a\}$. 
From (\ref{srb}), we derive that } $\lim_{T\rightarrow\infty}\frac{1}{T}\log \mathbb{P}(B_{T,\psi}^{a,+})=0$ $\forall a>\mu_{SRB}(\psi)$. 
where $\mathbb{P}=(1/m_n(U))m_n$ is a conditional probability measure on $U$. It is indeed possible to establish in general a large deviation principle for the finite-time averages of $\psi$. {\color{black} We obtain:  
\begin{equation}\label{largedeviationaxioma}
  \lim_{T\rightarrow\infty}\frac{1}{T}\log\left(\mathbb{P}(B_{T,\psi}^{a,+})\right)=
  -\inf_{z \in A} I_\psi(z) \qquad \mathrm{with\ }  A=[a,\infty).
\end{equation}
where $I_\psi(z)$ is the rate function. If, instead, we set $a<\mu_{SRB}(\psi)$ and  define the set $B_{T,\psi}^{a,-}=\{x\in U |\psi_{T,x}\leq a\}$, we obtain:
\begin{equation}\label{largedeviationaxiomb}
 \lim_{T\rightarrow\infty}\frac{1}{T}\log\left(\mathbb{P}(B_{T,\psi}^{a,-})\right)=
  -\inf_{z \in A} I_\psi(z) \qquad \mathrm{with\ }  A=(-\infty,a].
\end{equation}
These two results closely mirror what presented in (\ref{eq:ldp0.1})-(\ref{eq:ldp0.1b}). Note that the functional form of the rate function is known but is very non-trivial, as it must take into account the complex nonlinear correlations of the time evolving value of $\psi$ resulting from the chaotic dynamics. One can derive the following expression for the rate function:
\begin{equation}\label{largedeviationaxiomarate}
I_\psi(z)=-\sup_{\nu\in\mathcal{N},\nu(\psi)=z}\left(h(\nu) -\Sigma_\lambda^+(\nu)\right).
\end{equation}
where $\mathcal{N}$ is the set of the invariant measures of the system (if $\nu\in\mathcal{N}$ then $\Pi^t\nu=\nu$ $\forall t$). Note that we impose as a constraint that the expectation value of $\psi$ computed using the measure $\nu$ must be equal to $z$. 
In the previous expression $h(\nu)$ is the Kolmogorov-Sinai entropy, which measures the rate of creation of information, while $\Sigma_\lambda^+(\nu)$ indicates the sum of the positive Lyapunov exponents w.r.t. $\nu$. The positive Lyapunov exponents measure the possible asymptotic rates of stretching of infinitesimal perturbations aligned along the unstable manifold \cite{Eckmann1985}, see discussion in Sec. \ref{Lyapunov}. Note that $I_\psi(z)\geq 0$ because for all invariant measures $h(\nu) \leq\Sigma_\lambda^+(\nu)$. The rate function attains its unique minimum for $z=\mu_{SRB}(\psi)$, which is realised when $\nu=\mu_{SBR}$. 
Indeed, in this case one has that the Pesin identity is verified: $h(\mu_{SBR})=\Sigma_\lambda^+(\mu_{SBR})$ \cite{Ruelle1989}. Note that the Pesin identity is implicitly assumed as valid in most numerical applications where one wants to evaluate the Kolmogorov-Sinai entropy.

Summarizing, the large deviation law defined with respect to initial condition in the neighbourhood of the attractor $U$ \textit{is determined}  by all the non-SRB invariant measures supported on the attractor $\Lambda$.} 
While in the stochastic case the rate function is determined by the cost of deviating from the deterministic trajectory defined by the zero noise limit, in the deterministic case the rate function is, in some sense, determined by the cost of deviating from the reference SRB measure. In geophysical terms, such alternative invariant measures correspond to exceedingly unlikely, yet possible, exotic climates.

Following \cite{Chazottes2015}, an alternative way to derive a more direct and practically accessible definition of the rate function relies on using the scaled cumulant generating function described in Sec. \ref{sec:ldds_ta}. One defines:
\begin{equation}
\lambda_\psi(k) = \underset{T\rightarrow +\infty}{\lim} \frac{1}{T} \log{ \int \left[e^{k\int_0^T\mathrm{d}t \psi(S^tx)} \right] \,\mu_{SRB}(\mathrm{d}x) },
\end{equation}
Taking advantage of the G\"artner-Ellis theorem, whose hypotheses apply in the case of Axiom A systems, one derives the  rate function $I_\psi(\Delta)$ as Legendre transform of $\lambda_\psi(k)$ as follows: $I_\psi(z)=\sup_{k \in \mathbb{R}}[z k-\lambda_\psi(k)]$. The possibility of using the same construction for the rate function in both stochastic and Axiom A system clarifies the fundamental similarity, at statistical level, between the two.  

Near the minimum of the rate function $z_0$, the large deviation law describes the central limit theorem. The results mirror precisely what shown in the previous section. Indeed, mirroring 
(\ref{eq:approximation_rate_function_main}), for small values of $z$ one has $I_\psi(z)\approx (z-z_0)^2/(2\sigma_\psi^2\tau_\psi^c)$, where $\sigma_\psi^2=\int\mu_{SRB}(dy)(\psi(y)-\mu_{SRB}(\psi))^2$ is the variance of $\psi$ and $\tau_\psi^c=\int_{-\infty}^{\infty}\mathrm{d}t C_\psi(t)$ 
where $C_\psi(t)=1/\sigma_\psi^2\int\mu_{SRB}(dy)(\psi(y)-\mu_{SRB}(\psi))(\psi(S^t(y))-\mu_{SRB}(\psi))$ is the time-lagged correlation of the observable $\psi$ and $\tau_\psi^c$ is the integrated correlation time. We can interpret $\tau_\psi^c$ as a normalising factor for time in such a way that $M$ consecutive observations of the observable $\psi$ correspond to $M/\tau_\psi^c$ approximately independent stochastic variables. 

Note that the possibility of establishing large deviation laws for Axiom A systems is intimately related to the fact that for such systems one observes a rapid decay of correlations for observables (loss of memory being another characterization of chaotic dynamics). Additionally, under certain conditions it has been possible to prove the existence of large deviation laws also for systems obeying weaker notions of hyperbolicity with respect to the Axiom A case. Such large deviation laws might diverge from the exponential form described above in the case the system has slow decay of correlations \cite{rey-bellet_young_2008,Chazottes2015}.



\section{Applications of large deviation theory to geophysical systems}\label{sec:appl}

\subsection{Large deviation of time averaged observables and rare events}\label{sec:applpers}
The previous section clearly indicates that we can study persistent extremes in geophysical flows obliviously to the fact of whether we are considering a deterministic or stochastic framework for the dynamics. This is a very important point both on practical and epistemological grounds.

The basic idea that connects a persistent extreme event to large deviations is that the sample mean recorded during the persistent event can be regarded as a large deviation from the long term mean. Nonetheless, to be able to answer the question whether the respective sample mean, besides of being large, represents indeed a large deviation, one has to check the convergence to the large deviation limit based on the rate functions, as described in detail in Sec.~\ref{sec:ldds_ta}.

Geophysical observables are usually correlated in time and space, which should be considered in the computation of the rate functions (see also Sec.~\ref{sec:ldds_ta}). In case of weakly correlated observables (i.e. two values have an exponentially decreasing correlation if they are far enough from each other, in time or in space), it is recommended to re-normalise the rate function in (\ref{eq:rftau}) additionally by the integrated auto-correlation $\tau$ \cite{Galfi2019}
\begin{equation}\label{eq:rftau2}
 I_{n}(a)=-\frac{\ln p(A_{n}=a)}{n/\tau}.
\end{equation}
Thus, we take into consideration that the averaging block length $n$ consists of $n/\tau$ (nearly) independent data points. Please note that the notation in (\ref{eq:rftau2}) is slightly different from the one used for (\ref{eq:rftau}), with $n$ replacing $\tau_b$. The normalization by $n/\tau$ is useful especially for comparing rate functions of observables with different characteristic scales. 

\begin{figure}
    \centering
    \includegraphics[width=1\textwidth]{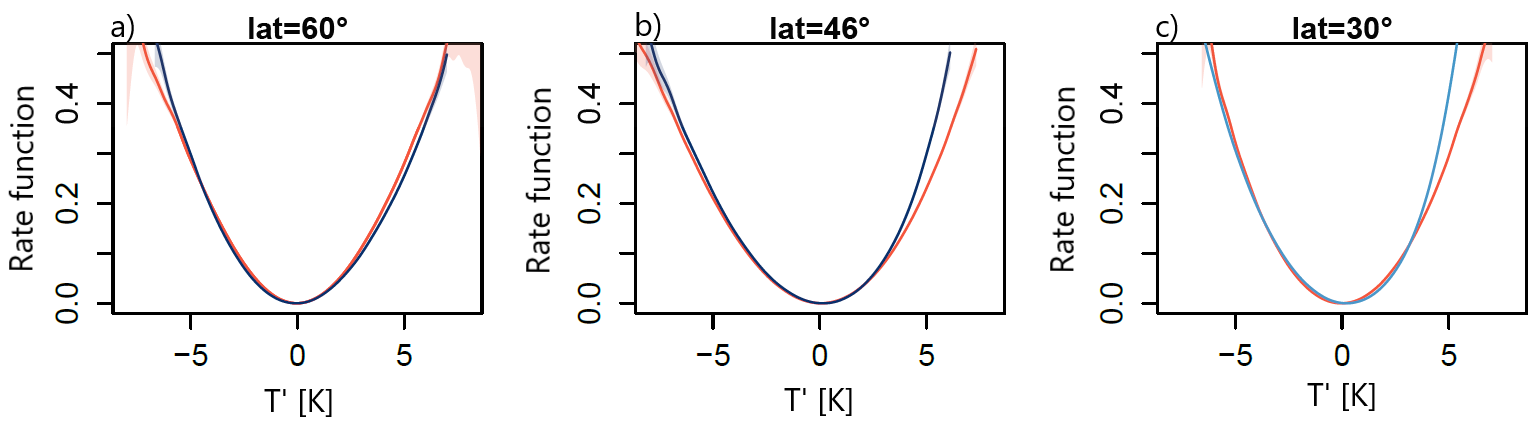}
    \caption{The rate functions obtained according to (\ref{eq:rftau2}) based on averaging the near-surface grid point temperature in time (red) and along a latitudinal band (blue) at latitudes a) $60^\circ$, b) $46^\circ$ and c) $30^\circ$ in the GCM PUMA illustrate the equivalence between temporal and spatial large deviations, as well as the increasing anisotropy as one goes from North to South, shown by a slight deterioration of the equivalence. From \cite{Galfi2019}. }
    \label{fig:univ}
\end{figure}

\cite{Galfi2019} compared rate functions of near-surface air temperature obtained based on (\ref{eq:rftau2}) and found that by looking locally in space at long time averages agrees with what is obtained, instead, by looking locally in time at large spatial averages along the latitude. They used the simplified General Circulation Model (GCM) of the atmosphere PUMA \cite{fraedrich1998} without orography and performed simulations in a nonequilibrium steady state. These results suggest that, in case of homogeneous statistics in both time and space, the apparent discrepancy between temporal and spatial large deviations is only due to the difference between temporal and spatial scales. If one normalises the rate functions based on the number of independent data, one finds a universal function describing both temporal and spatial large deviations (Fig.~\ref{fig:univ}). Hence, the correspondence between properly scaled temporal and spatial large deviations extends the universality emerging from the asymptotic nature of the LDP, discussed in Sec.~\ref{sec:introue}, by a further level connecting the dimensions of time and space. This connection can be useful for the analysis of observational data, for example. Let us consider the common situation when only time series at sporadic locations are available. If the universality between time and space is satisfied, one can derive the spatial large deviations based on the information about temporal averages and the characteristic spatial scale.

The above connection between spatial and temporal large deviations is valid for asymptotic scales, i.e. if the rate functions have converged to the asymptotic one. This is fulfilled in the atmospheric model used by \cite{Galfi2019} for temporal and spatial scales of around $20 \tau$, where $\tau$ is either the temporal or the spatial integrated auto-correlation. Obtaining the spatial large deviation law can represent a problem though, due to heterogeneous orography or even due to the limited size of the Earth, which is in certain cases simply not large enough to reach asymptotic levels. Hence, the possibility to make use of the asymptotic connection between temporal and spatial large deviations can be inhibited by a strong anisotropy between the temporal and spatial dimensions. This is illustrated by Fig.~\ref{fig:univ} showing that going from North towards the Equator, the correspondence between the temporal and spatial rate functions deteriorates slightly, due to the increase of characteristic spatial scales (391 km at $60^\circ $, 732 km at $45^\circ$, 1292 km at $30^\circ$) - and consequently a slower convergence in space - while the temporal scales do not change substantially (1.32 days at $60^\circ$, 1.05 days at $45^\circ$, 1.61 days at $30^\circ$). 
Besides the purely spatial and temporal large deviations of local temperature observables, the universal rate function describes also large deviations in time of spatially averaged observables, as long as the spatial averaging is performed along the same latitudinal band, and one disregards the effect of orography. However, the spatial averaging length is  crucial here. Due to strong spatial correlations on certain synoptic scales ($\approx 2000-4000$ km), the asymptotic rate function of spatial temperature averages is wider than the universal function, showing that large deviations on these spatial scales are more probable than on any other scales. Interestingly, these spatial scales correspond approximately with the ones of persistent synoptic disturbances, leading to high-impact heatwaves.

\begin{figure}
    \centering
    \includegraphics[width=1\textwidth]{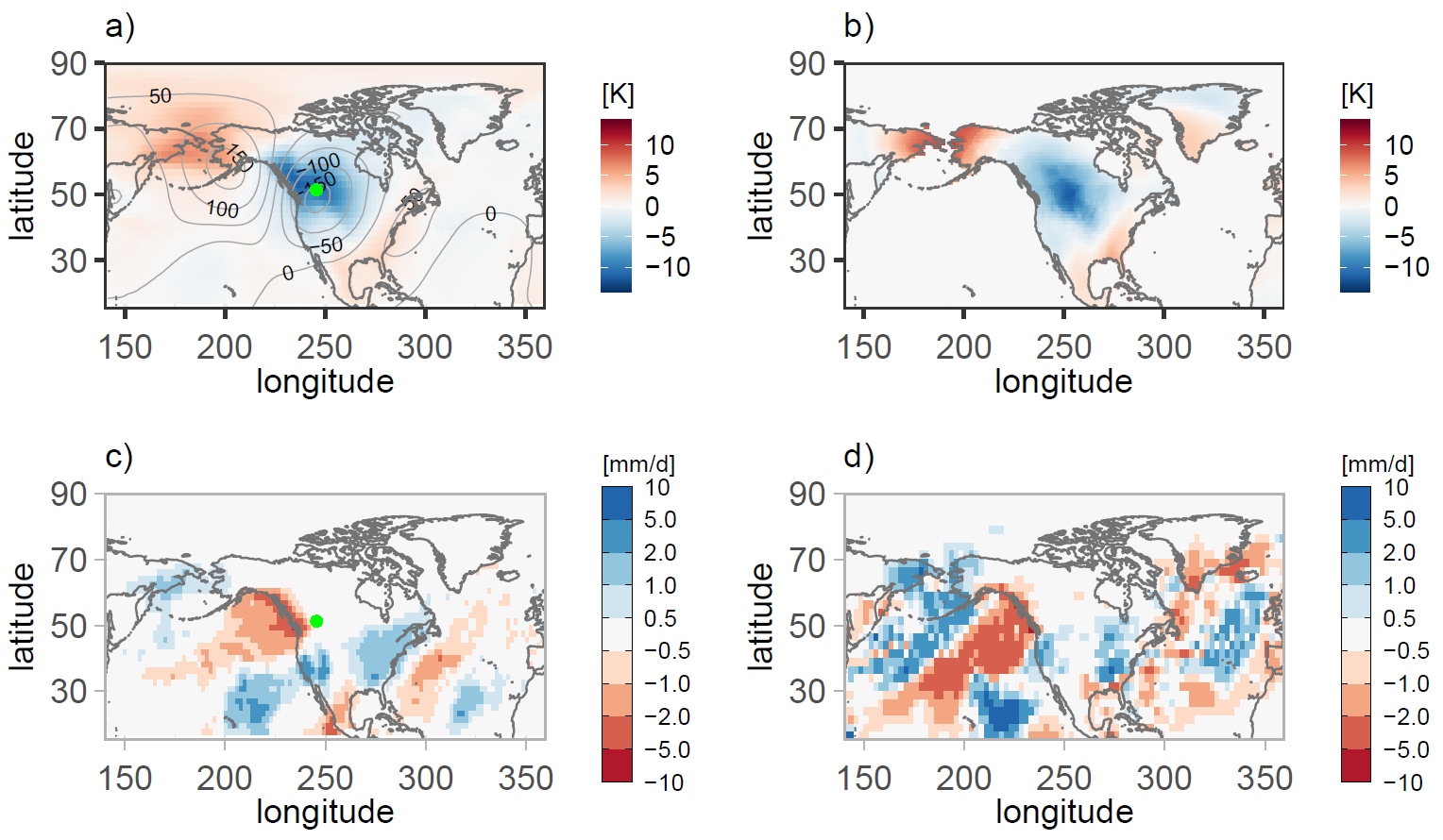}
    \caption{Composites of winter a) surface air temperature and 500 hPa geopotential height and c) precipitation anomaly fields corresponding to surface temperature anomalies of -10.5 K lasting 30 days in the locale indicated by the green dot from pre-industral CMIP6 simulations performed with the MPI-ESM-LR model. Observed b) surface temperature (CRU-TS 4.04) and d) precipitation (GPCP v2.3) anomalies for February 2019. The isolines in a) indicate 500 hPa geopotential height anomalies.}
    \label{fig:namcs}
\end{figure}

\cite{Galfi_Lucarini2020} studied the connection between large deviations of surface air temperature and persistent temperature events, like heatwaves or cold spells, more thoroughly. They used CMIP6 simulations \cite{Eyring2016} performed with the state-of-the-art Earth System Model (ESM) MPI-ESM-LR \cite{Giorgetta2013} with seasonal cycle and orography. Although, the universality in time and space of large deviations does not hold in the presence of orography, large deviations in time of local temperature are still connected to anomaly fields extended in space and persistent in time. They show that large deviations in time of surface temperature at one selected grid point (marked by the green dot) are related to spatially extended anomaly patterns (temporal averages of surface temperature and 500 hPa geopotential height anomaly fields), which resemble anomaly patterns from realanysis data during two high-impact persistent events, the 2010 Russian Heatwave and the 2010 Mongolian Dzud. For the model results, \cite{Galfi_Lucarini2020} use nonequilibrium steady state simulations with pre-industrial CO${_2}$ concentration, thus pointing out that both kind of events are manifestations of the natural variability of the climate system. Furthermore, these events seem to be typical persistent events from the perspective of LDT. {\color{black}The fact that by using LDT one is able to capture the dynamical features of typical events - within the class of the very unlikely ones - is further explored later in Sec. \ref{roguew}; see also discussion in \cite{Grafke2019}.}
Thus, it seems that, based on large deviations, we can identify spatial structures of typical persistent events of our system, and, at the same time, obtain probability estimates for their occurrence. We note that the term ``typical'' does not refer to the magnitude or severity of the event. Furthermore, precipitation anomaly patterns selected based on large deviations of temperature are also similar to the one during the 2010 Russian Heatwave (reanalysis data for August 2010) supporting the conclusion that the selection method based on large deviations is meaningful from a dynamical point of view. Remote effects are captured as well, as shown by the positive precipitation anomalies over the Indian subcontinent and Pakistan corresponding to the devastating floods in that region during the Russian Heatwave (see Fig.~2 in \cite{Galfi_Lucarini2020}). Another case study related to the 2019 North-American cold spell yields as well a noticeable similarity for both temperature and precipitation anomaly fields between the large deviation based selection of modelled (same model as in \cite{Galfi_Lucarini2020}) fields and reanalysis data for February 2019 (Fig.~\ref{fig:namcs}). For the composites in Fig.~\ref{fig:namcs}a,c we used CMIP6 pre-industrial control runs performed with the MPI-ESM-LR model, while Fig.~\ref{fig:namcs}b (c) represents observed ST (precipitation) anomalies from February 2019 with respect to the 1981-2010 long-term monthly mean based on the CRU-TS 4.04 \cite{Harris2020} (GPCP v2.3 \cite{Adler2003}) data set. 

\begin{figure}
    \centering
    \includegraphics[width=1\textwidth]{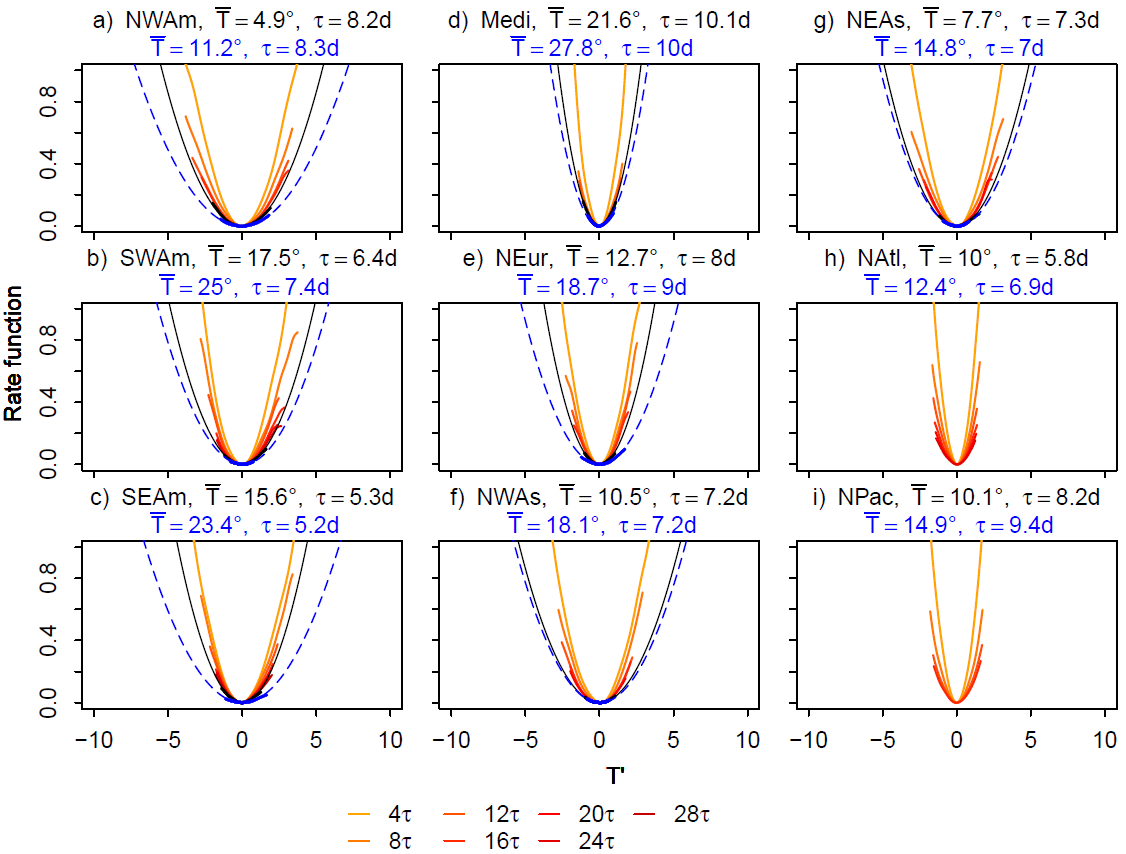}
    \caption{Summer rate functions for surface air temperature for increasing averaging windows (see legend) for a) Northwest, b) Southwest, and c) Southeast America, d) the Mediterranean, e) North Europe, f) Northwest and g) Northeast Asia, h) the North Atlantic, and i) the North Pacific. The black (blue dashed) line represents rate functions obtained via multi-seasonal averages for the pre-industrial (quadruple $\mathrm{CO_2}$) run. On top, the mean surface temperature and integrated auto-correlation, $\tau$, for the control (black) and quadruple $\mathrm{CO_2}$ (blue) runs. From \cite{Galfi_Lucarini2020}.}
    \label{fig:rf_gw}
\end{figure}

As we have seen, LDT provides the probability of large sample averages, which then can be related to high-impact persistent events. Fig.~\ref{fig:rf_gw} shows rate function estimates from \cite{Galfi_Lucarini2020} of summer surface air temperature for different regions over the Northern Hemisphere for pre-industrial and quadruple CO$_2$ concentration experiments (MPI-ESM-LR model). By comparing the best estimates of the pre-industrial (black lines) and quadruple CO$_2$ (blue dashed line) experiments, we notice that the rate function becomes wider over North-American and European regions as an effect of the increased CO$_2$ concentrations, suggesting that heatwaves in summer become more frequent and longer lasting. Opposite results are found for cold spells during winter \cite{Galfi_Lucarini2020}. Although these results are unsurprising considering the effects of global warming, they show the utility of large deviation rate functions to quantify the changing probability of the considered persistent events.

To a good approximation, rate functions of temperature found in \cite{Galfi2019,Galfi_Lucarini2020} are symmetric and close to a parabolic form. Consequently, \cite{Galfi_Lucarini2020} propose to use the formula for the Gaussian approximation to estimate the probability of observing an average anomaly of temperature $T_n$ of amplitude $a$ over a period of $n$ days: 
\begin{equation}\label{eq:qappr}
\log\left(p(T_n=a)\right)\approx - n I_T(a),\quad  I_T(a)=\frac{a^2}{2\tau_T\sigma_T^2}
\end{equation}
where $\sigma_T^2$ is the daily variance of the field and $\tau_T$ is the integrated auto-correlation of the variable $T$ in units of days, and $n\geq10\tau_T$. This approximate formula, can be useful for estimating the probability of occurrence of persistent large temperature deviations, i.e. heat wave or cold spells, from easily accessible statistical properties of the temperature fields. We note that the temporal resolution of the time series does not have to be daily, the formula can be used for any arbitrary resolution and for averaged series as well. However, one expects to obtain more robust results by using:
\begin{equation}\label{eq:qapprr}
p(T_{n'}=a')\approx 
p(T_n=a)\exp\left(\frac{na^2-n'a'^2}{2\tau_T\sigma_T^2}\right).
\end{equation}
Based on (\ref{eq:qapprr}) one can estimate the probability of occurrence of events of amplitude $a'>a$ and length $n'\ge n$ from the knowledge of events of amplitude $a$ and length $n$.
These results can be reframed in terms of average return periods of the events, which are simply the inverse of the occurrence probabilities. 


When dealing with finite sized data, it is advisable to define an optimal averaging block size $n^*$. This is usually the minimal block size, for which the rate function belongs to the asymptotic regime, i.e. $I(n^*) \approx I(n>n^*)$. In case of the GCM used in \cite{Galfi2019}, $n^* \approx 20 \tau$, where $\tau$ is the integrated auto-correlation. In case of the ESM from \cite{Galfi_Lucarini2020}, the optimal averaging length is between $12-20 \tau$ (1-2 months) for land areas, depending on the geographic region and the considered season. Over the oceans convergence, if any, takes longer than the duration of a season. As mentioned above, \cite{Galfi2019} and \cite{Galfi_Lucarini2020} estimate rate functions directly based on the probability density functions (pdf's) of sample averages according to (\ref{eq:rftau2}), which is a slightly modified version of (\ref{eq:rftau}). 
\cite{Ragone_Bouchet2020} follow a different strategy and estimate the rate function of spatially averaged temperature over Europe as Legendre transform of the scaled cumulant generating function (\ref{scaledcumulant}). They rely on an intermediate complexity climate model, run under perpetual summer conditions, and find a convergence to the large deviation limit at $n^*=3$ years.
Moreover, they find an asymmetric rate function, whereas the ones found by \cite{Galfi2019} and \cite{Galfi_Lucarini2020} are generally symmetric.

The studies conducted by \cite{Galfi2019} and \cite{Galfi_Lucarini2020}, on the one side, and \cite{Ragone_Bouchet2020}, on the other side, show some contradictory results although they both look at temperature observables. Reasons for the discrepancy could be related to differences in the used models, estimation methods, and spatial averaging regions. Let us shortly discuss the possible reasons one by one. Both \cite{Galfi2019} and \cite{Galfi_Lucarini2020} find that the convergence to the large deviation limit is quite fast, i.e. shorter than the length of a season, although the used models are extremely different in terms of model physics and complexity. Consequently, the model differences are probably not the main reason for the different results, unless the simplified ocean dynamics in the model used by \cite{Ragone_Bouchet2020} does not slow down the evolution of the system too much leading to slow convergence. Despite the fact that the rate function estimation methods based on pdf's and Legendre transforms of the scaled cumulant generating function (see Sec.~\ref{sec:ldds_ta}) can lead to slightly different results, they should provide similar rate functions \cite{Kwasniok2019}. The most feasible reason we can think of is related to the different spatial averaging areas. \cite{Ragone_Bouchet2020} average over Europe until longitude $\sim 25^\circ$ E, leaving out the most continental part of Europe and considering all coastal regions, which can be influenced by the slow ocean dynamics. The slowly decaying serial correlation of this spatially averaged observable can lead to the slow convergence of the rate functions estimates. 


\subsection{Large deviations of finite time Lyapunov exponents}\label{Lyapunov}
After discussing extremely persistent atmospheric states based on large deviations of temperature, we explore, in what follows, the finite time dynamics of the system at a more intrinsic level, based on large deviations of finite time Lyapunov exponents. {\color{black}The goal is to explore whether LDT can be of help for improving our ability to understand the predictability of geophysical flows \cite{Ghil2001d,Palmer2013,Krishnamurthy2019}.}

Lyapunov Exponents (LEs) describe the asymptotic growth or decay of infinitesimal perturbation acting on the trajectory of a dynamical system. Their finite time estimates, the so-called Finite Time Lyapunov Exponents (FTLEs), refer to stability properties of a specific state of the system with respect to a predefined predictability horizon \cite{Boffetta2002,kalnay2003,Vallejo2013,Pazo2013,VannitsemLucarini2016}. 
Large deviations of FTLEs point out extremely stable or unstable  states of the system \cite{Laffargue2013} and provide relevant information on its predictability on time scales that are intermediate between the one given by the inverse of the first LE and ultralong ones \cite{Kwasniok2019,LucariniG2020}.

We briefly and informally introduce the LEs and their finite-time version below. We consider the  autonomous deterministic dynamical system where evolution takes place in an $n$-dimensional compact manifold $\mathcal{M}$ obtained by removing the stochastic component from (\ref{eq:diffusion}):
\begin{equation}\label{eqdiff}
    \frac{\mathrm{d}x}{\mathrm{d}t}=b(x),
\end{equation}
where $x\in\mathbb{R}^n$ is the state vector and $b : \mathbb{R}^n \rightarrow \mathbb{R}^n$ is a smooth drift.
We assume that the system possesses an invariant measure $\rho(\mathrm{d}x)$ supported on a compact attractor $\Omega$. We define the orbit $x(t,x_0)=S^t x_0$ as the result of the evolution of the system after a time $t$ starting from the initial condition $x_0$ inside the basin of attraction of $\Omega$. 
An infinitesimal perturbation $\delta x\in\mathbb{R}^n$ evolves along the orbit according to the linearized equation
\begin{equation}
    \frac{\mathrm{d}}{\mathrm{d}t}\delta {x}(t)={\nabla}b \bigg\rvert_{x= x(t,x_0)} \delta x.
\end{equation}
The propagation of the infinitesimal perturbation between time $t_0$ and $t$ is described by
\begin{equation}\label{eq:tlin}
  \delta x(t)={M}(t-t_0,x_0)\delta {x}(t_0),
\end{equation}
where ${M}:\mathbb{R}^n\rightarrow\mathbb{R}^{n\times n}$ is the so-called tangent linear propagator or resolvent matrix. It follows from (\ref{eq:tlin}) that the growth in time of the Euclidean norm of tangent vectors is controlled by the matrix $M^\mathrm{T}M$. If the system is ergodic (as always assumed in this context), the limit
\begin{equation}
    \lim_{t\to\infty}W(t-t_0,x_0)=\lim_{t\to\infty}[M(t-t_0,x_0)^\mathrm{T}M(t-t_0,x_0)]^{\frac{1}{2(t-t_0)}}
\end{equation}
exists \cite{Oseledec1968} and does not depend on $x_0$. The logarithm of eigenvalues of ${W}$ are called (forward) Lyapunov exponents $\lambda_i$, $i=1,\ldots,K\leq n$.  Typically (in absence of symmetries), one has  $K=n$; additionally, in time-continuous systems one of the LEs has vanishing value as it corresponds to the direction of the flow. If only one vanishing LE is present, the system falls in the special category identified by Pesin as an extremely relevant case of (in general) nonuniform hyperbolicity \cite{Pesin1977}, which provides a natural generalisation to the stricter properties characterising the splitting between the stable and unstable manifold for Axiom A systems, described earlier in Sec.~\ref{sec:chaotic} \cite{Ruelle1989}. Indeed, the number of positive (negative) LEs, taken with their multiplicity if different from one, defines the dimension of the unstable (stable) manifold; see also discussion in Sec. \ref{sec:chaotic}. When sorted in a non-increasing order, the LEs form the so-called Lyapunov spectrum. The proposal by Pesin has had enormous importance for giving solidity and strong foundations to the exploration of the tangent space for many systems of practical relevance, and especially so in a geophysical context \cite{kalnay2003}.


In case one is interested in finite time stability properties instead of asymptotic growth/decay rates, one has to consider estimates over a finite time $\tau$ given by the FTLEs $\Lambda_i(\tau,{x}_0)$, which are the logarithms of eigenvalues of ${{{W}}}(\tau,{x}_0)$. By definition, we have that $\lim_{\tau\rightarrow\infty}\Lambda_i(\tau,{x}_0)=\lambda_i$. Additionally, the long-time averages of the FTLEs computed along the trajectory converge to the corresponding global LEs:
\begin{equation}\label{llle}
    \lambda_i=\lim_{T \to \infty}\frac{1}{T}\int_{t=0}^T \Lambda_i(\tau,S^t{x}_0)\mathrm{d}t=\int\rho(\mathrm{d}\mathrm{x})\Lambda_i(\tau,{x})
\end{equation}
where we have used ergodicity.
Unlike the LEs, the FTLEs are norm-dependent, thus they depend on the computation method used to obtain them, which can be, for example, based on forward integration (as presented above), backward integration, or following the intersections of Oseledec subspaces \cite{Ginelli2007,KuptsovParlitz2012}. 
(\ref{llle}) shows that averages in time of FTLEs converge to the global LEs, thus suggesting that finite time and global LEs can be connected by a large deviation law. This is indeed true for Axiom A systems, and, assuming the chaotic hypothesis, for a wide range of chaotic systems which are not Axiom A, as explained in Sec.~\ref{sec:ld_A}. 

\begin{figure}
    \centering
    a)\includegraphics[width=.47\textwidth]{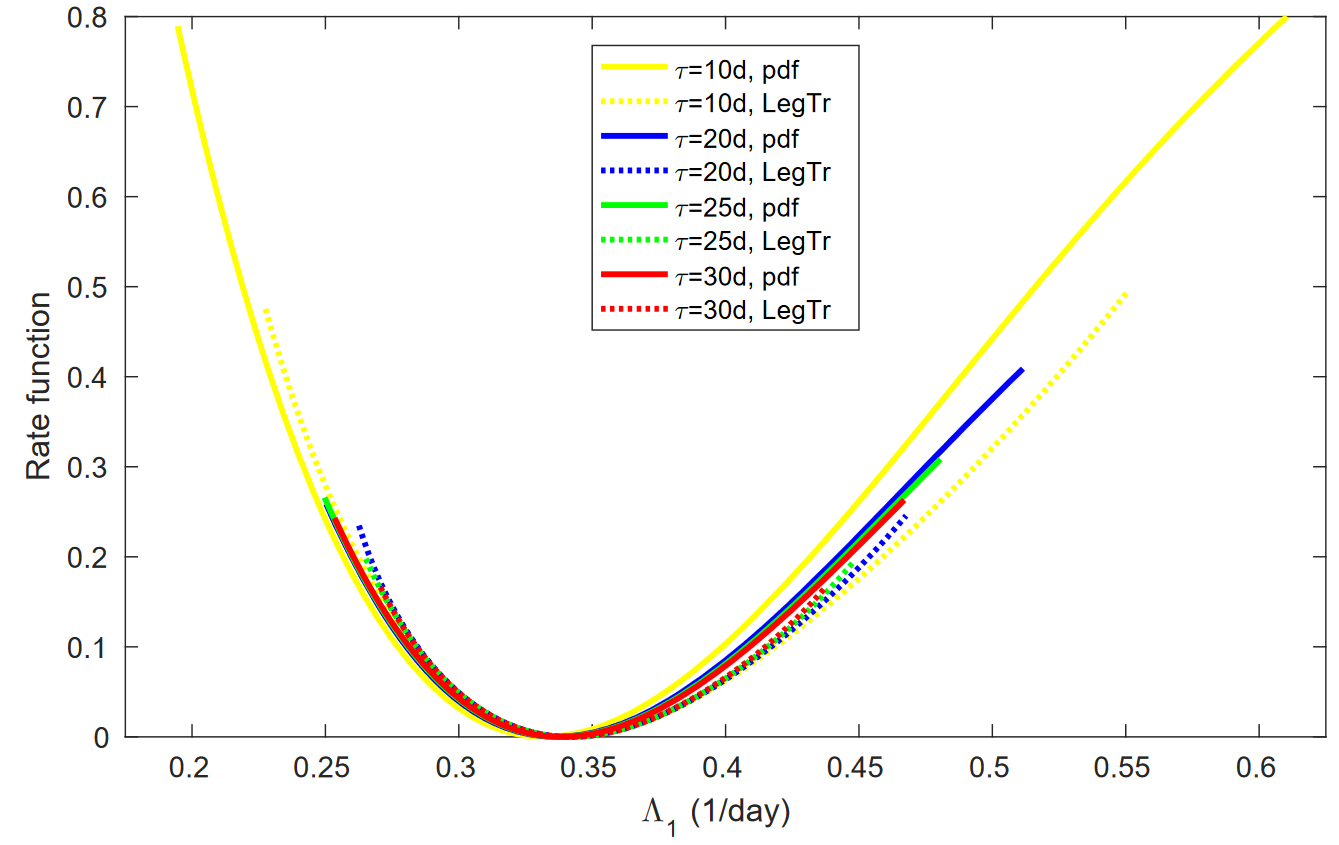}
    b)\includegraphics[width=.47\textwidth]{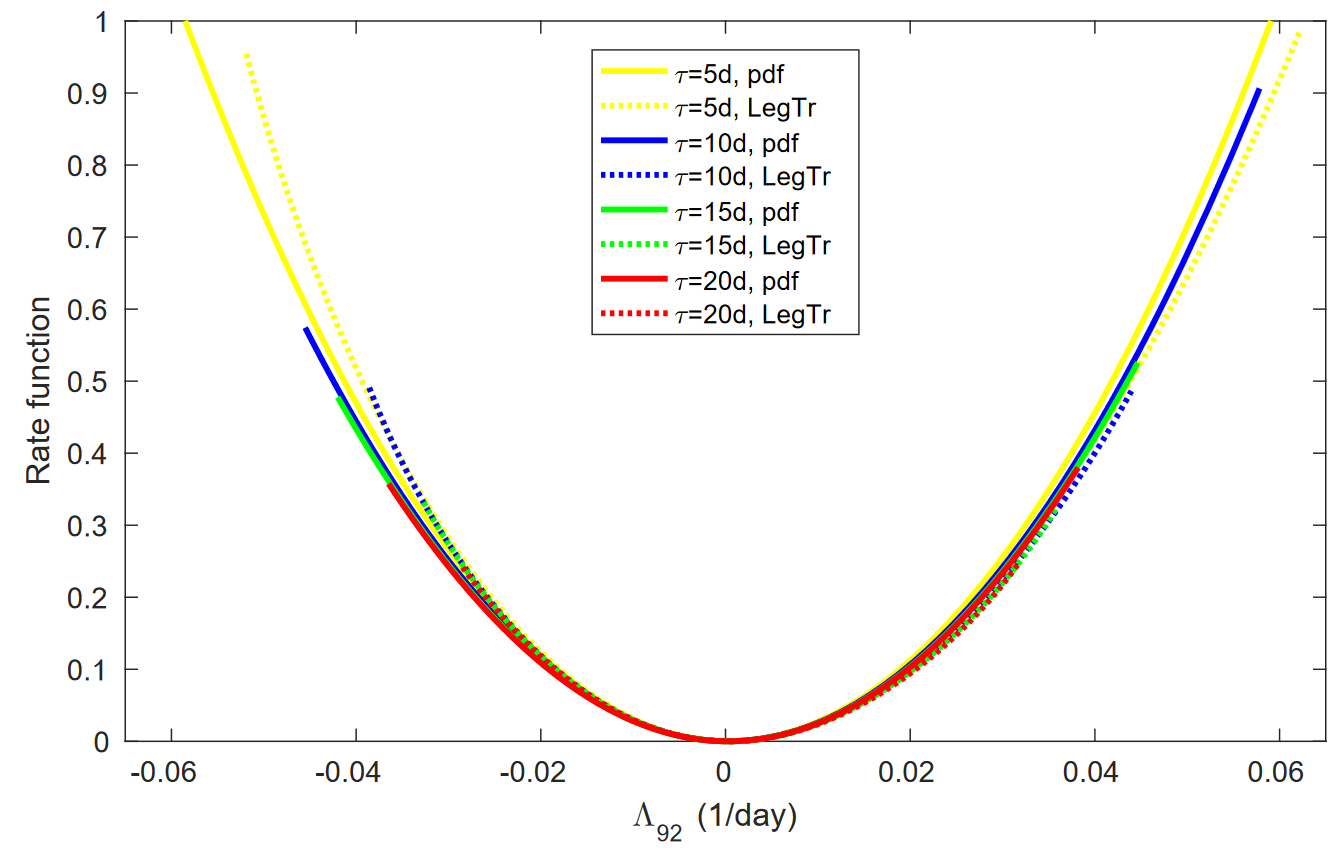}
    \caption{Rate function estimates of a) the first (most positive) FTLE  and b) the 92nd (zero) FTLE in an intermediate complexity atmospheric model. The colors represent the averaging block length, according to the legend. Continuous lines refer to probability density-based estimates, whereas the dashed lines to Legendre-transform based estimates. From \cite{Kwasniok2019}.}
    \label{fig:kwas}
\end{figure}

 A systematic study of large deviations of FTLEs has been performed on geophysical fluid systems with different degrees of complexity \cite{VannitsemLucarini2016,DeCruz2018,Kwasniok2019}. Using a three-layer quasi-geostrophic (QG) atmospheric model with $n=700$, in \cite{Kwasniok2019}, a relatively fast convergence was found to the large deviation limit for all LEs. Convergence was slightly slower and rate functions were asymmetric for the strongly positive/negative LEs (Fig.~\ref{fig:kwas}a), whereas convergence was very fast and rate functions were symmetric in case of the near-zero and weakly positive/negative LEs (Fig.~\ref{fig:kwas}b). 
\cite{VannitsemLucarini2016} studied the convergence of the FTLE in a coupled ocean-atmosphere QG model with $n=36$. They found a convergence and quadratic rate functions for positive and negative LEs, whereas very slow or even no convergence was found for near-zero LEs.

The apparently contradictory results of \cite{Kwasniok2019} and \cite{VannitsemLucarini2016} show that convergence properties of FTLEs strongly depend on the dynamical properties of the system. The Lyapunov spectra of the two systems clearly show the dynamical differences. Whereas the atmospheric model has a monotonically decreasing Lyapunov spectrum, the spectrum of the coupled ocean-atmosphere model has an extended range of slightly positive and negative LEs, which are indistinguishable from zero, corresponding to the so-called slow manifold, which appears due to the slow oceanic dynamics and the interactions between slows oceanic modes and atmospheric ones. 

Generally, for a physical observable, the convergence of finite time averages to the large deviation limit depends on two factors: the strength of the serial correlations and the asymmetry of the probability distribution. Due to serial correlations, larger averaging blocks are needed to reach the asymptotic limit, as compared to i.i.d random variables. Additionally, an eventual asymmetry (or departure from gaussianity) of the parent distribution delays the convergence of the small Gaussian deviations around the minimum of the rate function, as required by the central limit theorem \cite{Galfi2019,Kwasniok2019}. 
Accordingly, in case of the mentioned atmospheric model, on the one hand, serial correlations are so weak that the speed of convergence to the large deviation limit depends dominantly on the degree of non-Gaussianity of the FTLE's parent distribution. On the other hand, in case of the ocean-atmosphere model the strong, long-term correlations dominate the convergence (or non-convergence) to the large deviation limit, including the convergence of the FTLE's to the global LEs. 

\begin{figure}[t]
    \centering
  \includegraphics[width=.6\columnwidth]{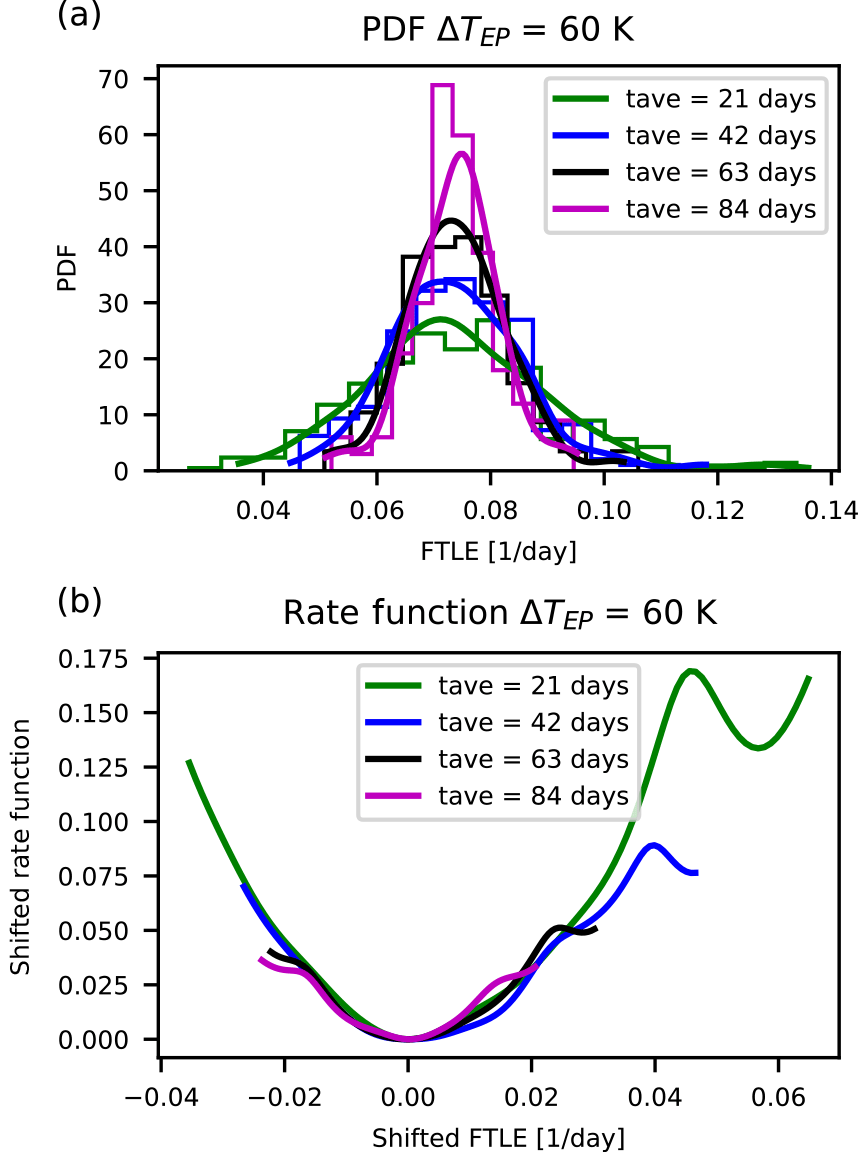}\\
   \caption{Statistics of the the largest FTLEs in a primitive equation model of the atmosphere. a) Probability density functions of the largest FTLE for different averaging times. b) Corresponding estimates of the rate function expressed in terms of the anomaly with respect to the asymptotic value of the first LE. From \cite{DeCruz2018}. \label{DeCruz}}
 \end{figure}

In \cite{DeCruz2018} it was shown that a large deviation law can be found for FTLEs in case of the  primitive equations model PUMA \cite{fraedrich1998} run in a standard setting allowing for the establishment of turbulent, Earth-like  conditions for the atmospheric flow. The primary injection of energy for PUMA occurs by imposing a fixed Equator-to-Pole temperature difference, as typically done for the so-called dynamical cores of atmospheric models \cite{Held1994}. When such temperature difference is set to the fairly realistic value of $60$ K (symmetry is taken between the two hemisphere), one obtains, in the the setting described in \cite{DeCruz2018}, 68 positive LEs which correspond to a very highly-dimensional attractor (Kaplan-Yorke dimension \cite{Eckmann1985} of 172.6). Figure \ref{DeCruz} shows the statistics of the largest FTLE computed with averaging times ranging from 21 to 64 days. One observes a good convergence to a well-define rate function when longer averaging times are considered. Comparably good properties of convergence have been found for all the FTLEs \cite{DeCruz2018}.

Conversely, in \cite{DeCruz2018} no convergence was found in a higher resolution version of the coupled ocean-atmosphere model mentioned above \cite{VannitsemLucarini2016}. The authors conclude that the length of the simulations with the coupled model (614 years) is not sufficient to obtain convergence due to the long-term correlations in the system. 

\subsection{Rare event sampling algorithms based on large deviation theory}\label{sec:applres}

Large deviations provide us with valuable insights into the probability with which and the way in which rare events occur. The theory is only valid in specific limits, for example taking the noise intensity in an SDE to zero, or the length of a time average to infinity. Nevertheless the results are in many cases still useful when this limit has not been reached, which is the realistic setting we are really interested in.

In this pre-limit cases we may want to estimate quantities that are not available from the instanton, for example
\begin{align*}
    \mathbb{P} (x (t) \in B|x (T) \in A) \text{ for } 0 < t < T 
\end{align*}
the probability of entering some set $B$ at an  intermediate time $0<t<T$, conditioned on arriving in the rare event set $A$ at the final time. To obtain such information numerical estimation is usually the only possible approach.

However, we face a difficult challenge in numerically sampling rare events. By definition any brute force Monte Carlo simulation will only contain a small sample of relevant events. This leaves us with large uncertainties on the quantities we want to estimate.

Numerical sampling methods known as rare event simulation (RES) methods have been successful in circumventing these sampling issues in many applications. Although many RES algorithms exist, the approaches are based on the same principle: we want to preferentially sample rare events, in such a way that it is possible to a posteriori determine how much more or less likely we have made the paths we have sampled. This information allows us to estimate probabilities of the system as though we had not interfered with its path distribution.

RES algorithms have been developed since the 50s \cite{KH} for a variety of applications in applied mathematics and statistical physics, and have been subject of recent mathematical interest \cite{delmoral_feynmankac_2004}. In recent years there has been several attempts to apply these methods for geophysical applications. They have been applied to Lorenz models \cite{WOUTERS:2016:A}, partial differential equations \cite{Rolland_2016}, turbulence problems
\cite{Grafke2015,Laurie,Grauer,Lestang_2018,lestang2020numerical},
geophysical fluid dynamics \cite{Bouchet_Rolland_Simonnet_2019:C}, heatwaves in general circulation models \cite{Ragone&al2018,Ragone_Bouchet2020,RagoneBouchet2021} and data-based stochastic weather generators \cite{Yiou2020}, and tropical cyclones in regional climate models \cite{webber_practical_2019,plotkin2019maximizing}. Here we give an overview of methods and their applications that have been used to study problems directly related to the dynamics of planetary atmospheres and making use of concepts from LDT. Other applications, including approaches through minimum action methods also related to LDT, are reviewed in \cite{grafkeNumericalComputationRare2019}. 

One of the applications we have discussed concerns the statistics of time averages of surface temperature and the study of heatwaves. In order to study long-lasting extremes of surface temperature, \cite{Ragone&al2018,Ragone_Bouchet2020} have applied a genealogical algorithm to the intermediate complexity climate model Plasim, and recently to the more complex, state-of-the-art model CESM \cite{RagoneBouchet2021}. The algorithm was specifically designed in \cite{Giardina_et_al_2006,Lecomte_et_al_2007,giardina_simulating_2011} to study large deviation functions of time averages, and is very similar to a class of methods described by \cite{del2005genealogical} and whose mathematical properties are studied in details in \cite{delmoral_feynmankac_2004}.

The algorithm consists in running an ensemble simulation of $N$  trajectories with a numerical model starting from different initial conditions. At constant intervals of a fixed resampling time $\tau$, each trajectory is assigned a weight, which determines if that trajectory is killed or if it continues its evolution generating copies of itself. The weight $w_i^n$ of trajectory $n$ at time $t_i=i \tau$ is computed in such a way that the weight is large when an appropriately chosen score function is large. Choosing well the score function in order to define the selection criteria is critical. In \cite{Ragone&al2018,Ragone_Bouchet2020,RagoneBouchet2021} the score function used was the time integral of the surface temperature over Europe. 
The weights are defined in such a way that, after a total running time $T_a$, the probability $\mathbb{P}_k(\{\vec{X}(t)\})$ of observing a trajectory in the ensemble generated by the algorithm is related to the probability $\mathbb{P}_0(\{\vec{X}(t)\})$ of observing the same trajectory in a normal ensemble simulation with no resampling as 
\begin{equation}\label{eq:importance_sampling}
\mathbb{P}_k(\{\vec{X}(t)\})=\frac{e^{k\int_{0}^{T_a} A(u)\,\mathrm{d}u}}{Z}\mathbb{P}_0(\{\vec{X}(t)\}),
\end{equation}
where $Z$ is a normalization term, $k$ is a biasing parameter of the algorithm that determines how strong is the tilt of the probability distribution, and the equation is valid in the limit of large $N$ with relative errors on the computation of expectation values of the order of $1/\sqrt{N}$ \cite{delmoral_feynmankac_2004}. 

\begin{figure}
    \centering
    \includegraphics[width=0.33\textwidth]{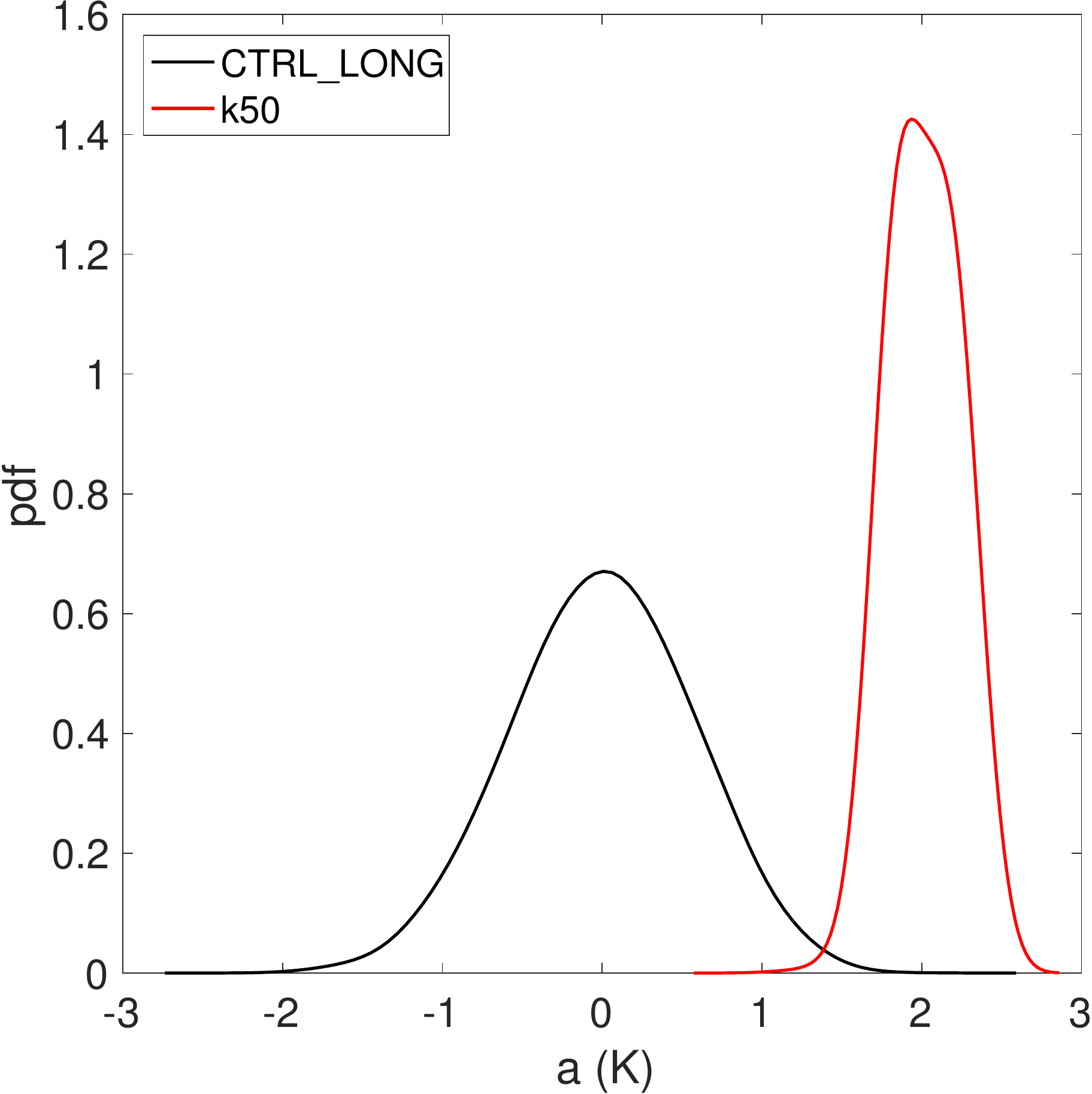}
    \includegraphics[width=0.33\textwidth]{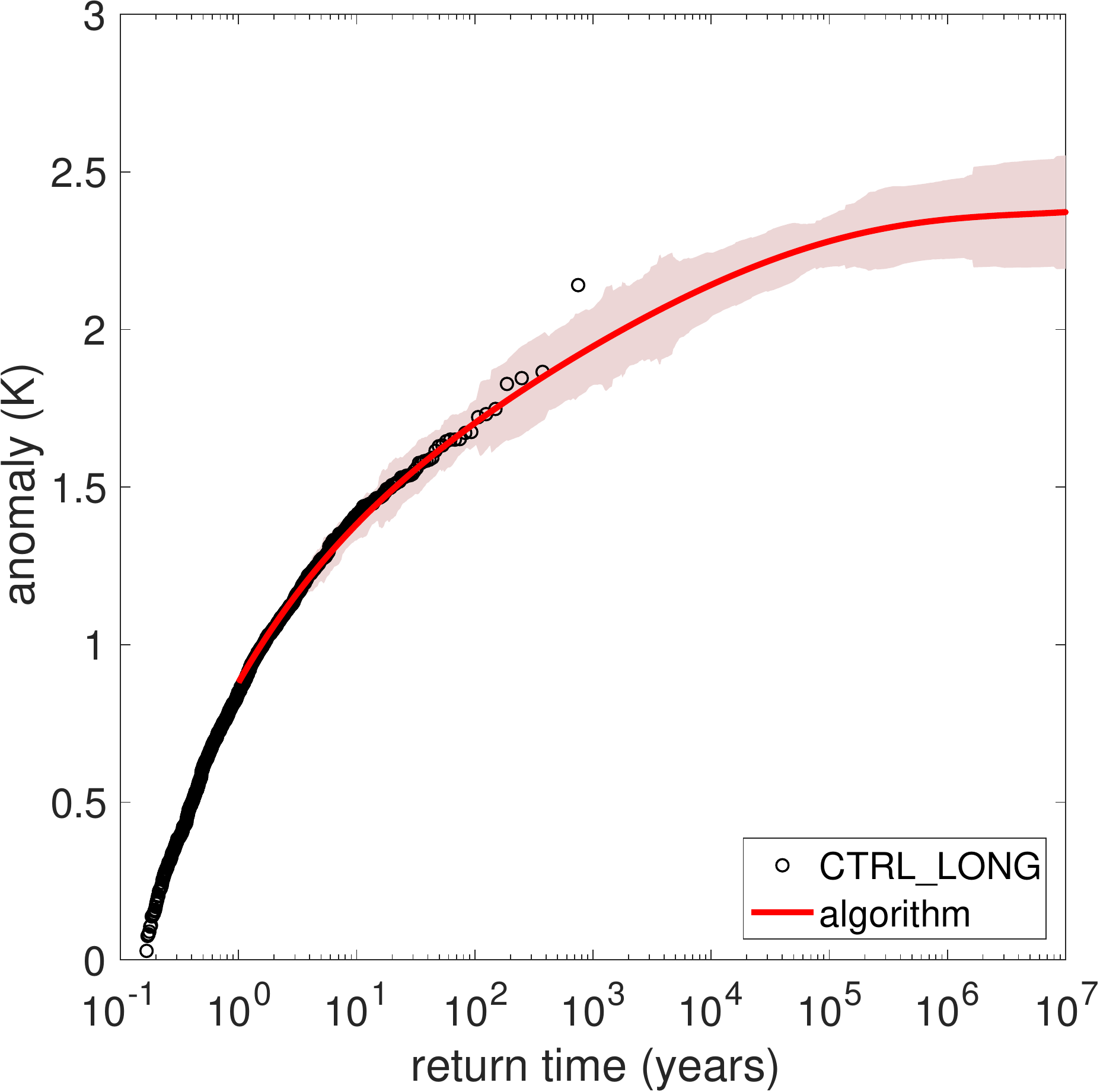}
    \includegraphics[width=0.28\textwidth]{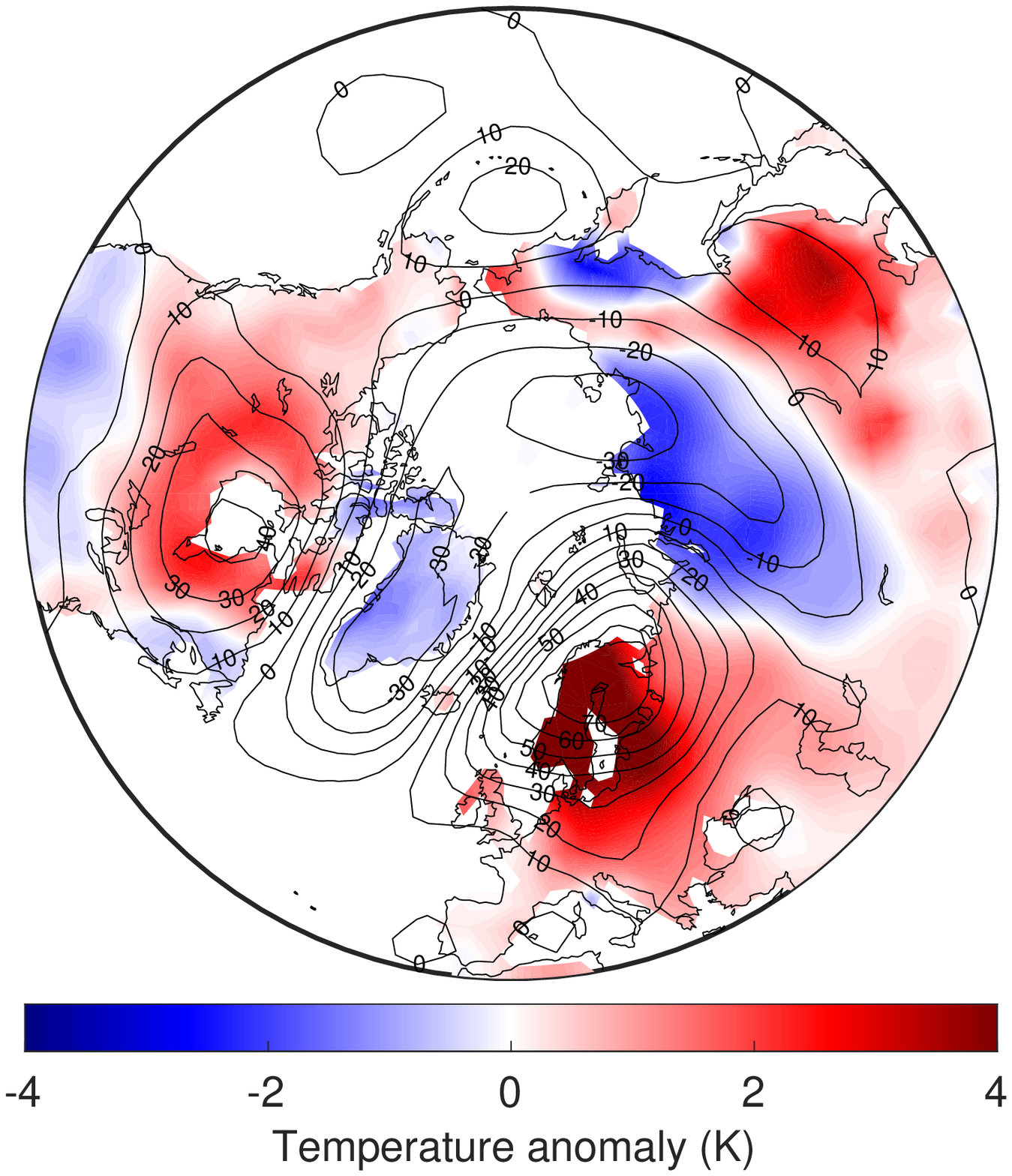}
    \caption{Left: importance sampling of average European surface temperature anomalies in ensemble simulations with a large deviation rare event algorithm applied to the general circulation model Plasim, with the distribution obtained with the algorithm (red) shifted to the tail of the distribution obtained with a direct Montecarlo approach (black), from \cite{Ragone&al2018}. Center: return times of simulated average European surface temperature anomalies from the same experiments, showing how the application of the rare event algorithm allows to simulate rare and ultra rare events (red) compared with what possible with a direct Montecarlo approach (black), from \cite{Ragone&al2018}. Right: low wavenumber teleconnection pattern of anomalies of surface temperature (red) and 500 hPa geopotential height (contours) in the composite averages of rare heatwaves (return time large than 1000 years) obtained with the large deviation rare event algorithm in the same experiments, from \cite{Ragone&al2018}.}
    \label{fig:rare_events_1}
\end{figure}

In ensemble simulations performed with the rare event algorithm, trajectories featuring a large value of the time average of the control observable over the entire simulation period are thus much more likely to be observed than in a normal simulation, and viceversa. This is called \textit{importance sampling}. It is important to note that equation \ref{eq:importance_sampling} applies to the probability of trajectories, not of the score function itself. It can thus be used to compute any statistical property of the system by averaging over the trajectories in the tilted ensemble and reweighting the contribution of each trajectory by the inverse of the factor that multiplies $\mathbb{P}_0(\{\vec{X}(t)\})$ in equation \ref{eq:importance_sampling}. Because of the tilting, statistical estimators based on (\ref{eq:importance_sampling}) of quantities conditional on the occurrence of extremes of the time average of $A$ have statistical uncertainties orders of magnitude smaller than with direct sampling, for a given computational cost.   

This method as developed by \cite{Giardina_et_al_2006,Lecomte_et_al_2007,giardina_simulating_2011} gives optimal estimates of the scaled cumulant generating function of the time average of the control observables for the value of $k$ used as biasing parameter. In fact the normalization term is the generating function $Z=\mathbb{E}[e^{k\int_0^TA(t)dt}]$, and an estimate of it is computed directly with a dedicated estimator based on the statistics of the weights responsible for the cloning of the trajectories. See \cite{Ragone_Bouchet2020} for a full analysis on this type of application. It is worth noting that the same method has been used by \cite{Tailleur} to compute the large deviation properties of the finite time Lyapunov exponents of a selection of simple dynamical system, highlighting peculiar properties of the chaotic dynamics associated to these systems.

\begin{figure}
    \centering
    \includegraphics[width=1\textwidth]{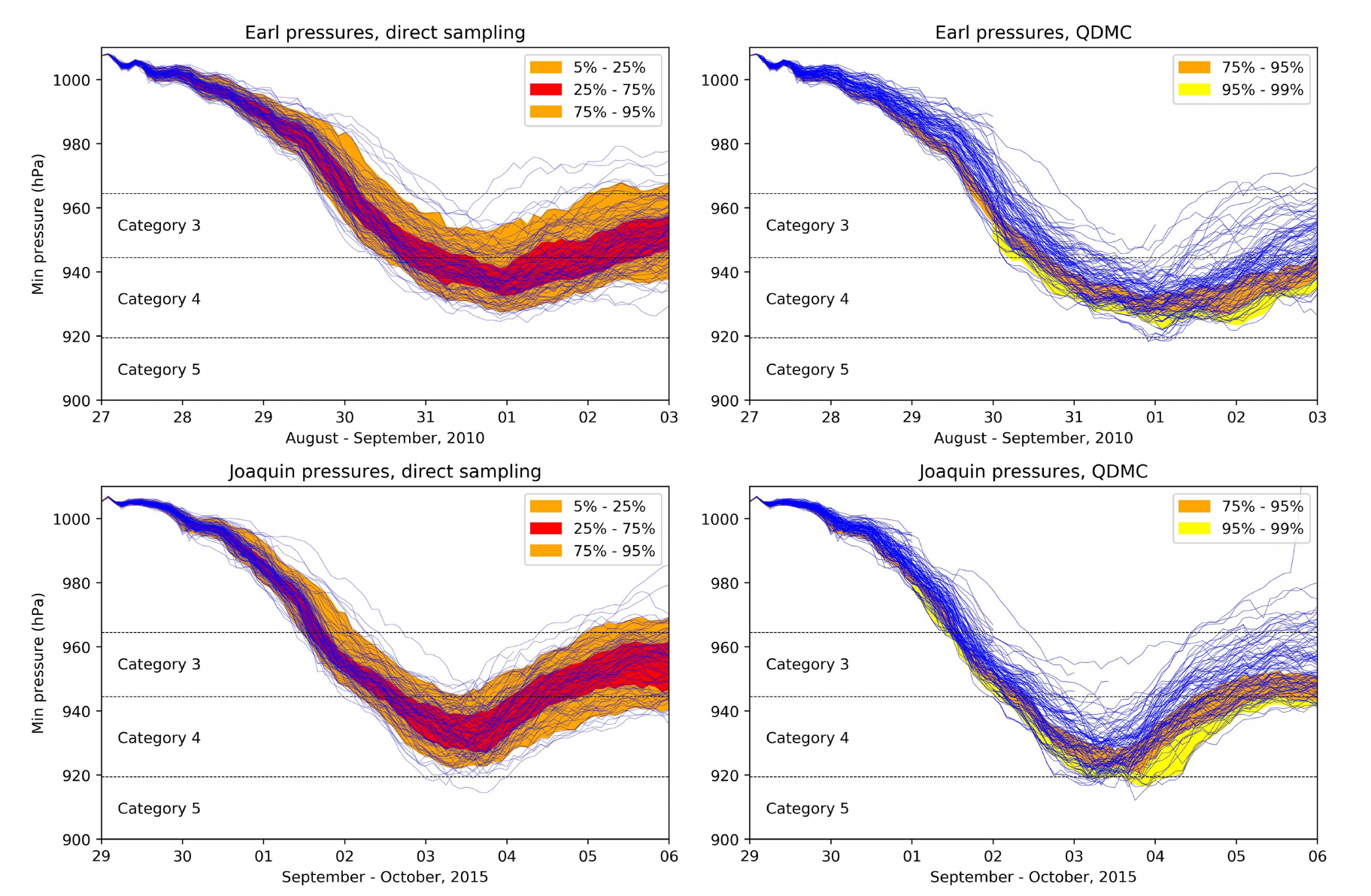}
    \caption{Intensity of tropical cyclones Earl (top) and Joaquin (bottom) in direct Montecarlo simulations with the regional climate model WRF (left) and in  ensemble simulations with a quantile Diffusion Montecarlo rare event algorithm applied to the same model, showing the shift of the trajectories to higher percentiles of the intensity range thanks to the rare event algorithm. Reproduced with permission from \cite{webber_practical_2019}.}
   \label{fig:rare_events_2}
\end{figure}

The method however proves extremely useful also to study the statistics of time averages not in the large deviation limit, as shown in \cite{Ragone&al2018,RagoneBouchet2021}. Simulations performed with the algorithm allowed to perform importance sampling of seasonal anomalies of the European surface temperature (left panel of Fig. \ref{fig:rare_events_1}), in a condition in which convergence to the large deviation limit was far from being achieved. This allowed to simulate and estimate precisely the return times of rare and ultra rare anomalies corresponding to seasonal heatwaves with return times up to millions of years, with computational costs 3 order of magnitude smaller than what would have been necessary with direct sampling (center panel of Fig.~\ref{fig:rare_events_1}). The access to the dynamical trajectories leading to rare heatwaves allowed to compute precisely composite statistics of anomalies of surface temperature and 500 hPa geopotential height conditional on the occurrence of heatwaves with return time larger than 1000 years. This shows how this very rare and intense events are related to the occurrence of a hemispheric teleconnection pattern of low wavenumber between 3 and 4 (right panel of Fig.~\ref{fig:rare_events_1}). The presence of these stationary patterns (although with varying wavenumbers) seems ubiquitous in analysis of extreme heatwaves with different approaches (see discussion in \ref{sec:appl}), which makes rare event algorithms a very promising tool to investigate the dynamic constituents and drivers of these events, thanks to the much better sampling they can provide. This approach has also been applied successfully to stochastic weather generators  based on circulation analogues to simulate persistent rare European heatwave \cite{Yiou2020}. Indeed, the fact that the algorithm is able to select the correct dynamical patterns that generate extremes of a target observable, makes its pairing with analogue classification and selection analysis potentially extremely interesting.

\begin{figure}
    \centering
    \includegraphics[width=0.53\textwidth]{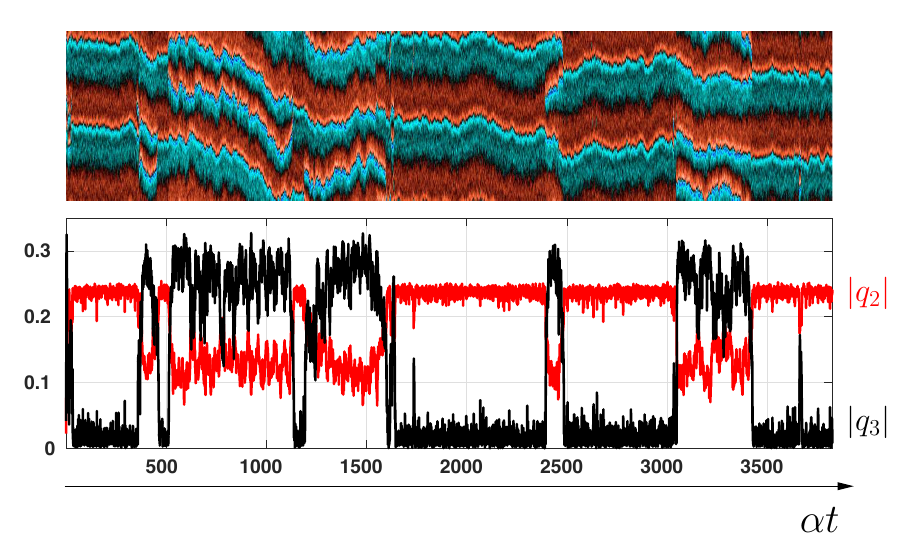}
    \includegraphics[width=0.46\textwidth]{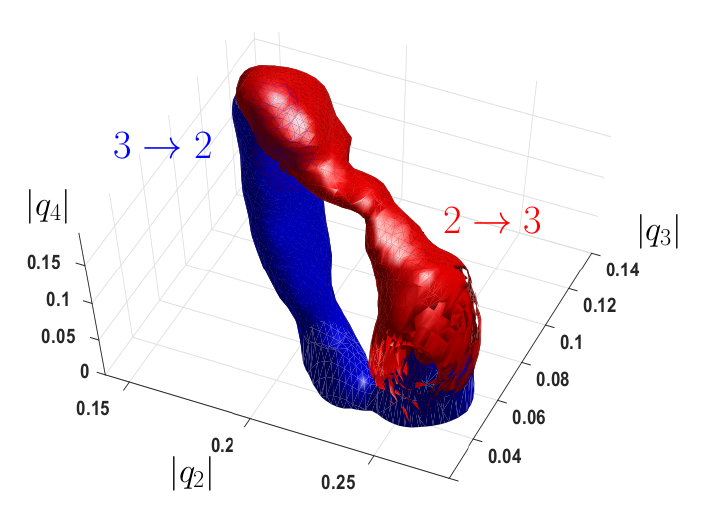}
    \caption{Left: transitions between attractors with 2 and 3 jets in a stochastic barotropic beta-plane quasi-geostrophic model of Jupiter atmosphere: Hovmoeller diagram of zonal mean vorticity (top) and time series of wavenumber 2 and 3 Fourier components of zonal mean vorticity (bottom). Right: reactive tubes showing the concentration of the transition paths around instantons for the transitions between 2 and 3 jets states in the model of Jupiter atmosphere, obtained with the Adaptive Multilevel Splitting rare event algorithm. Reproduced with permission from \cite{Bouchet_Rolland_Simonnet_2019:C}.}
    \label{fig:rare_events_2b}
\end{figure}

A similar method has been used by \cite{webber_practical_2019,plotkin2019maximizing} to study the intensification of tropical cyclones in the high resolution, non-hydrostatic regional climate model WRF. In this case the score function has been taken as the surface pressure anomaly at the center of the cyclone. Additionally, the trajectory selection and cloning procedure has been modified by mapping at each resampling step the distribution of the weights on a Gaussian distribution on an equal-quantile basis \cite{webber_practical_2019}. This procedure helps to avoid that for large values of $k$ the trajectory with the largest weight in the ensemble dominates all the others in the cloning rate, thus reducing the degree of degeneracy of the trajectories in the resampled ensemble. Applying this method \cite{webber_practical_2019,plotkin2019maximizing} were able to obtain a large number of trajectories of extremely intense tropical cyclones in simulations featuring boundary conditions corresponding to two historical tropical cyclones \ref{fig:rare_events_2}, thus showing the potential of this approach even with some of the most challenging applications in climate modelling.

It is worth noting that the resampling methods used in \cite{Ragone&al2018,webber_practical_2019,plotkin2019maximizing,Ragone_Bouchet2020,RagoneBouchet2021} belong to the same family of methods used in particle filtering by the data assimilation community \cite{Dubinkina_2013,Carrassi_et_al_2018,van_Leeuwen_2019}. Particle filtering is a notoriously difficult problem in data assimilation \cite{Snyder_et_al_2008} due to the difficulty to target specific trajectories in a very high dimensional space. Particle filtering has nonetheless shown promising advances in recent years \cite{van_Leeuwen_2019}. Although not directly related to extreme events or LDT, it would be interesting to see how developments in particle filtering and in the emerging field of the application of RES algorithms to climate modelling could possibly inform and support each other.

Another method that has been successfully applied to problems of interest for the geophysical and climate community is the Adaptive Multilevel Splitting (AMS) algorithm \cite{CerouGuya,cerou_adaptive_2019}. This method is particularly well suited to study rare transitions between two metastable states or attractors A and B. A score function or reaction coordinate is defined to measure the distance of a trajectory from say the target set B. An ensemble of $N$ trajectories is then simulated starting from the set A, until all of them end in either A or B. Then the worst performing trajectory (the one falling in A that was the furthest from B in its evolution) is replaced by a new trajectory whose initial condition is taken on one of the other better performing $N-1$ trajectories. This step is the resampling step of the algorithm, and is repeated $K$ times. The final ensemble of trajectories will have probability $(1-1/N)^K$, and it will be populated by many extremely rare transitions from A to B, which can be used to compute unbiased estimates of their probability. See \cite{Rolland_2015,Rolland_2016,Bouchet_Rolland_Simonnet_2019:C} and references therein for more details.

The AMS algorithm has been used among other applications to study transitions to turbulence \cite{Rolland_2018}, the forces acting on a solid object in a turbulent flow \cite{Lestang_2018,lestang2020numerical}, and transitions between metastable states in geophysical fluid dynamics \cite{Bouchet_Rolland_Simonnet_2019:C}. For example, \cite{Bouchet_Rolland_Simonnet_2019:C} studied the transitions between metastable states with two and three jets in a stochastic barotropic beta-plane quasi-geostrophic model  of  the atmosphere of Jupiter (left panel of Fig.~\ref{fig:rare_events_2b}). Thanks to the application of the AMS algorithm, \cite{Bouchet_Rolland_Simonnet_2019:C} were able to obtain thousands of trajectories representative of the rare transitions between the two turbulent attractors, showing how they cluster around preferential paths close to an instanton (right panel of Fig.~\ref{fig:rare_events_2b}). This was the first demonstration in a numerical simulation of a turbulent flow of this type of phenomenology. The AMS algorithm has thus proved to be a powerful tool to study rare transitions between different chaotic attractors, that could be applied also to climate studies where such transitions occur (see Sec. \ref{sec:metastability}).

As said several other RES methods exist, tailored to different types of studies. The application of RES algorithms to geophysical and climate problems is an emerging field of research with a very strong potential. The close connection between large deviations and rare event simulation, both providing information on rare event probabilities, begs the question whether they can be usefully combined to provide insights on each other in this context. In particular, it would be extremely interesting to understand for a wider class of processes and applications on one hand if RES algorithms can be useful in estimating large deviation rate functions in systems and models of the complexity of the ones typically involved in applications, and on the other hand if knowing the instanton to a rare event set could be useful to set up an efficient RES algorithm to study the corresponding physical process.



\subsection{Rogue Waves}\label{roguew}
Rogues waves are unusually large and virtually unpredictable surface waves that pose grave hazards to boats as well as to naval and coastal infrastructures, and to people living in coastal areas, as they can lead in open sea to tens of meters high water walls. They are traditionally defined as deep-water waves whose crest-to-trough height is at least twice as large as the significant wave height, which in turn is defined as four times the standard deviation of the ocean surface elevation \cite{Nikolkina2011,Slunyaev2011,Adcock2014,Didenkulova2020}. Traditionally, the understanding of such rogues waves, which are found with similar phenomenology in systems as different from the ocean as optical fibers \cite{Akhmediev2013} and plasmas \cite{Bailung2011}, has been approached according to two separate viewpoints. One sees rogue waves as emerging from linear superposition effects \cite{Lindgren1972}, the other one, instead, sees rogue waves as eminently nonlinear phenomena \cite{Didenkulova2011,Bertola2013} where a solitonic structure emerges via nonlinear focusing \cite{Tikan2017}. 

Recently, a unified theory of rogue waves based on LDT has been proposed \cite{Dematteis2018,Dematteis2019}. The fundamental idea is that rogue waves can be seen as instantons computed by minimizing the action associated with the dynamics described by a one-dimensional nonlinear Schr\"odinger equation (NLSE) with random initial data. The initial data are chosen in such a way that their spectrum agrees with the spectrum of ocean wave height measured in an oceanographic observational campaign in the North sea, namely the Joint North Sea Wave Project (JONSWAP) \cite{Hasselmann1973}. The idea is to study the probability of having a wave that increases to a height larger than a given threshold $z$ over a time $T$ from one of the possible initial conditions: 
\begin{equation}
    P_T(z)=\mathbb{P}(F(u(T)>z))
\end{equation}
where $[0,L]$ is the domain of the system, $F(u(T))=\max_{x\in[0,L]}(S^Tu_0)$,
$\mathbb{P}$ is the probability computed over the distribution defining the initial data $u_0$, and $S^T$ is the evolution operator up to time $T$ defined by the NLSE. The next step is to  
look at the  tail of the surface height distribution and write $P_T(z)$ as a large deviation law of the form $P_T(z)\approx \exp\left(- I_T(z)\right)$ and define, in a way analogous to what discussed in Sec.~\ref{dynoise}\footnote{But note that here the noise is included as random fluctuations of the initial conditions, whereas the evolution defined by the NLSE is fully deterministic.} the instanton as the trajectory minimizing the rate function $I_T(z)$.  

\begin{figure}[t]
    \centering
 a)\includegraphics[width=.45\columnwidth]{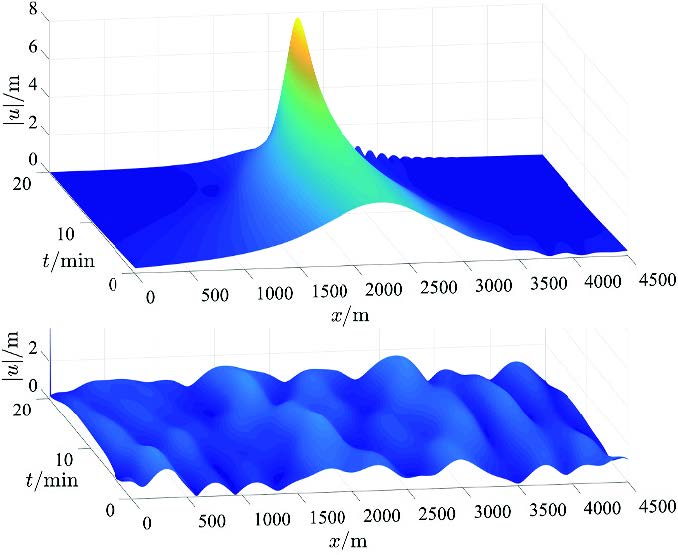}
   b) \includegraphics[width=.45\columnwidth]{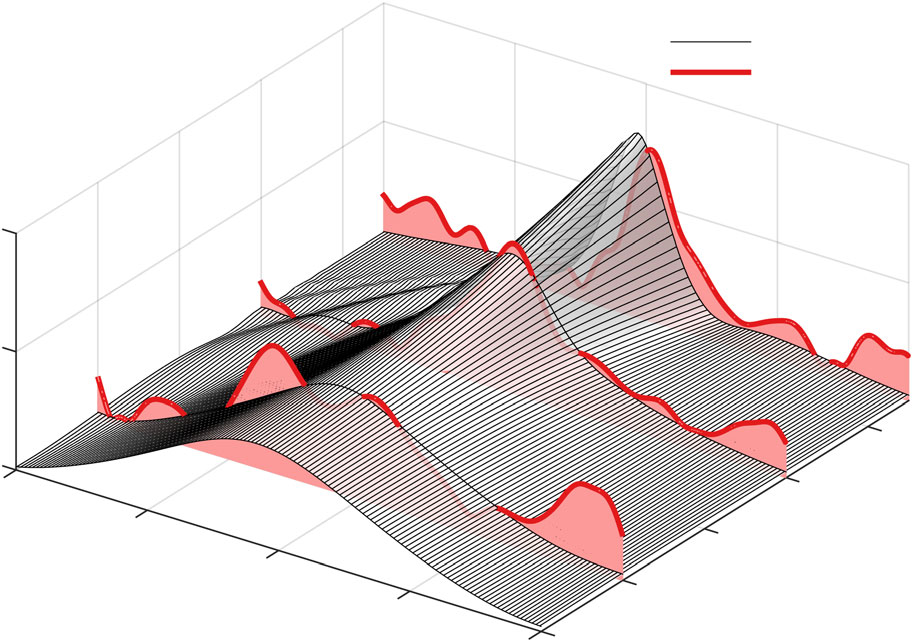}
    \caption{a) The rogue wave precursors are  wave patterns of regular height, but with a very specific shape. Upper panel: precursor leading to a rogue wave. Botton panel: regular wave field. From \cite{Dematteis2018}. b) The instanton computed via LDT (grey surface) resembles well individual rogue waves (light red lines) observed in a large water tank experiment. From \cite{Dematteis2019}. \label{rogue}}
 \end{figure}

Numerical simulations clearly indicate that rogues waves emerge from fields - precursors - that are typical in terms of intensity but special in terms of spatial pattern - see Fig. \ref{rogue}a. Especially impressive is the fact that experimental results obtained in large water tanks suggest that instantonic solutions do resemble well actual observed rogue waves - see Fig. \ref{rogue}b. This latter result is - conceptually - in agreement with what shown in Sec.\ref{sec:applpers} regarding the good correspondence between observed heatwaves and cold spells and LDT-tailored climate variability generated by an Earth system model, see Fig. \ref{fig:namcs}. Again, one can interpret such predictive power in terms of the universality resulting from the fact that LDT describes the most likely among events that are extreme and overall very unlikely.

\subsection{Metastability and Noise-induce Transitions Across Melancholia States}\label{sec:metastability}

  	\begin{figure}
 \includegraphics[width=0.9\columnwidth]{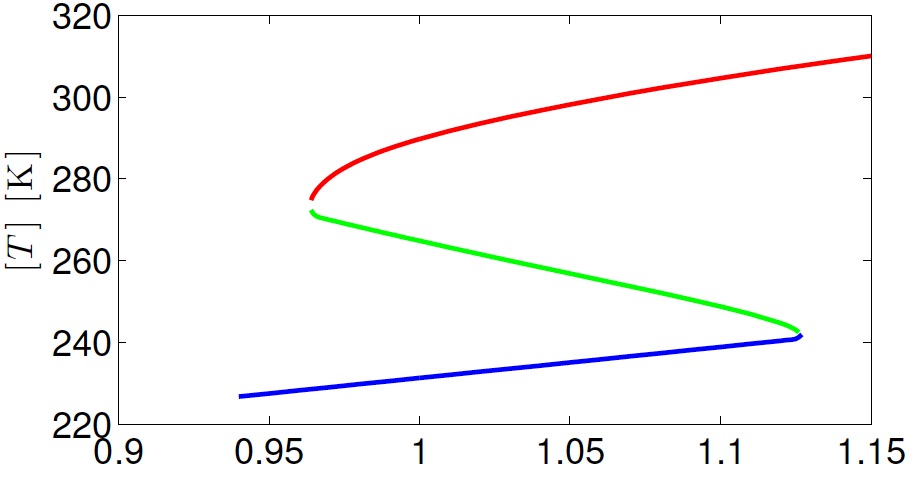}
 \caption{Bifurcation diagram of the Ghil-Sellers model \cite{Ghil1976} as a function of the ratio between the solar irradiance and the present-day one. The order parameter is the globally averaged surface temperature. The stabel W and SB states and the unstable M state are indicated by the red, blue, and green solid line, respectively. Reproduced with permission from \cite{bodai2015}.\label{snowball}}
 \end{figure}

\begin{figure}[t]
    \centering
  a)  \includegraphics[width=.7\columnwidth]{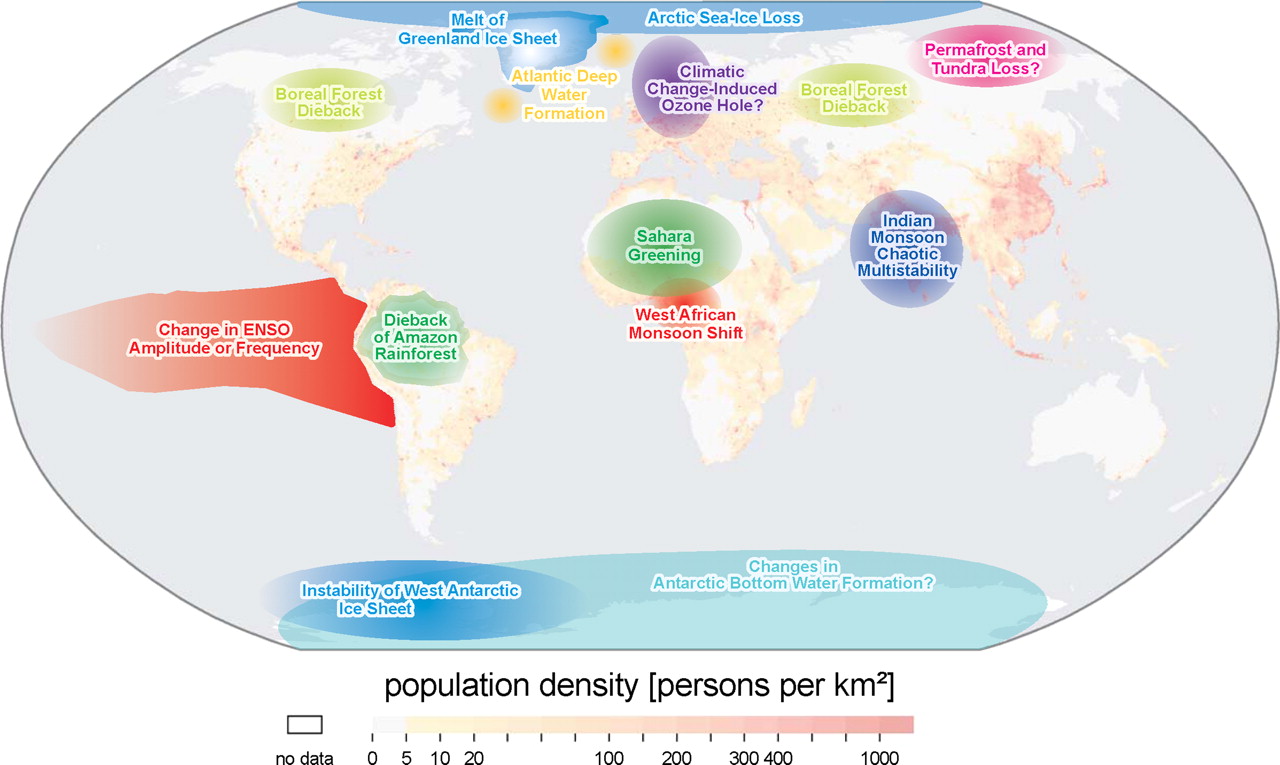}\\
    b) \includegraphics[width=.7\columnwidth]{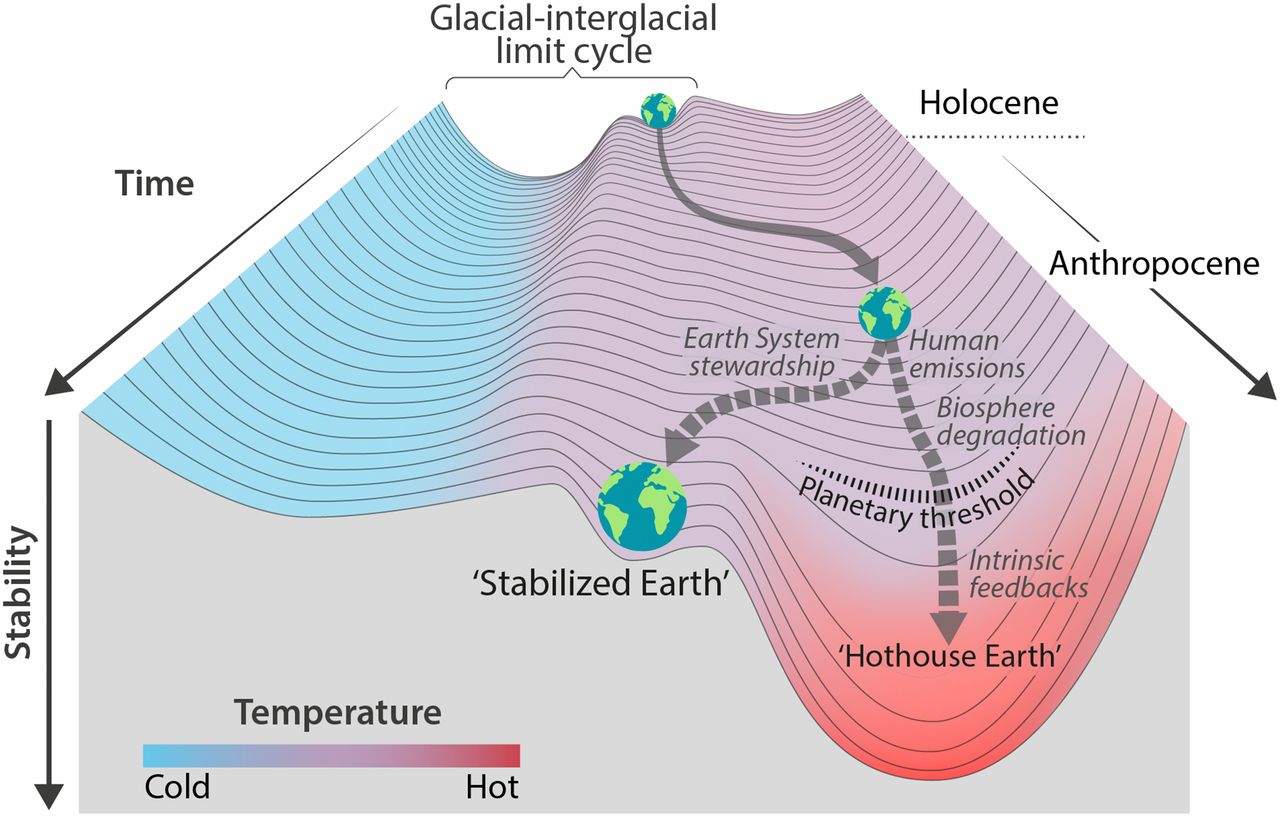}
    \caption{a) Tipping elements of the climate system. From \cite{Lenton2008}. b) Cartoon of how how the so-called \textit{stability landscape} (here: quasi-potential) changes as a result of anthropogenic forcings. From \cite{Steffen2018}. \label{tipping}}
 \end{figure}


As mentioned earlier in the paper, a key problem in geosciences is the investigation of metastable systems and of the properties of transitions between the competing modes of operation. And, indeed, there is a complex phenomenology of multistability for the Earth system at involving different temporal and spatial scales. We describe below how methods and ideas related to LDT can help us better understand such features. 

The current astronomical and astrophysical conditions support the co-existence of (at least) two competing states, corresponding to the snowball (SB) state and warm (W) state. Such states are so different that they seem to correspond, counterintuitively, to two entirely different planets. The multistability is, by and large, made possible by the competition of two feedbacks: the negative Boltzmann feedback (a warmer planet emits more radiation to space) and the positive ice-albedo feedback (a colder planet stores more water in ice form; the surface ice reflects more radiation to space). Figure \ref{snowball} portrays a bifurcation diagram taken from \cite{bodai2015} where multistability is given in terms of globally averaged surface temperature, while the control parameter is the ratio between the solar irradiance and its present value. The W solution is separated from the SB solution by an unstable solution - a saddle in the phase space, which we will refer to as Melancholia (M) state \cite{Lucarini2017N,Lucarini2019,Lucarini2020,Bodai2020basin}. The bifurcations of the system take place when one of the stable states meet the M state, according to the scenario of basin crisis \cite{Ott2002}. 
 
On a less dramatic scale, modulations in the parameters of the climate system can lead to the occurrence of smaller scale critical transitions that have very strong impacts on specific climatic subsystems, the tipping points \cite{Lenton2008}, see Fig.~\ref{tipping}a. Recently, it has been proposed that future scenarios of climate change  (\textit{trajectories of the Anthropocene}) might be loosely seen as dynamically determined by a motion on some energy potential, which might experience multiple local minima corresponding to competing metastable states \cite{Steffen2018}, see Fig. \ref{tipping}b.   
 
As we are considering a very high-dimensional system that cannot be reduced to a one-dimensional ordinary differential equation, linking minima of some dynamically-relevant potential and competing metastable states requires some nontrivial mathematical framework, which is close in spirit to Waddington's \textit{epigenetic landscape} metaphor in evolutionary biology~\cite{Waddington1957,Ao2009,Ferrell2012,Huang2012}. 

{\color{black} We take the point of view proposed by the Hasselmann programme \cite{Hasselmann1976} discussed in Sec.~\ref{sclimatem} and assume that the coarse-grained} evolution of the climate system can be written as in (\ref{eq:diffusion}). We also assume that the deterministic evolution law $\dot{x}=b(x)$  obtained by switching off the noise can be seen as given a smooth flow taking place in a compact $n$-dimensional manifold. 
If stochastic forcing is absent, the initial condition ${x}_0$  determines the asymptotic state of its orbit. If  several asymptotic states, defined by the attractors $\Omega_j$, $j=1,\ldots,J$, are present the system is multistable. The phase space is divided between the basins of attraction $B_j$ of the  attractors $\Omega_j$ and the basin boundaries $\partial B_l$, $l=1,\ldots,L$, which possess a set of saddle points $\Pi_{l_k}$, $l=1,\ldots,L$, $k=1,\ldots,k_{max}\geq1$. Such saddle points - the M states - attract initial conditions on the basin boundaries \cite{Grebogi1983,Vollmer2009,LT:2011} and can be computed using the so-called edge tracking algorithm \cite{Skufca2006}.


We now switch on the stochastic forcing. The presence of noise makes it possible for transitions between the competing metastable states to take place \cite{hanggi1986,Grassberger1989,freidlin1984}. 
	Under fairly general conditions, in the limit $\varepsilon\rightarrow 0$,  
	one can propose the following ansatz for the invariant measure in the form of a large deviation law
	\begin{equation}\label{eq:stationary_distr}
	\rho_\varepsilon({x}) \sim Z({x})\exp\left(-\frac{2 \Phi({x})}{\varepsilon}\right),
	\end{equation}
	where $\Phi({x})$, a nonequilibrium generalization of the notion of free energy, plays the role of a rate function. 
	while $Z({x})$ is a pre-exponential factor. 
	$\Phi$ obeys the following Hamilton-Jacobi equation \cite{Gaspard2002,Zhou2012}:
	\begin{equation}\label{eq:HJE}
	{F}_i({x}) \partial_i \Phi({x})+a_{ij}({x})  \partial_i \Phi({x}) \partial_j \Phi({x}) =0.
	\end{equation}
	where ${a}({x}) = {\sigma}({x}){\sigma}({x})^T:\mathbb{R}^n\rightarrow \mathbb{R}^{n\times n}$ is the noise covariance matrix.
	Additionally, $\Phi$ is a Lyapunov function whose decrease describes the convergence of an orbit to the attractor. Specifically, $\Phi({x})$ has local minima at the deterministic attractors $\Omega$'s, and has a saddle behaviour at the saddles  $\Pi$'s. If an attractor (saddle) is chaotic, $\Phi$ has constant value over its support, which can then be a strange set \cite{Graham1991,Hamm1994}. Note that these qualitative properties do not depend on spatial patterns of the noise. Instead, the simple fact of adding some noise to the otherwise deterministic dynamics (thus setting $\sigma>0$) allows for a global exploration of the phase space of the system. 
	
	

	 The instantons, which give, in the zero-noise limit, the most probable way to exit an attractor  \cite{Kautz1987,Grassberger1989}, can be constructed as minimizers of the Freidlin-Wentzell actions similarly to what has been shown in Sec. \ref{dynoise}\footnote{Some nontrivial issues emerge when the noise matrix is singular; see \cite{Margazoglou2021}.}.  
	 The instanton is intimately connected to the quasipotential $\Phi({x})$ in that the local quasipotential $\Phi_\Omega({x})$ within the basin of attraction of $\Omega$ is equal to the action for the instanton between $\Omega$ and ${x}$ \cite{Grafke2015,Bouchet2016,Grafke2019}. 
	 To recover the global quasipotential $\Phi({x})$, one needs to resort to a pruning-and-stitching strategy, glueing together the local portions $\Phi_{\Omega_j}, j=1,\ldots,J$, see \cite{Graham1991} and the careful description recently provided by~\cite{Zhou2016}. A separate view on this problem, based upon a different interpretation of the noise has been proposed in \cite{Ao2004,Yuan2017}.
	
	Escapes from an attractor $\Omega$ through a saddle $\Pi$ into a neighbouring basin are Poisson-distributed events, where the probability that an orbit does not transition up to time $t$ is, similarly to the classic Kramers' law \cite{Kramers1940}, given by: 
	\begin{equation}\label{eq:tt_distr}
	P(t) =\frac{1}{\bar{\tau}_\varepsilon}\exp\left(-\frac{t}{\bar{\tau}_\varepsilon}\right), \, \mathrm{with}\, \lim_{\varepsilon\rightarrow 0}\varepsilon \log\bar{\tau}_\varepsilon = 2\Delta \Phi_{\Omega\to \Pi},\quad
	\end{equation}
	being the expected escape time and $\Delta \Phi_{\Omega\to \Pi}=\Phi_\Omega(\Pi)-\Phi_\Omega(\Omega)$ is the quasipotential barrier height at the relevant saddle~\cite{LT:2011}. Unfortunately, due to the global stitching procedure, one cannot in general simply read off the barrier height $\Delta \Phi_{\Omega\to \Pi}$ from the $\Phi({x})$ of equation~\ref{eq:stationary_distr}. While the global quasipotential $\Phi({x})$ yields information about the relative probability of attractors, and is available e.g.~through global sampling of the system, the local notion of potential barriers, $\Delta\Phi_{\Omega\to \Pi}({x})$ is relevant for the time-scale of transition events, and can be obtained e.g.~by looking at transition times between attractors. 
	Equivalence between the information provided by the local and global quasipotentials is  realised  if the system is an equilibrium one or, more generally, if only two competing states are present with a single saddle embedded in the boundary between the two basins of attraction \cite{Lucarini2019,Lucarini2020}. 

	 	\begin{figure}
 \includegraphics[width=\columnwidth]{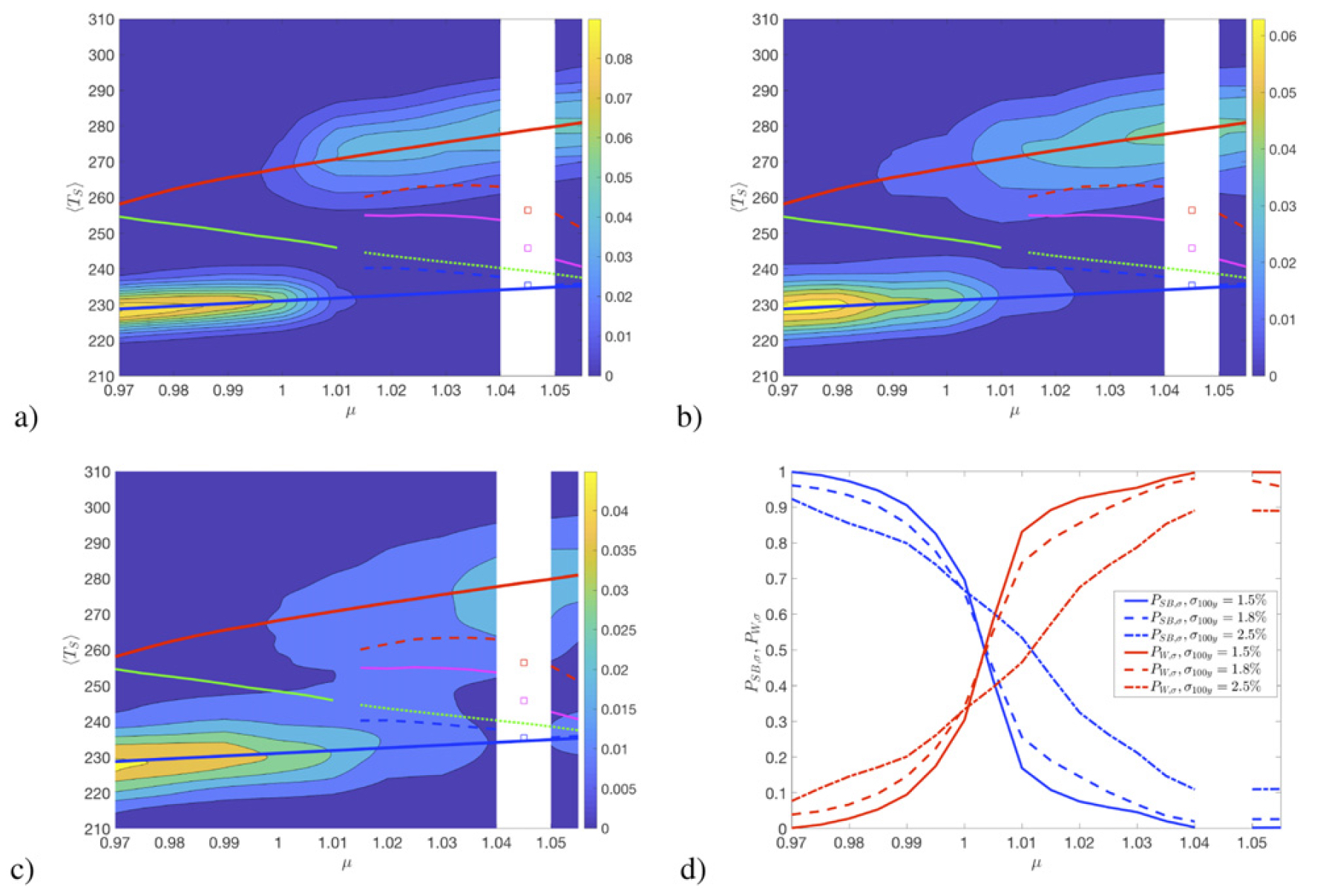} 
 \caption{Effect of changing the intensity of the noise on the invariant measure of a climate model. a)-c): y-axis: globally averaged surface temperature; x-axis: ratio between the solar irradiance and the present-day. Solid lines: Bifurcation diagram of the deterministic system (W state: red line; SB state: blue line; and M state: green line are of relevance here). Shading: projection of the invariant measure. Noise increases from a) to c). d) Population of the W state (red lines) and of the SB state (blue lines) for different noise intensity. Here $\sigma^2=\varepsilon$. From \cite{Lucarini2020}\label{phase}.}
 \end{figure}

\begin{figure}
 \includegraphics[width=\columnwidth]{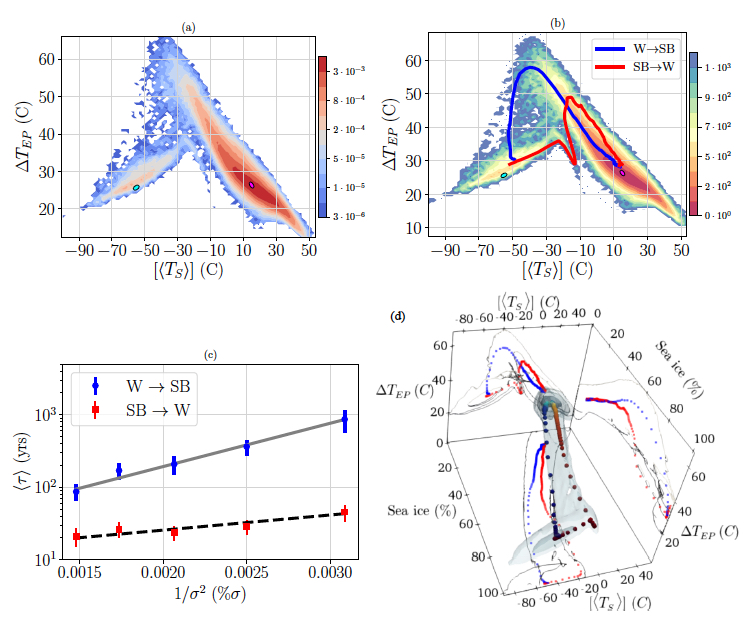} 
 \caption{Properties of the $W\rightarrow SB$ and $SB\rightarrow W$ transitions in a stochastically perturbed version of PLASIM \cite{Fraedrich2005,Lucarini2010}. a) Quasi-potential projected in the two-dimensional space given by the globally averaged surface temperature (x-axis) and difference between surface temperature at low and high latitudes (y-axis). b) Projection of the probability density function obtained for $1.8\%$ per century random fluctuation of the solar irradiance and estimates of the weak-noise limits of the $W\rightarrow SB$ and $SB\rightarrow W$ transition paths. c) Change in the $W\rightarrow SB$ and $SB\rightarrow W$  transition times as a function of the intensity of the noise (note the logarithmic scale in the y-axis). d) Three-dimensional representation of the $W\rightarrow SB$ and $SB\rightarrow W$ transition paths, where the sea-ice percentage acts as third dimension. Here $\sigma^2=\varepsilon$. From \cite{Margazoglou2021}\label{TransitionsMargazoglou}.}
 \end{figure}

Margazoglou et al. \cite{Margazoglou2021} have recently performed a thorough investigation of the noise-induced transitions between the competing SB and W states of the open-source climate model PLASIM \cite{Fraedrich2005,Lucarini2010}, see Fig.~\ref{TransitionsMargazoglou}. Panel b) shows the probability density function in the projected space given by the globally averaged surface temperature and by the difference between surface temperature at low and high latitudes, and the best estimates of the transition paths in the weak-noise limit, while panel a) shows the estimate of the projected quasi-potential $\Phi$. Additionally, panel c) shows the exponential dependence of the average transitions times with respect to the inverse of the variance of the noise, as indicated in (\ref{eq:tt_distr}). Note that in this case, because of the complex structure of the basin boundary, the $W\rightarrow SB$ and $SB\rightarrow W$ transitions take place through separate M states, as can be seen in Panel d) where a third dimension, corresponding to the sea ice percentage, in included.

Let's now look specifically at the case of bistable systems, where we have two attractors $\Omega_1$, $\Omega_2$, and one saddle $\Pi$. We can then express the average transitios times as follows: 
\begin{equation}\label{eq:tau2}
\lim_{\varepsilon\rightarrow0}\varepsilon\log\tau^{i\rightarrow j}_\varepsilon =2(\Phi(\Pi)-\Phi(\Omega_i), i\neq j=1,2
\end{equation}
so that 
\begin{equation}\label{eq:tau3}
\lim_{\varepsilon\rightarrow0}\varepsilon\log\left(\frac{\tau^{1\rightarrow2}_\varepsilon}{\tau^{2\rightarrow1}_\sigma }\right) = 2(\Phi(\Omega_2)-\Phi(\Omega_1)).
\end{equation}
This implies that, in the weak noise limit, both escape times diverge, but the escape time out of the attractor corresponding to the lower value of the quasi-potential diverges faster. 
By mass balance, at steady state the populations $P_{1,\varepsilon}$, $P_{2,\varepsilon}$ of the  neighbourhood of the two attractors 
obey the following relation: 
\begin{equation}\label{eq:pop}
\lim_{\varepsilon\rightarrow0}\varepsilon\log\left(\frac{{P}_{1,\varepsilon}}{{P}_{2,\varepsilon}}\right)=\lim_{\varepsilon\rightarrow0}\varepsilon\log\left(\frac{{P}_{1,\varepsilon}}{1-{P}_{1,\varepsilon}}\right)= 2(\Phi(\Omega_2)-\Phi(\Omega_1))\end{equation}
Equation~(\ref{eq:pop}) implies that in the zero-noise limit only one of the two deterministic attractors will be populated, and specifically the one where the quasi-potential has lower value.

Equation~(\ref{eq:pop}) can be obtained  by  integrating the invariant measure given in (\ref{eq:stationary_distr}) in the neighbourhood of the attractors and  taking a saddle point approximation. Hence, the statement above holds true for an arbitrary number of competing multistable states: in the zero-noise limit, as a result of the large deviation law, the measure will be concentrated only on the attractor corresponding to the lowest value of the quasi-potential. 
Note that, nonetheless, individual stochastic trajectories could be trapped for a very long time in secondary metastable states corresponding to the other attractors because low levels of noise could make it very time-consuming to reach the global minimum. We also note that two different noise laws differing for the correlation matrix $C$ acting on top of the same drift field will define two different quasi-potentials, see (\ref{eq:HJE}). As a result of that, they will in general feature a different selection of the limit measure in the zero noise limit. 

Panels a)-c) of Fig.~\ref{phase} provide an illustrative example of the limit behaviour of the measure for a multistable system undergoing stochastic forcing. The data are taken from the output of a simplified stochastic climate model having $\mathcal{O}(10^4)$ degrees of freedom constructed by coupling an atmosphere described by  primitive equations with an energy balance model - indeed, adapted from \cite{Ghil1976,bodai2015} - that describes in a parsimonious way the large scale heat transport performed by the global ocean \cite{Lucarini2019,Lucarini2020}. The deterministic version of this model features multistability of values of the ratio between the solar irradiance and the present-day one ($\mu$ in  Fig.~\ref{phase}) ranging from about 0.97 to about 1.06 \cite{Lucarini2017N}. Within this range, the system is bistable, except between 1.04 and 1.05, where a third competing state is present. For sake of simplicity - see \cite{Lucarini2017N,Lucarini2019,Lucarini2020} for further details - we focus on the W and SB states (indicated by the red and blue solid lines), which are separated by the unstable M state indicated by the green solid line, similarly to what shown in Fig.~\ref{snowball}. The shading in panels a)-c) of Fig.~\ref{phase} indicates for each value of $\mu$ the density of the 1D invariant measure projected on the globally averaged surface temperature. Going from c) to a), the intensity of the noise - here introduced as a yearly fluctuating value of the solar irradiance around the value indicated by $\mu$ - decreases. 

We observe that as the noise becomes weaker the measure peaks around the W (SB) attractor for $\mu$ larger than about 1.01 (smaller than about 0.99), with a  changeover around $\mu=\mu_{crit}\approx1.005$.  Panel d) shows the population corresponding to the W and SB states for various levels of noise.  We conclude that for $\mu<\mu_{crit}$ the quasi-potential has a lower value at the SB attractor than at the W attractor, while the opposite occurs for $\mu>\mu_{crit}$. We conclude that for $\mu=\mu_{crit}$ the system undergoes a nonequilibrium phase transition.

{\color{black}

\subsection{Transitions between Zonal Flow and Blockings}\label{block}

What has been presented here seems also relevant for investigating a separate, extremely relevant aspect of geophysical fluid dynamics, namely the existence in the atmosphere of different regimes of operation, which define the presence of substantial low-frequency variability on subseasonal time scales \cite{Ghil2001d,Ghil2020}. This boils down to the fact that, at coarse-grained level, due to extreme dynamical heterogeneity \cite{LucariniG2020}, one is practically looking at a multistable system \cite{Charney1979,Benzi1986,Mo.Ghil.1987,Itoh1996,Kondrashov2004,Ruti2006}, where one can define and detect transitions between different metastable states \cite{Bouchet2014,Bouchet2014a}; see discussion in Sec.~\ref{sclimatem}. In particular, the fundamental dichotomy for the mid-latitude atmosphere is between the \textit{standard} zonal flow and the blocked state \cite{Egger1978,Speranza1983,Lindzen1986,Ghil2001d}, which is characterized by large spatial extent and long persistence. As discussed in Sec.~\ref{sec:intro} and \ref{sec:applpers}, blockings are sometimes the leading cause of heatwaves and cold spells. In the coarse grained setting, the transition between the zonal and the blocked state are made possible by the stochastic forcing associated with higher frequency synoptic variability \cite{Benzi1986,Mo.Ghil.1987}. 
An accurate analysis of noise-induced transitions between the zonal and the blocked state in the celebrated Charney and DeVore \cite{Charney1979} minimal model of the atmosphere using the LDT framework  has recently been presented in \cite{grafke2019numerical}. There, the authors have been able to compute the optimal  paths for both zonal-to-blocking and blocking-to-zonal transitions, and have elucidated  a) that the two paths are different, as we are dealing with a nonequilibrium system; and b) that the two paths meet at the Melancholia state embedded in the  boundary separating the two basins of attraction, according to the scenario discussed in Sec.~\ref{sec:metastability}. This is illustrated in Fig. \ref{grafkecharney2019}, which portrays various snapshots of the streamfunction, which, in the Charney-DeVore model, is intended to approximate the 500 hPa geopotential height field. 

\begin{figure}
    \centering
   a)\includegraphics[width=0.45\textwidth]{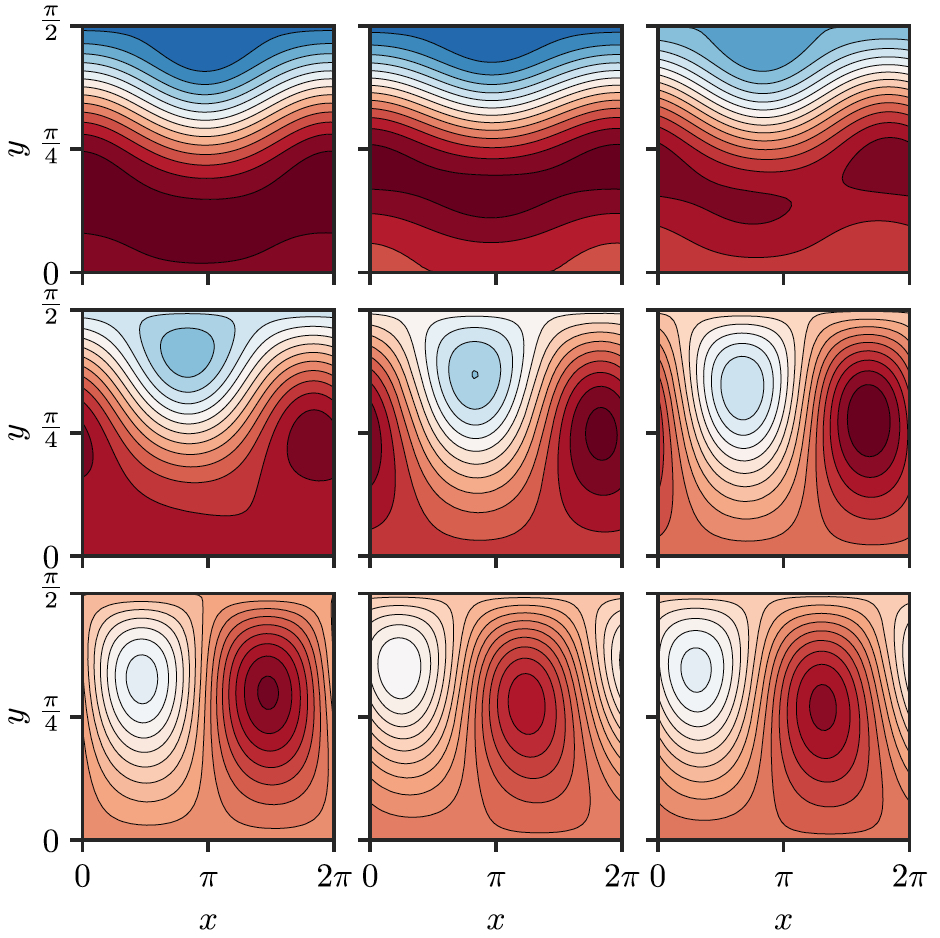}
     b)\includegraphics[width=0.46\textwidth]{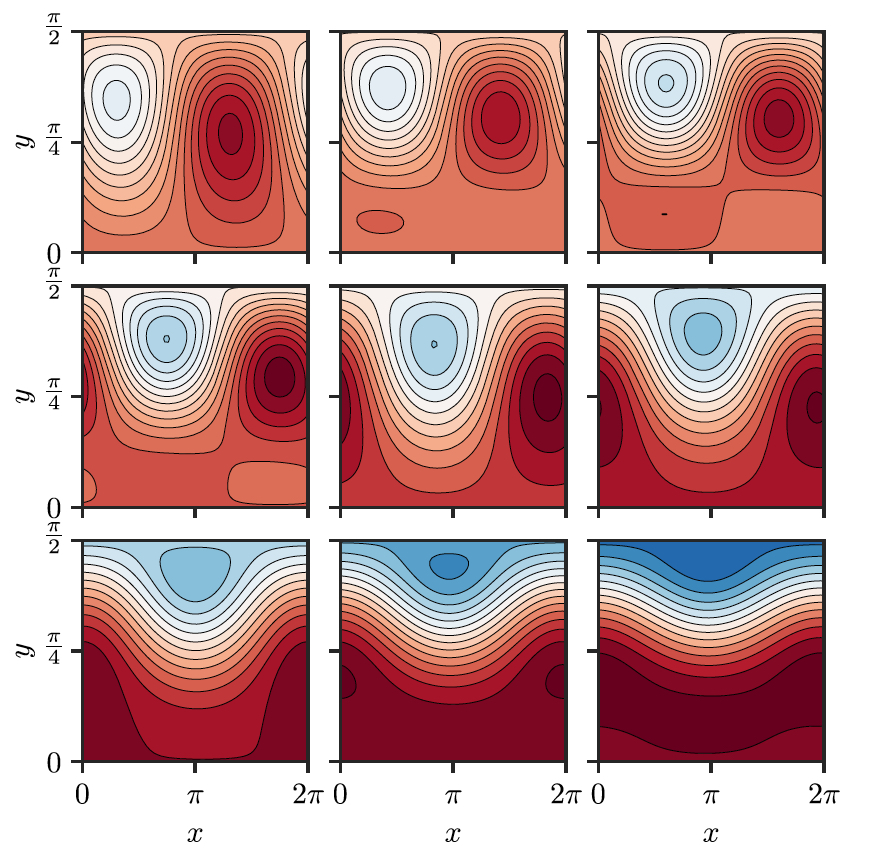}
    \caption{{\color{black}Transitions between the zonal flow and atmospheric blocking in the Charney-DeVore model \cite{Charney1979}. a) From top-left to bottom right: snapshots of the streamfunction along the transition from zonal flow to blocked state. b) Same as a), but for the reversed transition. The central panel is the same for both paths and corresponds to the Melancholia state. Reproduced with permission from \cite{grafke2019numerical}.}}
    \label{grafkecharney2019}
\end{figure}

}


\section{Conclusions and Perspectives}\label{conclusions}

Extreme Value Theory (EVT) has shown its potential for providing information on the fundamental properties of the dynamical system generating suitably defined extreme events \cite{Lucarini2016extremes}. As an example, this feature is now being extensively used for providing a fresh outlook on the problem of understanding and characterizing the predictability of the atmosphere \cite{Faranda2017,Messori2018,Bodai2017}, going beyond the more standard use of EVT for studying (rigorously) the tails of the distribution of meteo-climatic fields of interest \cite{katz2005statistics,Ghil2011}. When applying EVT to dynamical systems, universality emerges through the procedure of looking at smaller and smaller portions of the attractor, which makes it possible to analyse accurately the properties of the physical measure supported on it \cite{Lucarini2014,Galfi2017,PonsJSP,Caby2019}.   

Along similar lines, LDT provides powerful tools for addressing the complexity of geophysical flows and of the climate  as a whole, the basic idea being that by allowing one to focus on specific, non-standard events selected according to a - possibly universal - statistical procedure, it makes possible to elucidate fundamental properties of the system under investigation. {\color{black}Our viewpoint goes along the line of the scientific programme aimed at the development and interpretation of stochastic climate models proposed by Hasselmann \cite{Hasselmann1976}, as discussed in detail in Sec.~\ref{sclimatem}}. In this review we have provided a summary of some of the key ingredients of LDT  and have described the emergence and significance of large deviation laws in stochastic and deterministic chaotic systems. In a nutshell, since the procedure of computing large deviation laws relies on estimating the probability of occurrence of large anomalies in the finite size averages of stochastic variables with respect to their asymptotic values, one explores the combined effect of the static (invariant measure) and dynamic (statistics of correlations) features of the system of interest. LDT, by definition, captures the least unlikely of all the unlikely ways a given large and persistent fluctuation can take place \cite{Hollander2000}. Therefore, if we are in the correct asymptotic limit, LDT will define \textit{typical extreme events} (but, by definition, very \textit{atypical standard events}).  This does not exclude the possibility of - quantitatively much  unlikelier - freak events or \textit{dragon kings} \cite{Sornette2012}, which live outside the approximation associated with (finite size) LDT. The key presence of such a dynamic component marks the conceptual difference between LDT- and EVT-based approaches to the study of extremes. Indeed, EVT can deal with the problem of persistence of extreme events only in a rather indirect way, through the introduction of the so-called extremal index, which measures the inverse of the characteristic cluster length of consecutive extreme events \cite{Ferro2003,Lucarini2016extremes,Moloney2019}.

Hence, the ideas and methods proposed in this paper suggest new ways to look at the dynamics behind persistent and large fluctuations of the meteo-climatic field, in such a way that the underlying basic mechanisms behind them can be singled out and better understood. This paves the way for detailed analyses of the climate system starting both from observed data and reanalyses up to numerical models. Indeed, it seems urgent to specifically target audit and intercomparison studies of Earth system models to such special events, in order to highlight possible deficiencies that could hardly be seen by looking at standard statistical properties. This applies to the study of individual events like heatwaves and cold spells, {\color{black}to the - closely related - investigation of the transitions between different modes of operation of the atmosphere responsible for its low-frequency variability}, as well as to the analysis of the tipping elements of the climate system by looking at the noise-induced transitions between competing states.

It has been found that concurrent persistent extreme events, like heatwaves, droughts, or floods, are often related to specific almost stationary atmospheric wave patterns \cite{Petoukhov_2013,Screen2014,Coumou2014}. This has been shown for the summer of 2018, when several heat waves and rainfall extremes occurred almost simultaneously over different regions of the Northern Hemisphere \cite{Kornhuber2019}.  Persistent weak phases of the stratospheric polar vortex have been related to cold extremes in Northern continental regions, especially in central Asia \cite{Kretschmer2018}. Interactions between persistent stationary mid-latitude wave patterns and tropical variability is poorly understood, but is suspected to cause unusual persistent extremes, like the consecutive extreme flooding, heat wave and typhoon in Japan in 2018 \cite{Wang2019}. As explained in Sec.~\ref{sec:introue} when discussing the low-frequency variability of the atmosphere, local persistent anomalies of smooth observables, like temperature, are related to spatially extended large anomaly fields, and furthermore to persistent large-scale atmospheric fluctuations, i.e. quasi-stationary waves {\color{black}often associated with blocking events}. An explanation for the phenomena of recurrent wave patterns linked to persistent extreme events from the perspective of LDT follows the principle already stated several time above representing one of the key principles of this theory: \textit{there is one least unlikely way from all the unlikely ways that leads to a persistent extreme configuration of the atmosphere, which seems to manifest itself in the form of rather similar quasi-stationary wave patterns}. Thus, an interesting area of application would be to analyse persistent states of atmospheric circulation patterns, for example, as large deviations of jet indices, blocking indices, or of spatially averaged high-level wind fields. 

An impressive example of the relevance of the principle above can be found in the studies dealing with rogue waves, which have been briefly summarized in Sec.~\ref{roguew}. LDT allows to make sense of a very complex phenomenology and of competing theories by  providing a (possibly) universal characterization of rogue waves as instantons that minimize an action, in such a way that individual rogue waves do resemble the instantonic solutions, and that precursors can be identified \cite{Dematteis2018,Dematteis2019}. This seems an excellent meeting point between theoretical results and extremely important applications in the evaluation and anticipation of natural hazards.

A natural consequence of the connection between quasi-stationary large-scale wave patterns and persistent extreme events discussed above is that several different extreme events can appear at the same time. Thus we arrive at the concept of compound extremes, which received a lot of attention in the climate science community in the last years due to their devastating effect on both nature and human society \cite{Kornhuber2019,Lau2012,Hong2011,Boschi2019}. Following the computation of multivariate rate functions presented in \cite{Kwasniok2019}, large deviation rate functions could be used directly to study compound extreme events. However, the computation of multivariate rate functions would require a drastically larger amount of data than the already large amount of data needed to obtain univariate rate functions. \cite{Galfi_Lucarini2020} showed that the compound nature of the persistent event can be captured also by a simple composite approach based on large deviations. We note that a high-impact persistent extreme event is usually a compound extreme event at the same time, as the probability that it triggers the occurrence of other extreme events is very high. 

Summer weather persistence has been shown to increase as an effect of global warming \cite{Pfleiderer2018,Pfleiderer2019}. As shown in \cite{Galfi_Lucarini2020}, an increasing persistence of heatwaves in summer has been found in many regions of the Northern Hemisphere due to increasing CO$_2$ concentrations, based on a large deviation analysis. We point out that LDT provides a more natural way to define and analyse persistence than empirical approaches based on counting subsequent values above or below a certain pre-defined threshold. If the large deviations limit is reached at a certain averaging time scale, one can obtain the probability of every persistent event with duration equal to or longer than the respective averaging time, as explained in Sec~\ref{sec:ldds_ta} and \ref{sec:applpers}. Furthermore, it provides the probability of events of arbitrary intensity, within the limits set by the finite size of the available data. We further remark that LDT provides a macroscopic view on persistent events, in the sense that it focuses on deviations of the sample mean over a finite time period from the long term mean assumed to be equal to the expectation value, disregarding the amplitudes of individual instantaneous fluctuations. If one is interested to study instantaneous extremes instead of persistent ones, methods of EVT would be more appropriate to tackle the problem. We refer here also to the discussion about slow onset and fast onset events in Sec.~\ref{sec:intro}.

The studies mentioned in Sec.\ref{sec:applpers} and \ref{sec:applres} analysing persistent atmospheric extreme events, focus on persistent events of air temperature, i.e. heatwaves or cold spells. Nonetheless, by using the same large deviation based methodologies, one can study obviously in a similar way persistent events related to other geophysical observables too. Events like droughts, persistent rainfall events, floods, wind and solar lulls are just a few examples. However, the choice of the right observable is very important in order to be able to obtain a LDP, as explained in Sec.~\ref{sec:ldds_ta}. In case of some meteo-climatic observables the large deviation approaches presented in this article might not work due to long-term correlations. As an example, \cite{Galfi_Lucarini2020} finds that large deviation rate functions do not converge, at least on time scales shorter than several years, in case of air temperature anomalies over oceanic regions. However, it is possible that anomalous scaling laws \cite{rey-bellet_young_2008,Nickelsen2018,Gradenigo_et_al_2013,Harris_and_Touchette_2009,Chazottes2015,Derrida_et_al_2017}, mentioned in Sec.~\ref{sec:ldds_ta}, act in some cases in which the standard large deviation scaling fails due to the presence of some form of long-term memory. In a very recent paper \cite{Alqahtani2021}, a nonlinear reparametrization of the scaled cumulant generating functions is proposed to properly compute instantons in case of observables with heavy tailed distributions.

	 	\begin{figure}
 \includegraphics[width=\columnwidth]{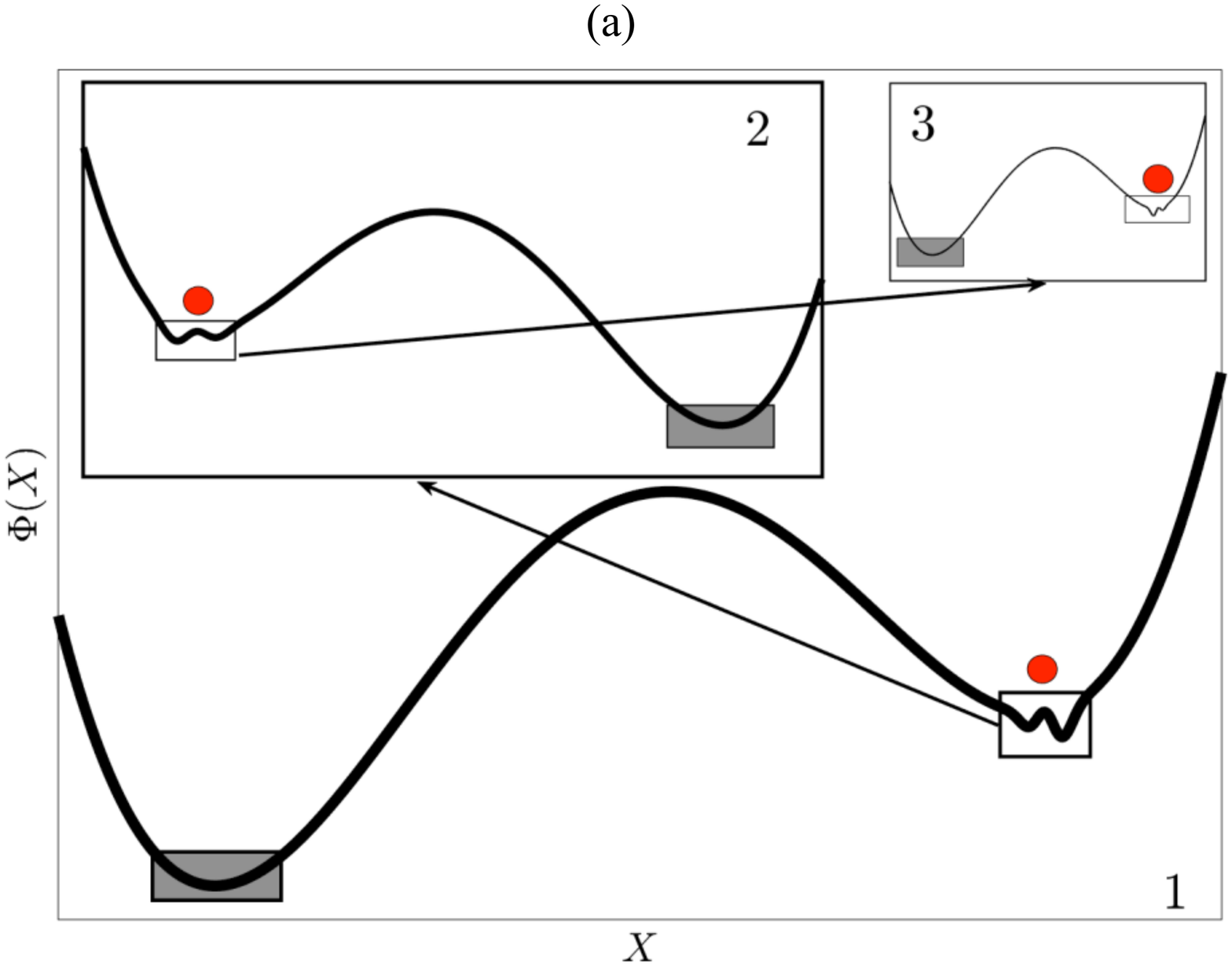}     
 \caption{Cartoon showing an idealised  quasi-potential where multistability occurs at multiple scales. Going from 1) to 3) one looks at smaller and smaller scales. Local maxima, minima, and saddles decorate the quasi-potential at all scales. From \cite{Margazoglou2021}\label{multiscale}.}
 \end{figure}

Besides serial correlations, there are also other factors that can hinder the applicability of LDT to geophysical time series. Non-stationarity is a serious issue, which confines the application of LDT to out-of-equilibrium steady state model simulations or to simulations with many ensemble members, unless one follows a pragmatic approach and pre-processes the data, for example by removing trends and the seasonal cycle, as discussed in Sec.~\ref{sec:introue}. Strong heterogeneity of the data, i.e. the existence of several different parent distributions, can be another problem with respect to obtaining a large deviation principle. Nonetheless, LDT can be extremely helpful for several geophysical applications. Computing large deviation rate functions to obtain probabilities of persistent events could be useful for attribution studies, which try to answer the question whether and to which extent it is possible to attribute the probability of occurrence of individual extreme events to climate change, and rely at the moment mainly on empirical frequentist approaches \cite{Allen2003,Otto2012,Otto2019,Dole2011,Rahmstorf2011}. Large ensemble climate model simulations \cite{VanderWiel2019,Milinski2020} could be used to study the change in time of persistent event probabilities based on large deviations. 

In case the data availability is poor for a proper computation of rate functions, one could rely on rare event sampling algorithms, discussed in Sec.~\ref{sec:applres}. Rare events simulation techniques have a long history and have been applied in several fields of applied mathematics and statistical physics in the past decades. The application to problems specific to geophysical fluid dynamics and climate science are however very recent. Their potential in the field of climate modelling in particular is huge, as it allows to tackle two crucial problems emerging in the study of extreme events in the climate system, either from observations or numerical simulations.

First, extreme events are rare, which means that statistical errors are large, and robust analysis are very hard to perform, in particular if one is interested in understanding the dynamical properties of the events. This is one of the reasons for which dynamical properties are often analysed on individual case studies. This is particularly problematic when one wants to understand the impact of climate change on the statistics and dynamics of extremes (it is difficult enough to obtain estimates for a given condition, to compute variations is even worse). This however is one of the main concerns and areas of interest in the climate change debate, both at the scientific and policy level, and in the communication of the issue to the general public. 

A second problem is that, given the length of the observational records (which is what it is) and given the length of the numerical experiments that is possible to perform with a given amount of computational power (which again, it is what it is), one can only study the events that have been sampled. As we discussed in this review, techniques like EVT and LDT can be used to extend the estimates of return periods or other statistical properties to events beyond what has been observed, making use of limit theorems in their corresponding areas of application. However, they do not provide the specific dynamics in the full phase space of the very rare, unobserved events, which may be relevant to study their predictability, climatic drivers and impacts. Also, the reliability of the extensions provided by EVT and LDT is dependent on the robustness of the statistics of the observed extremes. If the available statistics of extreme events is just barely enough to use them to fit the limit distributions, the uncertainties on the estimates of the properties of the unobserved events may be so large to make these estimates effectively useless. Worse, the very convergence to the limit distributions may be not properly obtained with the available samples, this may be difficult to assess, and the application of these limit theories on data not properly approaching the asymptotic regime could give misleading results. 

Rare event simulations techniques can provide solutions to these issues in numerical simulations, by improving drastically the statistics of extreme events, and allowing to explore ranges of events that would be simply impossible to observe in direct numerical simulations. Promising results have been obtained with genealogical algorithms for  heatwaves (and in general persistent events, also not in the LDT regime) in general circulation models and data-based stochastic weather generators, and for the intensification of tropical cyclones in regional climate models. The application to more general problems will require different methods, which will be an exciting area of research in the future years. It is worth noticing that some of the methods used so far share a similar DNA with the methods used in particle filtering by the data assimilation community. Therefore, they also share similar problems and challenges when applied to complex, high dimensional systems (the \textit{curse of dimensionality} problem). It is thus very encouraging that positive results have already been obtained with general circulation and regional climate models of state-of-the-art, nearly operational level of complexity. 

The events we are able to investigate using LDT, while rare, are of disproportionately great relevance in terms of impacts on human and environmental welfare and need to be carefully taken into account when planning long-term infrastructures. Additionally, in many cases, the presence of a correspondence between long temporal persistence and large spatial coherence means that such events can pose situations of systemic risk, as their impacts can be very relevant for large regions. Hence, the results contained in this paper might be of extreme practical use for addressing climate risk in present and future, because LDT provides robust ways to estimate return times of yet unobserved events if one is in the right asymptotic regime. The possibility of establishing accurate climatologies of, e.g., heatwaves and cold spells in various regions of the globe seems particularly relevant for climate service centres, public agencies and private actors active, for example, in the energy and in agricultural sectors as well as in finance, and re-insurances. Indeed, in order to achieve full real-life applicability in the context of the presence of the seasonal cycle and of a changing climate, the results presented in this paper should be extended in such a way to incorporate the case of non-stationary time series of observables of non-autonomous dynamical systems, {\color{black} along the lines of what has been proposed in the context of EVT \cite{Coles2001,Niu1997,Felici2007b,Nakajima2012,KonzenNeves2021}.} This is another great example of a very fruitful meeting point between applied and basic research. 

\begin{figure}
    \centering
    \includegraphics[width=1\textwidth]{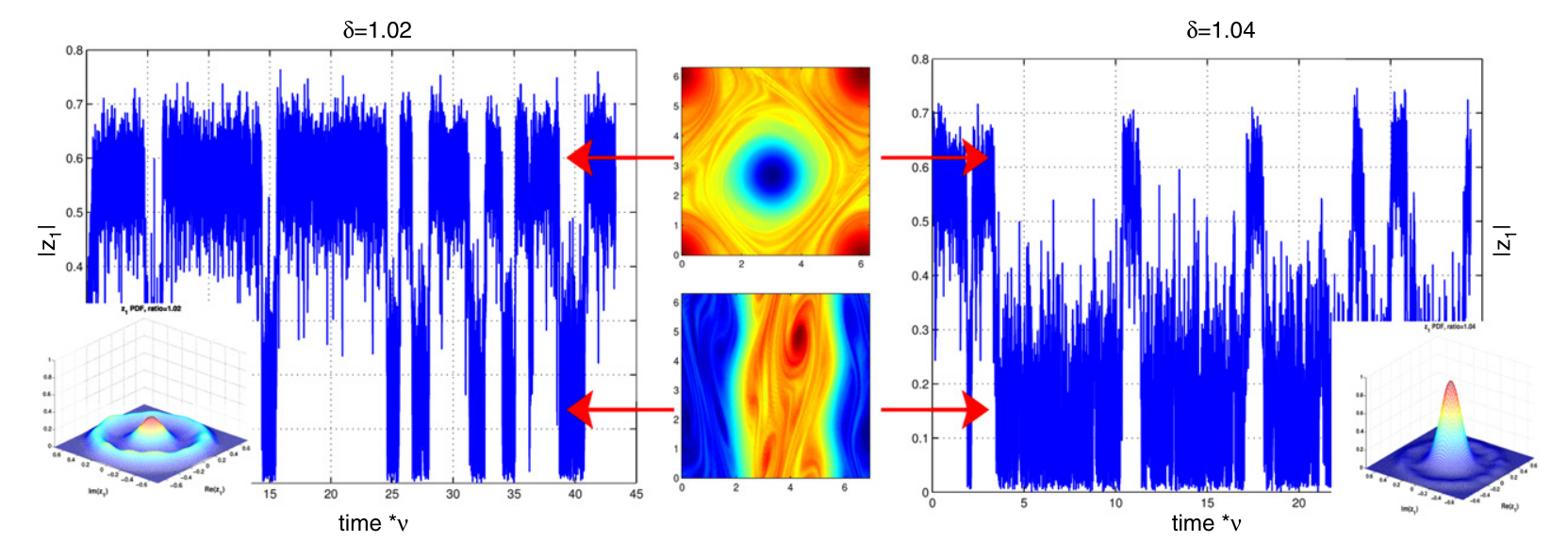}
    \caption{Phase transitions in the stochastic 2D Navier–Stokes equations in a doubly periodic domain: time series and probability density functions of the amplitude of the wavenumber 1 Fourier component in the y direction of vorticity, showing transitions between dipoles and unidirectional flows. Reproduced with permission from \cite{Bouchet2012}.}
    \label{fig:rare_events_3}
\end{figure}

{\color{black}LDT is also extremely useful for investigating some key dynamical properties of geophysical flows. As discussed in Sec.~\ref{Lyapunov}, it provides a way to frame the very important problem of evaluating the fluctuations of the predictability of the atmosphere \cite{Ghil2001d,Palmer2013,Krishnamurthy2019} on different temporal scales by computing rate functions of various finite time Lyapunov exponents \cite{DeCruz2018}. Indeed, one finds confirmation that the atmosphere is extremely \textit{heterogenous} in terms of its  predictability \cite{LucariniG2020}. 

LDT is also helpful for quantifying predictability in a statistical sense and on climatic time scales.} While tools like response theory allows one to study the impact of perturbations on the statistical properties of a complex system like the climate \cite{Ghil2020}, such an exploration is, by definition, a local one. In order to understand the global stability properties of the climate system one needs to resort to a different viewpoint. We have shown that LDT is a key tool for studying noise-induced transitions between competing metastable states for the climate system as a whole. Indeed, one can introduce a rate function, the quasi-potential, which condenses information from the deterministic vector field - the drift - and the properties of the noise. This viewpoint might prove of great usefulness for better understanding the dynamics of tipping points. This is another task of great urgency for Earth science, as the unraveling of critical transitions associated with tipping points poses great danger for human and environmental welfare. Indeed, one can represent the global stability properties of the climate system as being given by a multistable and multiscale quasi-potential - see Fig.~\ref{multiscale}. The quasi-potential features troughs, saddles, and ridges at different scales and will be characterized by a relatively few large basins (e.g. corresponding to the warm and snowball states), separated by high barriers, decorated by smaller local minima corresponding to hierarchically lower multistability features, e.g. associated to the tipping elements shown in Fig.~\ref{tipping}a. As a result of the existence of large deviation laws describing the invariant measure and the transition probabilities between the competing metastable states, the presence of multiscale features in the quasi-potential makes it hard to perform an accurate exploration of the dynamical landscape of the system for any given choice of the noise intensity: some large scale feature of the dynamical landscape might be unattainable in any reasonable time because the noise is too weak to allow for the system to escape for a given region, whereas, conversely, other smaller scale features might be entirely washed out because the noise is too strong.

 Previous investigations performed in relatively simple yet extremely relevant near-equilibrium physical systems  like the stochastically perturbed two-dimensional Navier-Stokes equations have shown that LDT makes it possible to predict the various competing metastable states \cite{Bouchet2012}; see Fig.~\ref{fig:rare_events_3}. Along these lines, one hopes to be able to make the quasi-potential formalism a more constructive tool, rather than a descriptive one, in order to be able to use it to predict at qualitative and quantitative level the multistability properties of the climate system and of its subsystems. {\color{black}At this regard, one would like to extend the very promising results \cite{grafke2019numerical} obtained on the transitions between zonal and blocked state in the Charney-DeVore model reported in Fig.~\ref{grafkecharney2019} to more complex models of the atmosphere.} A limitation for the numerical calculation of quantities such as minimum action paths and the quasi-potential for complex climate models is the lack of availability of derivatives of the deterministic fields determining the model evolution. Such derivatives are required for example when implementing Hamilton's equation for the instanton, as discussed in Section \ref{dynoise}. The development of climate models in flexible and high-performance programming languages that provide automatic differentiation, as pursued by the SciML \cite{rackauckas2018comparison} and ClimateMachine.jl \cite{clima_jl} projects in the Julia programming language, may soon provide new ways of tackling this problem.

\acknowledgments
VL acknowledges the support received from the EPSRC project EP/T018178/1 and from he EU Horizon 2020 project TiPES (Grant no. 820970). This review is TiPES contribution no. 79. VMG and VL acknowledge the support of the Collaborative Research Centre TRR 181 ‘Energy Transfer in the Atmosphere and Ocean’ funded by the Deutsche Forschungsgemeinschaft (DFG, German Research Foundation, project number 274762653). The authors wish to thank F. Bouchet, D. Faranda, G. Gallavotti, M. Ghil, T. Grafke, A. Laio, B. Hoskins, G. Margazoglou, G. Messori, C. Penland, S. Vaienti, and S. Vannitsem for many scientific exchanges on the topics covered in this paper.

\appendix

\section{Generalizations of large deviation laws: from pair empirical measures to empirical processes}\label{app1}
In the following, we summarize some generalizations of Theorem~\ref{th:cramer} and Theorem~\ref{th:sanov} to higher dimensions, based on \cite{Hollander2000}. We consider the same situation as in case of Theorem~\ref{th:sanov} where $X_i$ takes values in a finite set $\Gamma =\{1,...,r\} \subset \mathbb{N}$, and look at large deviations of pair empirical measures recording two successive values of $X_1, X_2,...$ at each instant of time, $L_n^2=\frac{1}{n}\sum_{i=1}^n \delta_{(X_i,X_{i+1})}$. The random measure $L_n^2$ belongs now to the set of two dimensional probability measures $M(\Gamma \times \Gamma)$, the total variation distance is reformulated accordingly: $d(\mu,\nu)=\frac{1}{2}\sum_{s,t} |\mu_{st}-\nu_{st}|$. The analogue of Theorem~\ref{th:sanov} describes the large deviations of the pair empirical measure $L_n^2$ away from the true measure $\rho^2=\rho \times \rho$, whose componentwise expression is $\rho^2_{st}=\rho_s\rho_t$:

\begin{theorem}\label{th:sanov2}
Let ($X_i$) be i.i.d. random variables satisfying the conditions above, and $L_n^2=\frac{1}{n}\sum_{i=1}^n \delta_{(X_i,X_{i+1})}$ with periodic boundary conditions. Then, the family $(P_n^X)$ defined by $P_n^X(\cdot)=\mathbb{P}^X(L_n^2\in \cdot)$ satisfies the large deviation principle with rate $n$ and with rate function 
\begin{equation}\label{eq:rf2}
    I_\rho^2(\nu)=\sum_{s,t}\nu_{st}\log\left(\frac{\nu_{st}}{\bar{\nu}_s\rho_t}\right) \ := H(\nu|\bar{\nu}\times \rho) ,
\end{equation}
with $\bar{\nu}_s=\sum_t \nu_{st}$.
\end{theorem}

Equation (\ref{eq:rf2}) tells us that the rate function $I^2_\rho(\nu)$ of the pair empirical measure $\nu$ is the relative entropy $H(\nu|\bar{\nu}\times \rho)$ of $\nu$ with respect to $\bar{\nu}\times \rho$. 
Large deviations of pair empirical measures are especially useful if one considers Markov sequences. We stay in the finite state space setting with random variables taking values on a finite set $X_i \in \Gamma =\{1,...,r\} \subset \mathbb{N}$, but this time $X_1, X_2,...$ is Markov with transition matrix $P=(P_{st})_{s,t \in \Gamma}$, $P_{st}>0$ for all $s,t \in \Gamma$. We denote by $\pi=(\pi_s)$ the unique stationary distribution of the Markov chain satisfying $\pi_s>0$, for all $s \in \Gamma$. 

\begin{theorem}\label{th:mc}
Let ($X_i$) be a Markov chain satisfying the conditions above, and $L_n^2=\frac{1}{n}\sum_{i=1}^n \delta_{(X_i,X_{i+1})}$ with periodic boundary conditions. Then, the family $(P_n^X)$ defined by $P_n^X(\cdot)=\mathbb{P}^X(L_n^2\in \cdot)$ satisfies the large deviation principle with rate $n$ and with rate function
\begin{equation}\label{eq:rf3}
    I_P^2(\nu)=\sum_{s,t}\nu_{st}\log\left(\frac{\nu_{st}}{\bar{\nu}_s P_{st}}\right) := H(\nu|\bar{\nu}\otimes P),
\end{equation}
with $\bar{\nu}_s=\sum_t \nu_{st}$.
\end{theorem}
By comparing Theorem~\ref{th:sanov2} with Theorem~\ref{th:mc} becomes clear that large deviations of pair empirical measures are strongly related to those of pair dependence in Markov chains. In particular, note that if we have that $P_{st}=\pi_t$ $\forall s,t\in\Gamma$, \textit{i.e.} all the rows of the transition matrix are identical and are equal to the invariant measure, the Markov chain describes a sequence of i.i.d. random variables distributed according to the measure $\pi$. In this case, indeed, (\ref{eq:rf3}) takes the form of  (\ref{eq:rf2}). Additionally, one can deduce directly large deviation properties of Markov chains from those of i.i.d. sequences based on the Radon-Nikodym formula
\begin{equation}\label{eq:rn}
    \frac{d\mathbb{P}^X}{d\mathbb{P}^Y}[\cdot]=O(1)e^{n F(L_n^2[\cdot])},
\end{equation}
with 
\begin{equation}
    F(\nu)=\sum_{s,t}\nu_{st}\log\left(\frac{P_{st}}{\pi_t}\right), \qquad \nu \in M(\Gamma \times \Gamma),
\end{equation}
where $X$ represents the Markov chain and $Y$ denotes i.i.d. $\Gamma$-valued random variables with distribution $\pi$.  
The rate function in (\ref{eq:rf3}) can be obtained directly from the rate function for the pair empirical measure of $Y$ by using $F(\nu)$: $I_P^2(\nu)=I_\pi^2(\nu)-F(\nu)$.

We have started the discussion of large deviations of Markov sequences with the pair empirical measure since pairs are innate for Markov chains. Nonetheless, the empirical measure $L_n$ can still be relevant since it shows how often the Markov chain visits different points of the sate space. Hence, it is sometimes termed as  the occupation measure. The LDP for the empirical measure can be derived based on the pair empirical measure by contraction, similarly to the case of i.i.d. random variables presented above in Sec.~\ref{sec:theo_iid}.

We assume that $I_P^2$ is finite, continuous and strictly convex. Let $P_n^X(\cdot)=\mathbb{P}(L_n\in\cdot)$. Then $(P_n^X)$ satisfies a LDP on $\mathcal{M}(\Gamma)$ with rate $n$ and with rate function
\begin{equation}
    I_P(\mu)=\inf_{\nu \in \mathcal{M}(\Gamma\times \Gamma):\ \bar{\nu}=\mu} I_P^2(\nu).
\end{equation}
At the end of this section, we present another strategy to derive rate functions for Markov sequences based on the G\"artner-Ellis theorem. 

One can pursue the generalization of the above large deviation laws further in a quite straight-forward way by extending the length of the successive values of $X_1, X_2,...$ to $N$ successive values.
The rate functions obtained in this finite state space setting have the very convenient properties of being finite, continuous and strictly convex. Furthermore, by letting $N\to \infty$ one can derive the rate function of the empirical process.
Due to $N\to \infty$, the mathematically rigorous way to formulate a LDP involves the use of upper and lower limits. 
It is further possible to relax the condition of the finite state space to countable state space. In that case, $\Gamma =\mathbb{N}$, 
and the rate function looses the everywhere finiteness and continuity properties. 
It is clear that Theorem~\ref{th:sanov2} is a generalization of Theorem~\ref{th:sanov}, and the later one is a generalization of Theorem~\ref{th:cramer}. Higher level large deviations laws imply the laws of lower levels, thus one can follow the link back provided by the contraction principle. 

\section{Quadratic approximation of rate functions of time averaged observables}\label{app2}
Here we provide a step by step derivation of the approximate form of the rate function for Gaussian fluctuations. Let us consider the scaled cumulant generating function and autocorrelation time of a process with mean $\mu$, variance $\sigma^2$ and temporal covariance  $C(t,s)=\mathbb{E}[(A(t)-\mu)(A(s)-\mu)]$
\begin{equation}
\lambda(k)=\lim_{T \rightarrow +\infty}\frac{1}{T}\log{\mathbb{E}\left[e^{k\int_0^T A(t)dt} \right]},
\,\,\,\,\,\,\,\,\,\,\,\,\,
\tau_c=\frac{1}{\sigma^{2}}\int_{-\infty}^{+\infty} C(\tau,0)d\tau.
\end{equation}
The scaled cumulant generating function can be rewritten as
\begin{equation}
\lambda(k)=\mu k + \lim_{T \rightarrow +\infty}\frac{1}{T}\log{\mathbb{E}\left[e^{k\int_0^T (A(t)-\mu)dt} \right]}.
\end{equation}
Expanding the exponential around $k=0$ and neglecting terms $O(k^3)$ we have
\begin{equation}
\lambda(k)\approx\ \mu k + \lim_{T \rightarrow +\infty}\frac{1}{T}\log{\mathbb{E}\left[1 + k\int_0^T (A(t)-\mu)dt + \frac{k^2}{2}\left(\int_0^T (A(t)-\mu)dt\right)^2  \right]}.
\end{equation}
Using $\mathbb{E}[A(t)]=\mu$ and rewriting the last term as a double integral on two variables we have
\begin{equation}
\lambda(k)\approx \mu k + \lim_{T \rightarrow +\infty}\frac{1}{T}\log\left[1+ \frac{k^2}{2}\int_0^T\int_0^T \mathbb{E}[(A(t)-\mu)(A(s)-\mu)]dtds\right].
\end{equation}
Assuming again $k$ small enough and expanding the logarithm we obtain
\begin{equation}
\lambda(k)\approx \mu k+  \frac{k^2}{2}\lim_{T \rightarrow +\infty} \frac{1}{T} \int_0^T\int_0^T C(t,s)dtds.
\end{equation}
Since the process is stationary and the covariance is a symmetric function of $t-s$ we have
\begin{equation}
\lim_{T \rightarrow +\infty} \frac{1}{T} \int_0^T\int_0^T C(t,s)dtds=\int_{-\infty}^{+\infty}  C(\tau,0)d\tau=\sigma^2\tau_c,
\end{equation}
and the scaled cumulant generating function is eventually
\begin{equation}
\lambda(k)\approx \mu k+\frac{\sigma^2\tau_c}{2} k^2.
\end{equation}\label{approximation_SCGF}
Taking the Legendre transform we obtain the equivalent approximation for the rate function
\begin{equation}
I(a)\approx\frac{(a-\mu)^2}{2\sigma^2\tau_c}.
\end{equation}\label{approximation_rate_function}
From the way we derived the approximation it is easy to see that $\sigma^2\tau_c k^2/2 < 1/T$ is a necessary condition on the values of $k$ given a value of $T$ for the approximation to hold. The corresponding condition on the values of $a$ can be obtained using the approximate form of $\lambda(k)$ in the solution of the variational problem involved in the Legendre transformation, $a=\lambda'(k)$, that gives $a=\mu+\sigma^2\tau_c k$. The condition then becomes
\begin{equation}
\frac{|a-\mu|}{\sqrt{2\sigma^2\tau_c}}<\frac{1}{\sqrt{T}}\label{CLTstoc},
\end{equation}
consistently with the expected behaviour of Gaussian fluctuations.

\end{document}